\documentclass[a4paper,11pt]{article}

\usepackage{geometry}										
\usepackage[english]{babel}									
\usepackage[utf8]{inputenc}									
\usepackage[T1]{fontenc}									

\usepackage{lmodern}										

\usepackage[normalem]{ulem} 
\usepackage{xcolor}
\usepackage{siunitx}
\usepackage{placeins} 
\usepackage{bbold} 
\usepackage{algorithm}
\usepackage{algpseudocode}
\usepackage{graphicx}

\usepackage{amsmath}										
\usepackage{amstext}										
\usepackage{amsfonts}										
\usepackage{amssymb}										

\usepackage{authblk}										

\title{Automatic selection of basis-adaptive sparse polynomial chaos expansions for engineering applications}
\author[1]{Nora L\"uthen}
\author[1]{Stefano Marelli}
\author[1]{Bruno Sudret}
\affil[1]{Chair of Risk, Safety and Uncertainty Quantification, ETH Z\"{u}rich, Stefano-Franscini-Platz 5, 8093 Z\"{u}rich, Switzerland}

\date{\today}

\usepackage[format=plain, labelfont={bf,it},textfont=it]{caption}	

\usepackage{natbib} 
\usepackage{hyperref}
\hypersetup{
	pdftitle = {Automatic selection of basis-adaptive sparse polynomial chaos expansions for engineering applications},
	pdfauthor = {Nora L\"uthen, Stefano Marelli, Bruno Sudret},
	pdfsubject = {Manuscript submitted to International Journal of Uncertainty Quantification},
	pdfkeywords = {uncertainty quantification, surrogate modelling, sparse regression, sparse polynomial chaos expansions, compressive sensing, experimental design, literature review, basis adaptivity, meta-selection},
	colorlinks = true,										
	linkcolor={red!90!black},
	citecolor={green!50!black},
	urlcolor={red!50!black}
}

\usepackage{subcaption}
\usepackage{enumitem} 
\usepackage{siunitx}
\usepackage{placeins} 
\usepackage{amsfonts, amssymb}
\usepackage{bbold} 
\usepackage{algorithm}[0.1]
\usepackage{algpseudocode}
\newcommand{\Graycomment}[1]{ \hfill \textcolor{gray}{  $\triangleright$ #1} }
\usepackage{tablefootnote} 

\newcommand{\ve}[1]{\boldsymbol{#1}}			
\newcommand{\cA}{\mathcal{A}}
\newcommand{\cd}{{\mathcal D}}
\newcommand{\cm}{{\mathcal M}}
\newcommand{\cx}{{\mathcal X}}

\newcommand{\Nn}{{\mathbb N}}
\newcommand{\Rr}{{\mathbb R}}
\newcommand{\Espe}[2]{{\mathbb E}_{#1}\left[#2\right]}
\newcommand{\Vare}[2]{{\rm Var}_{#1}\left[#2\right]}
\newcommand{\enum}{ , \, \dots \,,}

\newcommand{\SPorig}{$\text{SP}_{k=5}${}}
\newcommand{\SPloo}{$\text{SP}_\text{LOO}${}}
\newcommand{\LtwofX}{{L^2_{f_{\ve X}}(\cd_{\ve X})}}
\newcommand{\norme}[2]{\left\| #1 \right\|_{#2}}
\newcommand{\alp}{{\ve{\alpha}}}

\begin{document}
\maketitle

\begin{abstract}
		Sparse polynomial chaos expansions (PCE) are an efficient and widely used surrogate modeling method in uncertainty quantification for engineering problems with computationally expensive models. 
		To make use of the available information in the most efficient way, several approaches for so-called basis-adaptive sparse PCE have been proposed to determine the set of polynomial regressors (``basis'') for PCE adaptively.
		
		The goal of this paper is to help practitioners identify the most suitable methods for constructing a surrogate PCE for their model. We describe three state-of-the-art basis-adaptive approaches from the recent sparse PCE literature and conduct an extensive benchmark in terms of global approximation accuracy on a large set of computational models.
		Investigating the synergies between sparse regression solvers and basis adaptivity schemes, we find that the choice of the proper solver and basis-adaptive scheme is very important, as it can result in more than one order of magnitude difference in performance. 
		No single method significantly outperforms the others, but dividing the analysis into classes (regarding input dimension and experimental design size), we are able to identify specific sparse solver and basis adaptivity combinations for each class that show comparatively good performance.
			
		To further improve on these findings, we introduce a novel solver and basis adaptivity selection scheme guided by cross-validation error. We demonstrate that this automatic selection procedure provides close-to-optimal results in terms of accuracy, and significantly more robust solutions, while being more general than the case-by-case recommendations obtained by the benchmark. 
		
\end{abstract}

\section{Introduction}
Surrogate modeling techniques are a popular tool in applied sciences and engineering, because they can significantly reduce the computational cost of uncertainty quantification analysis for costly real-world computational models. Here, the computational model is approximated by a cheaper-to-evaluate function, which is created based on a small number of model evaluations, the so-called experimental design.
One well-known and popular surrogate modeling technique is polynomial chaos expansion (PCE), which approximates the output of a computational model by a spectral expansion in terms of an orthonormal polynomial basis in the input random variables \citep{Xiu2002}. PCE is particularly well suited for surrogating smooth models in low to medium dimension, and for the efficient computation of moments and Sobol' indices \citep{SudretRESS2008b,SudretHandbookUQ}.
Engineering models are often challenging due to their computational cost: complex models often depend on a large number of input parameters, but we can only afford a few tens or hundreds of model evaluations. This is the so-called low-data regime.
Sparse PCE techniques, which aim to compute an expansion involving only few terms, have proven especially powerful and cost-efficient for real-world engineering problems
such as, among many others, 
surrogate-assisted robust design optimization \citep{Chatterjee2019},
hybrid simulation for earthquake engineering \citep{AbbiatiMSSP2021},
dam engineering \citep{Guo2019, Harari2020},
and wind turbine design \citep{SlotWind2020}.
Note that real-world applications are typically not exactly sparse; however, sparse regression-based PCE is a useful tool to find good approximations at low computational cost. The aim in this context is not sparsity, but the accurate approximation of the computational model. In particular, we are interested in global approximation accuracy, as measured by the relative mean-squared error. Applications with specific requirements on accuracy, such as optimization or reliability analysis, might rely instead on more specialized surrogate techniques.

In the last decade, a large amount of literature on sparse PCE has been published, proposing methods that make sparse PCE more accurate, efficient and applicable to high-dimensional problems.
However, it is often not obvious how well these methods perform when compared to and combined with each other, especially on real-world engineering problems.
In an attempt to fill this gap, the authors recently conducted a literature survey on sparse PCE and a classification of the available methods into a general framework, as well as a benchmark of selected methods on a broad class of computational models \citep{LuethenSIAMJUQ2021}. 
This benchmark extensively compared sparse regression solvers and experimental design sampling schemes, using a fixed 
polynomial basis to focus on the effect of those two classes of methods.
We found that the choice of sparse solver and sampling scheme can make a difference of up to several orders of magnitude in the resulting approximation error, and that different methods are preferable in different regimes of ED size. The performance of solvers and sampling schemes seemed to be mostly independent from one another.

The goal of the present paper is to build on and complement this earlier benchmark by exploring the promising field of \textit{basis-adaptive sparse PCE}, in which the basis is iteratively augmented and refined.
So far, novel basis-adaptive methods have been proposed in isolation and not been compared to one another.
We want to help practitioners choose the most suitable methods for constructing a PCE surrogate for their applications by answering the following questions: (1) Is there a significant difference between different combinations of sparse solvers and basis adaptivity strategies, and does the proper choice matter in actual applications? (2) In the case that no combination clearly emerges as superior, is there a smart strategy to automatically select a good combination?
To answer the first question, we describe three basis-adaptive schemes from the sparse PCE literature in detail, namely, degree and q-norm adaptivity \citep{BlatmanJCP2011, UQdoc_14_104}, forward neighbor degree adaptivity \citep{Jakeman2015}, and anisotropic degree basis adaptivity \citep{Hampton2018}. 
We then evaluate the performance and synergies of combinations of these basis-adaptive schemes with several sparse regression solvers in terms of validation error%
\footnote{We focus on global approximation accuracy instead of the computational cost of training, since the main cost in surrogate modeling for engineering applications is typically incurred by the model evaluations themselves. 
Indeed, the training of each of the surrogates presented in this work requires between a few seconds and a few minutes on a standard business laptop. Even orders of magnitude difference in the training costs of each surrogate do not matter as long as the resulting accuracy is appreciably better.}
on a library of 11 computational models of varying complexity, representative of a broad class of engineering models. These range from three- to 100-dimensional and include both analytical and numerical models. 
Since no combination of solver and basis adaptivity significantly outperforms the others, we address the second question by introducing
an additional selection step, choosing one among several candidate PCEs computed by different combinations of methods on the same experimental design, using a cross-validation estimate for the generalization error.

The paper is structured as follows.
In Section \ref{sec:theory}, we recall the definition of (sparse) PCE and the computing framework introduced in \citet{LuethenSIAMJUQ2021}. 
We discuss various estimators for the generalization error%
, sparse regression solvers%
, and basis adaptivity%
, and introduce automatic selection.
The associated benchmark results for basis adaptivity and automatic selection are presented in Section \ref{sec:numerical_results}. 
Finally, a discussion and summary of our results is provided in Section \ref{sec:conclusion}.
Additional information and results can be found in the Appendix.

\section{Sparse polynomial chaos expansions}
\label{sec:theory}

Let $\ve X$ be a $d$-dimensional random vector with mutually independent components and probability density function $f_{\ve X}$. Denote by $\cd_{\ve X}$ the domain of the random vector $\ve X$. 
Define the space $\LtwofX = \{h:\cd_{\ve X} \to \Rr \ | \ \Vare{\ve X}{h(\ve X)} < +\infty \}$ of all scalar valued models with finite variance under $f_{\ve X}$.
Under very general conditions on the input distribution $f_{\ve X}$, there exists a polynomial basis $\{\psi_\alp:\alp \in \Nn^d\}$ of $\LtwofX$ \citep{Xiu2002, Ernst2012}. 
Since the components of $\ve X$ are mutually independent, each polynomial basis function can be built as a product of univariate polynomials in $X_1 \enum X_d$ and characterized by the multi-index $\alp \in\Nn^d$ whose entries are equal to the degrees of the univariate terms.

For a computational model $\cm \in \LtwofX$, let $Y = \cm(\ve X)$ denote the output random variable. $Y$ can be cast as the following spectral expansion:
	\begin{equation}
	Y = \cm(\ve X) = \sum_{\alp \in \Nn^d} c_\alp \psi_\alp(\ve X).
	\end{equation}
In practice, a finite, \textit{truncated polynomial chaos expansion}
\begin{equation}
Y = \cm(\ve X) \approx \cm^\text{PCE}(\ve X) = \sum_{\alp \in \cA} c_\alp \psi_\alp(\ve X)
\end{equation}
is computed, 
where $\cA \subset \Nn^d$ is the truncation set defining the basis elements used in the expansion. The accuracy of the expansion depends on $\cA$ and the coefficient values $(c_\alp)_{\alp \in \cA} =: \ve c \in \Rr^P$, with $P = \text{card}(\cA)$.

Among several methods for computing the coefficient vector $\ve c$ for a given truncation set, \textit{sparse regression} is a particularly powerful and efficient method \citep{Doostan2011, BlatmanJCP2011}. In this approach, the model is evaluated at a number of points $\ve\cx = \{\ve x^{(1)} \enum \ve x^{(N)} \} \subset \cd_{\ve X}$ called the \textit{experimental design} (ED), yielding the vector of model responses $\ve y = (\cm(\ve x^{(1)}) \enum \cm(\ve x^{(N)}))^T$. Let $\{\alp_j\}_{j=1}^P$ be an arbitrary ordering of the multi-indices in the truncation set and define the regression matrix $\ve \Psi \in \Rr^{N\times P}$ with entries $\Psi_{ij} = \psi_{\alp_j}(\ve x^{(i)})$. Sparse regression methods determine a vector $\ve c$ that minimizes the residual norm $\norme{\ve\Psi \ve c - \ve y}{2}$ under the constraint that it has only few nonzero entries i.e., it is \textit{sparse}. 
This is usually achieved by regularized regression, resulting e.g.\ in the LASSO formulation
\begin{equation}
\hat{\ve c} = \min_{\ve c \in \Rr^P} \norme{\ve\Psi\ve c - \ve y}{2}^2 \quad \text{s.t. } \ \norme{\ve c}{1} \leq \tau,
\label{eq:LASSO}
\end{equation}
where $\tau$ is a parameter regulating the sparsity of $\ve c$.
A PCE with a sparse coefficient vector $\ve c$ is called \textit{sparse PCE}.
Provided that the regression matrix fulfills certain properties \citep{Candes2008a,Bruckstein2009, Candes2011}, sparse regression can find robust solutions to underdetermined systems of linear equations, which means that the experimental design can be smaller than the number of unknown coefficients.

\begin{figure}
	\centering
	\includegraphics[width=.5\textwidth]{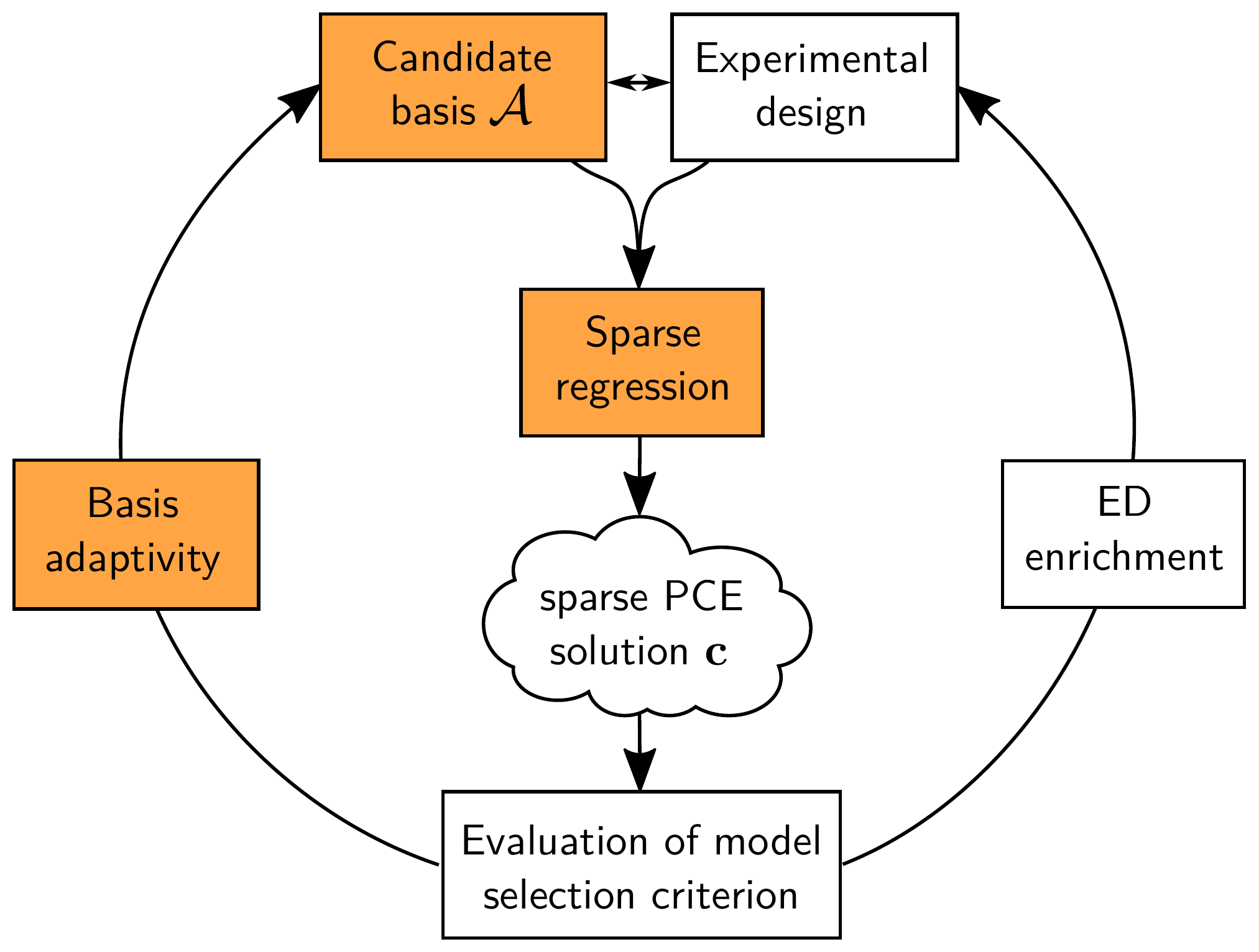}
	\caption{Framework for computing sparse PCE introduced by \citet{LuethenSIAMJUQ2021}, who conducted a literature survey and a benchmark for the central components ``Experimental design'' and ``Sparse regression'' (adapted from \citet{LuethenSIAMJUQ2021}). In the present work, we discuss and benchmark the components marked in orange. In particular, we discuss the component ``Basis adaptivity'' and explore its relationship to sparse regression solvers.}
	\label{fig:framework}
\end{figure}

The quality of the solution depends on the choice of the basis $\cA$, on the experimental design $\ve\cx$, and on the method used for computing the coefficients. For each of these components, many different methods have been proposed in recent years, including iterative methods which adaptively determine $\cA$, or the experimental design $\ve\cx$. 
These methods were recently surveyed and classified into the framework shown in Fig.~\ref{fig:framework} (modified from \citet{LuethenSIAMJUQ2021}). 
In the present contribution, we focus on the question of how to determine a suitable basis $\cA$. To this end, we benchmark several basis-adaptive approaches and explore their interplay with sparse regression solvers.

\subsection{Error estimation and model selection}
\label{sec:model_selection}
Model selection is applied on several levels in the sparse PCE framework:
\begin{itemize}
	\item Within the sparse solver to select its hyperparameter (see Section~\ref{sec:solvers})
	\item Within the basis adaptivity scheme to select a basis (see Section~\ref{sec:basis_adaptive_schemes})
	\item Finally, between solvers and basis adaptivity schemes, to automatically select a combination that is close to best (see Section~\ref{sec:autom_selection}).
\end{itemize}
Our main quantity of interest is the generalization error, in the form of the relative mean squared error normalized by the model variance
\begin{equation}
E_\text{gen} = \frac{\Espe{\ve X}{(\cm(\ve X) - \cm^\text{PCE}(\ve X))^2}}{\Vare{\ve X}{\cm(\ve X)}}.
\label{eq:generalization_error}
\end{equation}
It can be approximated by the validation error in the form of the relative mean squared error (MSE)
\begin{equation}
\text{RelMSE} = 
\frac{\sum_{i = 1}^{N_\text{val}} \left(\cm(\ve x^{(i)}_\text{val}) - \cm^\text{PCE}(\ve x^{(i)}_\text{val})\right)^2 }
{\sum_{i = 1}^{N_\text{val}} \left(\cm(\ve x^{(i)}_\text{val}) 
	- \frac{1}{N_\text{val}} \sum_{j=1}^{N_\text{val}} \cm(\ve x^{(j)}_\text{val})\right)^2 }
\label{eq:relMSE}
\end{equation}
computed on a large validation set $\{\ve x^{(i)}_\text{val}\}_{i=1}^{N_\text{val}} \sim_\text{i.i.d.} f_{\ve X}$.
We only consider model selection criteria that are estimates of the generalization error.

To get an accurate estimate of the generalization error, without using an additional validation set, a widely used method is \textit{cross-validation} \citep{Vapnik:1995}. 
Here, the available data is repeatedly divided into two disjoint sets, one for computing the solution (training) and the other for evaluating the error (validation). Aggregating the error estimates from the different splits, we get a cross-validation estimate of the generalization error.

One cross-validation strategy is to divide the data randomly into $k$ disjoint, roughly equally large parts, and use each of the parts in turn as validation set. This is called \textit{$k$-fold cross-validation}. 
If $k$ is chosen to be equal to the size of the experimental design, the strategy is known as \textit{leave-one-out cross-validation} (LOO). This is closest to using all data points to compute a solution, since in each iteration, only one point is left out from the training set. 

In general, LOO can be quite costly, since for an experimental design of size $N$, the method under consideration has to be applied $N$ times. 
A cheap approximation to the LOO error, which requires only one application of the method, is available for certain sparse regression solvers, namely for those which in their last step recompute the solution with ordinary least-squares (OLS) on the set of regressors with nonzero coefficient (called \textit{active basis}) \citep{BlatmanPEM2010}. In particular, this is the case for the solvers hybrid least angle regression (LARS) \citep{BlatmanJCP2011}, orthogonal matching pursuit (OMP) \citep{Pati1993, Jakeman2015}, and subspace pursuit (SP) \citep{Diaz2018} (see Section \ref{sec:solvers}).
For these solvers, the OLS-based LOO estimate coincides with the true LOO error if the following holds: regardless of which experimental design point is left out, the sparse regression solver consistently yields the same active basis.

Since the repeated use of LOO for model selection often results in the generalization error being underestimated, especially on small experimental designs, \citet{BlatmanJCP2011} proposed to use a modification factor originally developed for the empirical error \citep{Chapelle2002}, defined by 
\begin{equation}
T = \frac{N}{N-P_\text{active}}\left(1 + \mathrm{tr}((\ve\Psi_\text{active}^T\ve\Psi_\text{active})^{-1})\right),
\label{eq:modification_factor}
\end{equation}
where $P_\text{active}$ denotes the number of nonzero coefficients (the corresponding basis functions are called \textit{active}), and $\ve\Psi_\text{active}$ denotes the regression matrix containing only the active regressors. The product of the modification factor $T$ with the LOO error is called \textit{modified LOO error}.

\subsection{Sparse regression solvers}
\label{sec:solvers}
Various sparse regression solvers available for solving sparse PCE were described in detail in \citet{LuethenSIAMJUQ2021}. We give here only a short overview of the solvers used in our benchmark, 
and refer to \citet{LuethenSIAMJUQ2021} for further details.
These solvers are common choices in the sparse PCE literature. 
\begin{itemize}
	\item Hybrid least angle regression (LARS) \citep{BlatmanJCP2011}: adding regressors one-by-one, following a least-angle strategy. The hybrid approach recomputes the final coefficient values by ordinary least squares (OLS) on the selected regressors.
	\item Orthogonal matching pursuit (OMP) \citep{Pati1993,Jakeman2015}: greedily adding orthogonal regressors one-by-one,
	computing their coefficients by OLS.
	\item Subspace pursuit (SP) \citep{Diaz2018}: searching iteratively for a solution with a certain $\ell^0$-norm by adding and removing regressors from the active set. Coefficients are computed by OLS.
	We include two variants of SP in this benchmark, one which determines its hyperparameter by 5-fold cross-validation similar to the implementation in \citep{DOPTPCE}, which we denote by \SPorig{}, and one where it is determined by OLS-based LOO, introduced in \citep{LuethenSIAMJUQ2021} and denoted by \SPloo{}.
	\item Bayesian compressive sensing (BCS) \citep{Babacan2010}: using a Bayesian framework to enforce sparsity of the coefficients.
	\item SPGL1 \citep{Vandenberg2008}: a convex optimization solver following the Pareto front of the residual-sparsity trade-off.
\end{itemize}
Each of the solvers features at least one hyperparameter whose value needs to be determined via cross-validation in order to get a good solution. For LARS, OMP, \SPorig{}, and \SPloo{}, this hyperparameter is the number of active basis functions (nonzero coefficients) of the final solution. For BCS and SPGL1, it is the bound on the residual norm in the sparse regression formulation.

The benchmark in \citet{LuethenSIAMJUQ2021} of sparse regression solvers on a non-adaptive polynomial basis showed that BCS and \SPloo{} generally outperform other sparse solvers for low-dimensional models, while for high-dimensional models, BCS is by far the best sparse solver.

\subsection{Basis adaptivity}
\label{sec:basis_adaptive_schemes}
The sparse solver and the experimental design, which were benchmarked in \citep{LuethenSIAMJUQ2021}, are not the only ingredients to a sparse PCE. 
The choice of the \textit{candidate basis}, from which the sparse solver determines the \textit{active basis} (i.e., the set of regressors with nonzero coefficient), is another important ingredient: 
if the candidate basis is chosen too small, important terms might be missing, which might lead to a large model error. On the other hand, if the candidate basis is too large, the ratio of the number of experimental design points to the number of coefficients is small, which causes some properties of the regression matrix to deteriorate and can result in a large approximation error.

Of course, it is not possible to know a-priori the best choice of the candidate basis. 
Basis-adaptive schemes start with an initial candidate basis and adapt it iteratively, adding or removing basis functions in each iteration according to various heuristics. The performance of the bases in the different iterations is evaluated using a model selection criterion, typically an estimate of the generalization error.

Several procedures for basis-adaptive sparse PCE have been proposed in the literature. 
We discuss three approaches in detail, namely 
\begin{itemize}
	\item degree and q-norm (``p\&q'') basis adaptivity, as implemented in UQLab \citep{UQdoc_14_104}, see Section~\ref{sec:degree&qnorm_adaptivity};
	\item forward neighbor basis adaptivity \citep{Jakeman2015}, see Section~\ref{sec:forward_neighbor_BA}; and 
	\item anisotropic degree basis adaptivity \citep{Hampton2018}, from the software BASE\_PC \citep{BASEPC}, see Section~\ref{sec:anisotropic_degree_BA}.
\end{itemize}
We also briefly mention several other approaches found in the literature.

\subsubsection{Degree and q-norm (``p\&q'') adaptivity}
\label{sec:degree&qnorm_adaptivity}
A typical choice for a PCE basis is the \textit{basis of total degree $p$} defined by the set of multi-indices
\begin{equation}
\cA^p = \{\alp \in \Nn^d: \sum_{i=1}^d \alpha_i \leq p\}.
\label{eq:total_degree}
\end{equation}
Furthermore, \textit{hyperbolic truncation} \citep{BlatmanJCP2011} uses the q-(quasi-)norm
\begin{equation}
\norme{\ve x}{q} = \left( \sum_{i=1}^d |x_i|^q\right)^{\frac{1}{q}}
\label{eq:qnorm}
\end{equation}
with $q \in (0,1]$ to truncate a total-degree basis further:
\begin{equation}
\cA^{p,q} = \{\alp \in \Nn^d: \norme{\alp}{q} \leq p \}.
\label{eq:hyperbolic_truncation}
\end{equation}
Hyperbolic truncation has the effect of excluding some of the basis functions with high degree and high interaction order. This is particularly effective for high-dimensional problems.

A simple basis-adaptive scheme is \textit{degree adaptivity} \citep{BlatmanJCP2011}, which computes a number of PCEs on a sequence of total-degree candidate bases of increasing size, and returns the PCE minimizing a certain error estimate as final solution.
Analogously, a q-norm adaptive scheme can be developed, and easily be combined with degree adaptivity \citep{UQdoc_14_104}, yielding \textit{degree and q-norm (p\&q) adaptivity}.
Degree and q-norm adaptivity is solution-agnostic, i.e., it does not take the solution computed in the previous iteration into account.

\subsubsection{Forward neighbor basis adaptivity}
\label{sec:forward_neighbor_BA}
\begin{figure}[htbp]
	\centering
	\includegraphics[width=.5\textwidth]{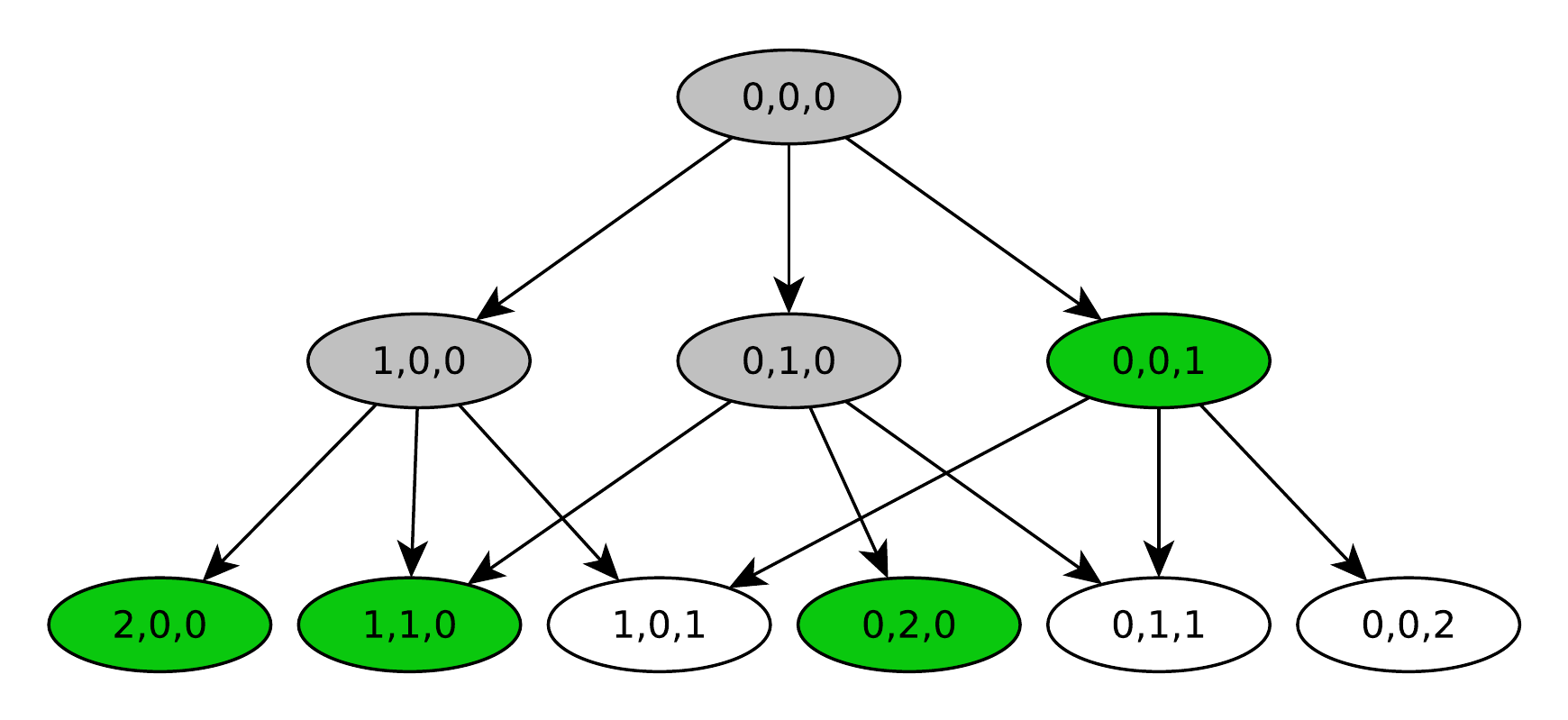}
	\caption{Illustration of forward neighbor basis adaptivity \citep{Jakeman2015} in $d = 3$ dimensions. The regressors are seen as nodes in a graph, where a directed edge connects regressors whose multi-index differs by 1 in exactly one dimension. Assume that the set of active regressors $\cA^\text{active}$ initially consists of the gray nodes. All \textit{admissible} nodes are highlighted in green. Together with the gray nodes, they constitute the set $\cA^{(1)}$ and would be added to the model in the next expansion step of forward neighbor basis adaptivity (Step 3 in Algorithm~\ref{alg:FN-BA}).}
	\label{fig:forward_neighbor}
\end{figure}

\citet{Jakeman2015} suggest a basis-adaptive algorithm based on a graph view of the PCE regressors (see also \citet{Gerstner:Griebel:2003, Narayan2014}). 
Since the input random vector is assumed to have independent components, the basis functions have tensor product structure. The basis functions can be considered nodes of a directed graph constructed as follows (see also Fig.~\ref{fig:forward_neighbor}): two regressors are considered neighbors if their multi-index of degrees differs only in one dimension by 1, i.e., there is a directed edge from $\psi_\alp$ to $\psi_{\ve \beta}$ iff $\ve\gamma :=\ve \beta - \ve \alpha$ is a multi-index with $\gamma_i = 1$ for one $i\in \{1\enum d\}$ and $\gamma_j = 0, j \neq i$. 
\textit{Forward neighbors} of a regressor are regressors reachable by an outgoing edge, and \textit{backwards neighbors} are regressors connected by an incoming edge.

In the context of basis-adaptive PCE, this construction is used to determine a number of candidate regressors to be added to the currently active basis, starting from an initial basis of small total degree.
Assume that an important high-degree or high-order regressor is not yet part of the candidate basis but some of its backwards neighbors are.
The fact that it is missing should be visible in the coefficients of its backwards neighbors, which can be expected to have a significant nonzero coefficient to compensate for the fact that the important high-degree regressor is not yet part of the basis.

This heuristic is the foundation of the algorithm whose pseudocode is summarized in Algorithm~\ref{alg:FN-BA}.
In each iteration, the current set of active basis functions is determined (restriction step). 
All forward neighbors of these active basis functions are surveyed and added to the candidate basis if they are \textit{admissible}, i.e., if all of their backwards neighbors are in the active set (expansion step).
\citet{Jakeman2015} employ several expansion steps to prevent premature termination, and use cross-validation error estimates.
We call this algorithm \textit{forward neighbor basis-adaptive}.

\begin{algorithm}
	\caption{Forward neighbor basis adaptivity \citep{Jakeman2015}}
	\label{alg:FN-BA}
	\begin{algorithmic}[1]
		\State Initial PCE (basis chosen a-priori, typically small total-degree basis)
		\State Restriction: retain only the active regressors $\cA^\text{active}$ \label{enum:restrict}
		\State Expansion: let $\cA^{(0)} = \cA^\text{active}$. For $t = 1 \enum T$, obtain the set $\cA^{(t)}$ by augmenting $\cA^{(t-1)}$ by all its admissible forward neighbors. 
		\label{enum:expand}
		\State Compute a PCE and its error estimate for each augmented basis $\cA^{(1)} \enum \cA^{(T)}$.
		\State Choose the PCE with the lowest error estimate among the $T$ candidates. Stop if this error estimate is larger than the previously obtained best error estimate. Else, continue iteration with restriction step (Step \ref{enum:restrict})
	\end{algorithmic}
\end{algorithm}

This algorithm is implemented in the software Dakota \citep{Dakota2015}. 
We use our own Matlab-based implementation of the algorithm.
Consistently with \citet{Jakeman2015}, we set $T = 3$.

\subsubsection{Anisotropic degree basis adaptivity}
\label{sec:anisotropic_degree_BA}
\citet{Hampton2018} propose an algorithm called BASE-PC, which combines a basis-adaptive scheme with sequential enrichment of a coherence-optimal experimental design. The two steps are executed alternatingly: based on the current ED, a suitable basis is chosen; and according to the currently active basis functions, the ED is enriched and the weights are updated.

In the present work, we solely consider the basis adaptivity part of the BASE-PC algorithm.
The main idea of this algorithm is to use an \textit{anisotropic degree basis}, defined by a \textit{degree vector} $\ve p \in \Nn^d$. 
A related idea was explored earlier by \citet{BlatmanICOSSAR2009}. 
Similarly to a total-order basis, an anisotropic degree basis is defined by
\begin{equation}
\cA^{\ve p} = \{\alp \in \Nn^d: \sum_{i = 1}^d \frac{\alpha_i}{p_i} \leq 1\}.
\label{eq:anisotropic_degree}
\end{equation}
If all entries of $\ve p$ are the same, i.e., $p_1 = p_2 = \ldots = p_d = p$, this definition reduces to a total-order basis of degree $p$.
The equation $\sum_{i = 1}^d \frac{\alpha_i}{p_i} = 1$ defines a hyperplane that cuts the $i$-th coordinate axis at $\alpha_i = p_i$. 

In each iteration, the algorithm determines the current anisotropic degree vector based on the currently active basis. A new, larger candidate basis is then constructed by considering the anisotropic degree basis corresponding to a uniformly increased anisotropic degree vector.

We use our own, slightly modified implementation of the basis adaptivity part of BASE-PC, as summarized in Algorithm \ref{alg:AD-BA}. 
The most costly operation is the computation of the anisotropic-degree basis (line~\ref{enum:BASEPCexpand}). \citet{Hampton2018} developed a specialized efficient algorithm to generate the multi-indices of an anisotropic degree basis, which we utilize in our implementation.
We call Algorithm~\ref{alg:AD-BA} \textit{anisotropic degree basis-adaptive}. 

\begin{algorithm}
	\caption{Anisotropic degree basis adaptivity \citep{Hampton2018}}
	\label{alg:AD-BA}
	\begin{algorithmic}[1]
		\State Initial PCE (fixed basis of low order)
		\For{$o = 1 \enum 10$}
		\Graycomment{outer loop}
		\State Restriction: denote by $\cA^\text{active}$ the set of active regressors of the last selected PCE 
		\label{enum:BASEPCrestrict}
		\For{$i = 1 \enum 10$} \label{enum:BASEPCloop}
		\Graycomment{inner loop}
		\State Additional restriction: remove $\frac{i-1}{10}$ regressors from the set $\cA^\text{active}$ starting with the ones with smallest-in-magnitude coefficients, obtaining a set $\cA^i \subseteq \cA^\text{active}$
		\label{enum:BASEPCaddrestrict}
		\State Expansion: compute the dimension-wise maximal degree of the regressors in $\cA^i$, denoted by degree vector $\ve{p}^\text{max}$.
		\label{enum:BASEPCmaxdegree}
		Compute $\ve{p}^\text{new} = (p^\text{max}_1 + 1, p^\text{max}_2 + 1 \enum p^\text{max}_d + 1)$ and the associated anisotropic-degree basis $\cA^{i, \text{new}}$
		\label{enum:BASEPCexpand}
		\State Compute a PCE on the basis $\cA^{i, \text{new}}$ and its error estimator $e^i$
		\State If $e^i \geq e^{i-1}$, increase the so-called \textit{strike counter} by 1. Break from the inner loop if the strike counter is $\geq 3$
		\EndFor
		\State From the $10$ candidate PCEs, select the PCE with the lowest error estimate 
		\State Break from the outer loop if the error estimate has increased three times in subsequent iterations (ExpansionEarlyStop; same idea as inner loop strike counter)
		\label{enum:BASEPCstop}
		\EndFor
		\State Return the PCE with the lowest error estimate among all PCEs selected in the outer loop.
		\label{enum:BASEPCreturn}
	\end{algorithmic}
\end{algorithm}

\subsubsection{Other basis adaptivity algorithms}
We briefly summarize other approaches for basis adaptivity for sparse PCE.
These approaches will not be investigated in our benchmark.

Many algorithms which use stepwise regression to build up a sparse basis regressor-by-regressor can be classified both as sparse regression solvers (as done in \citet{LuethenSIAMJUQ2021}) and as basis-adaptive algorithms.
As an example, the approach by \citet{BlatmanPEM2010} adds and removes regressors one-by-one based on the induced change in LOO error, thereby building up a small set of active basis functions which is sparse in a larger total-degree basis. In this case, the candidate basis coincides with the active basis.
\citet{MaiUNCECOMP2015} use the ``principle of heredity'' together with LARS to determine additional bivariate interaction terms to be added to the model, once a univariate term is identified as relevant.
We do not consider these approaches, since we are interested in algorithms modifying the candidate basis not only one regressor at a time, but at a larger scale.

\citet{Alemazkoor2017} propose a basis-adaptive algorithm relying on sparsity and step-by-step augmentation of the basis. In each step, the candidate basis is a total-order basis for a subset of input random variables, while the remaining input random variables are considered constant.
The initial basis consists only of the constant term. In each iteration, either one of the constant dimensions is activated, or the total degree of the candidate basis is increased by one (active dimensions only), depending on the resulting error estimate or sparsity of the solution. 
The coefficients are solved by OLS until a pre-defined threshold of residual norm is reached. Then, the sparse regression solver SPGL1 is used, which identifies the sparsest solution whose residual norm is smaller than the given threshold.
We do not consider this method due to its high computational cost and its strong tie to the solver SPGL1, making it less effective when paired with other sparse regression solvers.

\citet{Loukrezis2019} propose a basis-adaptive algorithm for interpolating PCE on Leja-based sparse grids. Their algorithm is a more cautious version of forward neighbor basis adaptivity (Section~\ref{sec:forward_neighbor_BA}): after adding all admissible forward neighbors to the basis and computing their coefficients, all of the recently added regressors are removed again, except for the one that achieved the largest-in-magnitude coefficient.
\citet{Thapa2020} suggest a basis-adaptive procedure that relies on total-order bases of increasing degree. In contrast with degree adaptivity (Section~\ref{sec:degree&qnorm_adaptivity}), the basis functions of degree $p+1$ are not added all at once, but in chunks of a certain size dependent on the dimension and the degree $p$. After adding a chunk of basis functions, the PCE is recomputed and regressors with a coefficient smaller than a certain threshold are removed from the basis. 
We do not consider these two approaches because they are similar to the previously presented approaches while being more costly.

\subsection{Automatic selection of a sparse PCE solution from several candidate solutions}
\label{sec:autom_selection}
For realistic simulators used in engineering and applied sciences, evaluating the computational model is the expensive part of the surrogate modeling process. 
Once the model evaluations are obtained, all further computations are post-processing steps, and are computationally cheap in comparison to the model evaluations.
Thus, it is feasible to apply several adaptive sparse PCE methods and choose the best one.

We therefore propose the use of an additional layer of selection which we call here \textit{automatic selection}. It can be seen as a simple case of \textit{ensemble modeling} \citep{Sagi2018} where a single metamodel (``strong learner'') is chosen based on its cross-validation performance -- or, in other words, model selection.
For a given experimental design, we compute several sparse PCEs through different combinations of sparse solvers and basis adaptivity schemes. From the resulting PCE solutions, we choose one based on the value of a suitable estimate of generalization error as model selection criterion.
There are several possibilities.
One possible choice is the estimator already used for selecting the hyperparameter of the solver and for basis adaptivity, i.e., (modified) LOO for LARS, OMP and \SPloo{}, and $k$-fold cross-validation for \SPorig{}, BCS and SPGL1. However, this estimator might not be consistent between solvers, which is however necessary for such an automatic selection. Furthermore, this estimate might be biased due to its repeated use. 

A second class of estimators is given by the so-called \textit{hybrid cross-validation error estimators}. The word \textit{hybrid} is a reference to \citet{Efron2004}, who created the hybrid version of least-angle regression (LARS) which uses LARS only to identify the set of active basis functions, and then recomputes the coefficients by ordinary least-squares (OLS) on this active set. 
To compute the hybrid leave-one-out (LOO) or hybrid $k$-fold cross-validation error estimate for any PCE solution, the same procedure is used: first, the active basis functions are identified using the whole experimental design. 
Then, the coefficients are recomputed several times using OLS, following the chosen cross-validation framework. This requires solving a linear system of equations $k$ times in the case of $k$-fold cross-validation. In case of LOO, only one linear system of equations needs to be solved \citep{BlatmanPEM2010}.
Furthermore, we can make use of the modification factor of Eq.~\eqref{eq:modification_factor} to compute the hybrid modified LOO error estimate.

As baseline cases, we will also select a solution 1) randomly from a set of generally well-performing methods, and 2) corresponding to the best combination identified in the benchmark of basis adaptivity schemes (Section \ref{sec:BA_benchmark}).

\section{Numerical results}
\label{sec:numerical_results}
In this benchmark, we compare the performance of various combinations of sparse regression solvers with basis-adaptive schemes. We use the following methods and associated implementations.
The sparse solvers (wrapped to fit into our benchmark framework) are: 
\begin{itemize}
	\item Hybrid-LARS: UQLab \citep{UQdoc_14_104, MarelliUQLab2014}
	\item OMP: UQLab \citep{UQdoc_14_104, MarelliUQLab2014}
	\item \SPorig{}: own implementation of \citet{Diaz2018}, available in UQLab \citep{UQdoc_14_104, MarelliUQLab2014}
	\item \SPloo{}: own adaptation of \citet{Diaz2018}, available in UQLab \citep{UQdoc_14_104, MarelliUQLab2014}
	\item BCS: own implementation of FastLaplace \citep{Babacan2010}, available in UQLab \citep{UQdoc_14_104, MarelliUQLab2014}
	\item SPGL1: SPGL1 v1.9%
	\footnote{Our benchmark is performed with SPGL1 v1.9. A new version of SPGL1, v2.1, is available since June 2020. In our tests, the new version (with ``hybrid'' mode) did not perform significantly better than the old version. The numerical results show therefore results for SPGL1 v1.9 with default parameters.}
	\citep{Vandenberg2008,SPGL1}
\end{itemize}
The basis adaptivity schemes are: 
\begin{itemize}
	\item p\&q adaptivity: UQLab \citep{UQdoc_14_104, MarelliUQLab2014}
	\item forward neighbor basis adaptivity: own implementation of the algorithm based on the description in \citet{Jakeman2015}
	\item anisotropic degree basis adaptivity: own implementation of an algorithm adapted from the basis-adaptive part of BASE-PC (v1) \citep{Hampton2018}, using the function \texttt{basis\_anisotropic\_total\_order.m} from \citet{BASEPC} to generate the multi-indices of an anisotropic degree basis (see Section~\ref{sec:anisotropic_degree_BA})
\end{itemize}
To reduce the complexity of our benchmark, we choose Latin hypercube sampling with maximin distance optimization to generate the experimental design (ED) since \citet{LuethenSIAMJUQ2021} demonstrated that the choices of sparse solver and sampling scheme are mostly independent from each other.

We use the following model selection methods:
\begin{itemize}
	\item For the selection of the hyperparameters of the sparse regression solvers, we use
	\begin{itemize}
		\item modified OLS-based LOO for the solvers LARS, OMP and \SPloo{}
		\item $k$-fold CV for the solvers \SPorig{} ($k = 5$), BCS ($k = 10$) and SPGL1 ($k = 10$)
	\end{itemize}
	\item The basis adaptivity schemes use the same criterion as the respective solver uses.
	\item We investigate in Section~\ref{sec:results_final_model_selection} which criterion is suited best for the final model selection.
\end{itemize}
I.e., instead of prescribing fixed values for the hyperparameters, we let the parameters be determined adaptively. In this sense, we give each method equal opportunity to produce the best possible solution (assuming that the CV error is a good proxy for solution quality).

For our benchmark, we use 11 benchmark models ranging from low-dimensional analytical models to high-dimensional differential equations.
All of these models have previously been used as numerical examples for surrogate modeling or reliability analysis. 
None of these models has an exactly sparse representation. Note that we do not aim at benchmarking the ability of the methods to recover ``true'' sparse solutions, but instead their approximation capabilities on engineering models, which are typically not exactly sparse but \textit{compressible}, meaning that the magnitude of their PC coefficients decays rapidly.
Table~\ref{table:models} provides an overview of the benchmark models. For a more detailed presentation, we refer the interested reader to the respective publications (see column "Reference" of Table~\ref{table:models}).

\begin{table}[htbp]
	\footnotesize
	\centering
	\caption{Overview of the 11 computational models used in our benchmark. \textit{Finite element} models are marked in italic font, all other models are analytical. The column ``Reference'' provides the literature in which the models and their input are described in detail. The column ``ED sizes'' contains the two sizes of experimental design (small and large) used in the basis adaptivity benchmark. 
	}
	\label{table:models}
	\renewcommand{\arraystretch}{1.5}
	\begin{tabular}{p{.21\textwidth}c>{\centering}p{.24\textwidth}>{\centering}p{.15\textwidth}>{\centering\arraybackslash}p{.12\textwidth}}
		\hline
		Model & Dimension & Distributions & Reference & ED sizes \newline (small, large)\\
		\hline
		Ishigami function & 3 & uniform & \citet{BlatmanJCP2011} & 50, 150 \\
		Undamped oscillator & 6 & Gaussian & \citet{Echard2013} & 60, 120\\
		Borehole function & 8 & Gaussian, lognormal, uniform & \citet{Morris1993} & 100, 250\\ 
		Damped oscillator & 8 & lognormal & \citet{DubourgThesis} & 150, 350 \\
		Wingweight function & 10 & uniform & \citet{Forrester2008} & 100, 250\\
		\textit{Truss model} & 10 & lognormal, Gumbel & \citet{BlatmanJCP2011} & 100, 250 \\
		\hline
		Morris function & 20 & uniform & \citet{BlatmanRESS2010} & 150, 350 \\ 
		\textit{Structural frame model} & 21 & lognormal, Gaussian; {dependent input variables} & \citet{BlatmanPEM2010} & 150, 350 \\
		\textit{2-dim diffusion model} & 53 & Gaussian & \citet{KonakliRESS2016} & 100, 400 \\
		\textit{1-dim diffusion model} & 62 & Gaussian & \citet{FajraouiMarelli2017} & 100, 400 \\
		100D function & 100 & uniform & UQLab example\tablefootnote{\url{https://www.uqlab.com/sensitivity-high-dimension}} & 400, 1200\\
		\hline	
	\end{tabular}
\end{table}

\subsection{Basis adaptivity}
\label{sec:BA_benchmark}
We benchmark the sparse solvers LARS, OMP, \SPorig{}, \SPloo{}, BCS, and SPGL1 combined with the basis-adaptive schemes described in Section~\ref{sec:basis_adaptive_schemes}:
\begin{enumerate}
	\item degree and q-norm (p\&q) adaptivity (abbreviation: PQ)
	\item forward neighbor basis adaptivity (FN)
	\item anisotropic degree basis adaptivity (AD)
\end{enumerate}
As a base case, we include a \textit{static basis} following the rule $P \approx \frac{10}{3}N$  (abbreviation: ST), where we use hyperbolic truncation with $q=0.5$ for high-dimensional models ($d \geq 20$).

The benchmark is performed on all 11 models presented in Table~\ref{table:models}.
The experimental design (ED) is created by Latin hypercube sampling (LHS) with optimized maximin distance. 
We investigate one ``small'' and one ``large'' ED size per model (see last column of Table~\ref{table:models}), which correspond to the second-smallest and second-largest experimental design size, respectively, of an earlier benchmark \citep{LuethenSIAMJUQ2021} dedicated to investigating solvers and sampling schemes on a static basis. 
The small ED size is chosen to be at the lower end of the range of reasonable ED sizes. The large ED size represents the ``highly informative'' regime (in the engineering sense), which for costly engineering models is in the order of a few hundred model evaluations.
Here we restrict ourselves to two ED sizes to control the complexity of the results; however, two ED sizes are needed since the earlier benchmark showed that the solver behavior is different in the two regimes.
Since almost no engineering model is exactly sparse in the PCE basis, adaptivity can be expected to help identify relevant higher-order terms in all regimes of ED size -- as soon as the ED is large enough to contain some information about the model.
For each model and experimental design size, we repeat the analysis $R = 30$ times to account for the stochasticity of the sampling method. 
Due to their excessive computational cost%
\footnote{Anisotropic degree basis adaptivity increases the degree of the basis uniformly by 1 in each dimension, which for high-dimensional models often results in infeasibly large bases.
SPGL1, or the implementation we are using, 
takes considerably longer than the other solvers for an increasing number of basis elements $P$. Since our benchmark setup requires several thousand runs of the sparse solver for the case of high-dimensional models, the computational cost of the full benchmark is infeasible given our computational resources.}%
, we omit SPGL1 and anisotropic degree basis adaptivity from the benchmark for high-dimensional models ($d \geq 20$).
More details on the settings for the basis adaptivity schemes (e.g., initial basis and investigated degree ranges) can be found in \ref{app:basis_adapt_details}.

\subsubsection{Boxplots of results for the Ishigami function}
\begin{figure}[htbp]
	\subcaptionbox{Ishigami function, $N = 50$ \label{fig:BA_ishigami_lowdim}} {\includegraphics[width=.48\textwidth, trim=0 .3cm 1.6cm .4cm, clip]{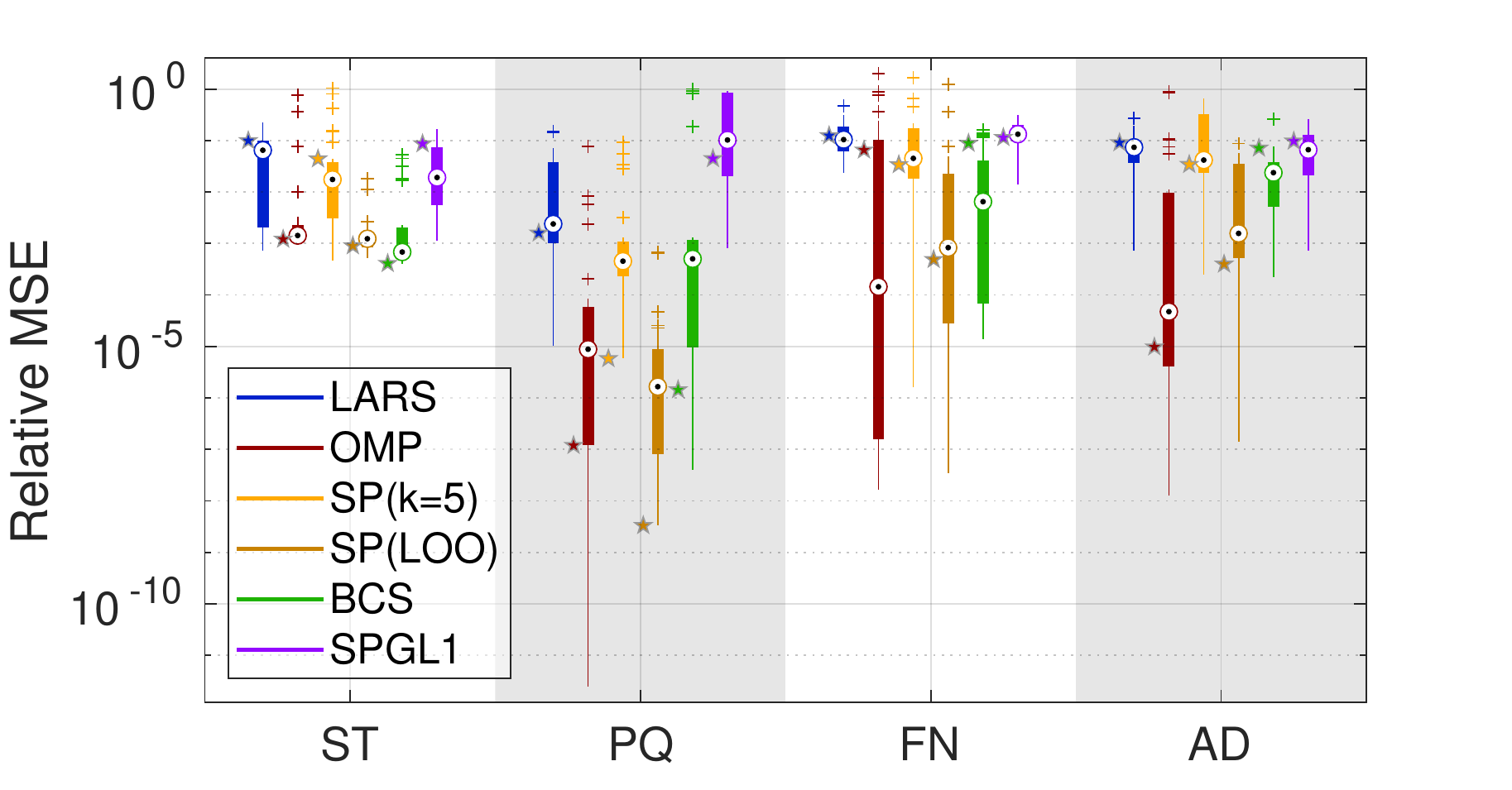}}
	\hfill%
	\subcaptionbox{Ishigami function, $N = 150$} {\includegraphics[width=.48\textwidth, trim=0 .3cm 1.6cm .4cm, clip]{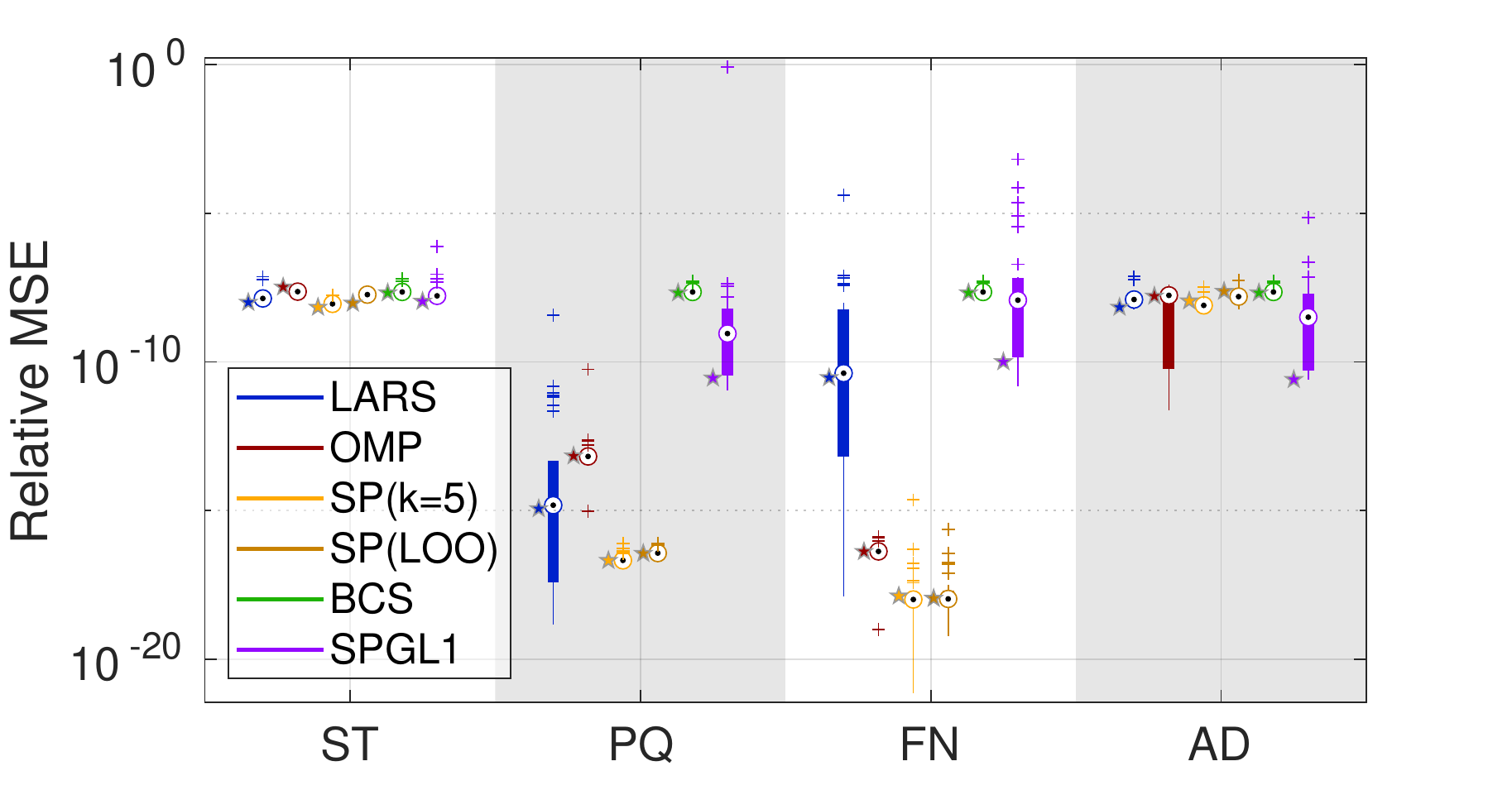}}%
	\caption{Comparison of all possible combinations of solvers and basis adaptivity schemes for the Ishigami function.
	Abbreviations of basis adaptivity schemes: ST -- static basis, PQ -- degree- and q-norm, FN -- forward neighbor, AD -- anisotropic degree. 
	We show validation errors attained by all 24 combinations of methods on 30 realizations of experimental design. The boxplots visualize the attained errors across all 30 realizations. 
	The star-shaped markers denote for one arbitrarily selected ED realization the errors that the different combinations attain (highlighting one set of data points which is also contained in the boxplots).
	Left: small ED; right: large ED.
	}
	\label{fig:BA_ishigami}
\end{figure}

The results from this benchmark for the Ishigami function are displayed in Fig.~\ref{fig:BA_ishigami} in the form of boxplots. Results for the remaining models are shown in \ref{app:raw_data}. The boxplots visualize the results for all combinations on 30 independent ED realizations. 
The star-shaped markers denote for one (arbitrarily selected) ED realization the respective validation errors of each of the combinations, highlighting one set of data points which is also contained in the boxplots. 
We observe that the ranking based on statistics (e.g., median as denoted by the white circle) and the ranking based on the actually attained error on a specific ED can be quite different (see e.g.\ \SPorig{} \& FN vs.\ BCS \& FN in Fig.~\ref{fig:BA_ishigami_lowdim}). From the star-shaped markers, it is obvious that the solvers and basis-adaptive schemes do not exploit the available information in the same ways: while some combinations show their best performance on the selected ED, others perform average or worse on the same ED. 
We want to find the ``best'' method, i.e., a solver-basis adaptivity combination that consistently, on each different model and ED realization, achieves a close-to-optimal error. In other words, in this comparative study we are less interested in the absolute values of the error (since we assume that the tested methods have been validated before in the associated literature), but rather in the \textit{relative performance} of the methods.
The boxplots alone do not give the full picture, since they do not show which errors correspond to the same ED realization. Also, looking at the results for this and other benchmark models in \ref{app:raw_data}, it is difficult to visually extract information about the overall performance of the methods.

\FloatBarrier

\subsubsection{Bar plots -- results aggregated over models and replications, separately per model dimensionality and ED size}
\begin{figure}[htbp]
	\centering
	\subcaptionbox{low-dimensional models, small ED {($6\times30$ runs)} \label{fig:aggregated_results_BA_a}}
	{\includegraphics[width=.48\textwidth, trim=0.25cm 0cm 0.6cm 0.75cm, clip]{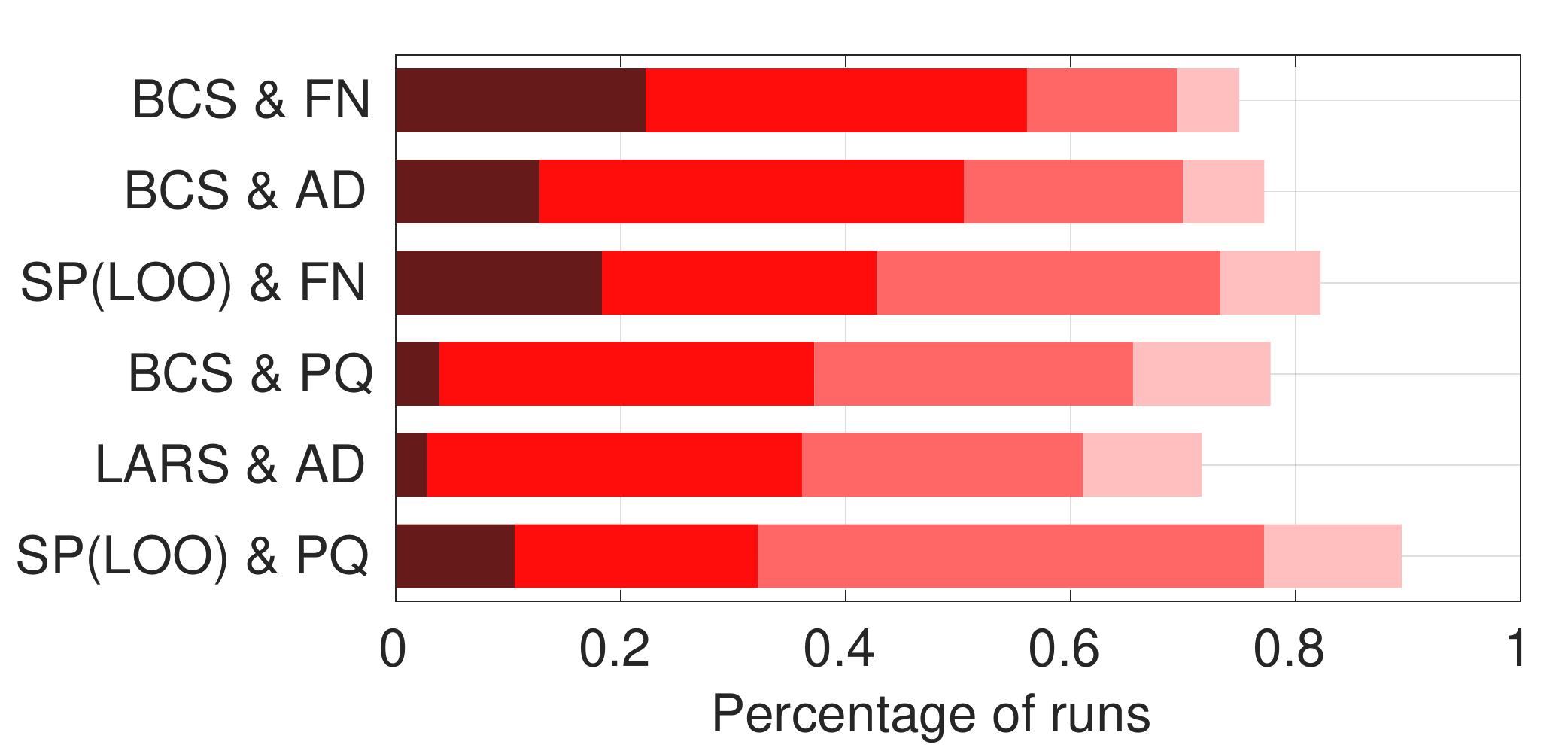}}
	\hfill
	\subcaptionbox{low-dimensional models, large ED {($6\times30$ runs)} \label{fig:aggregated_results_BA_b}}
	{\includegraphics[width=.48\textwidth, trim=0.25cm 0cm 0.6cm 0.75cm, clip]{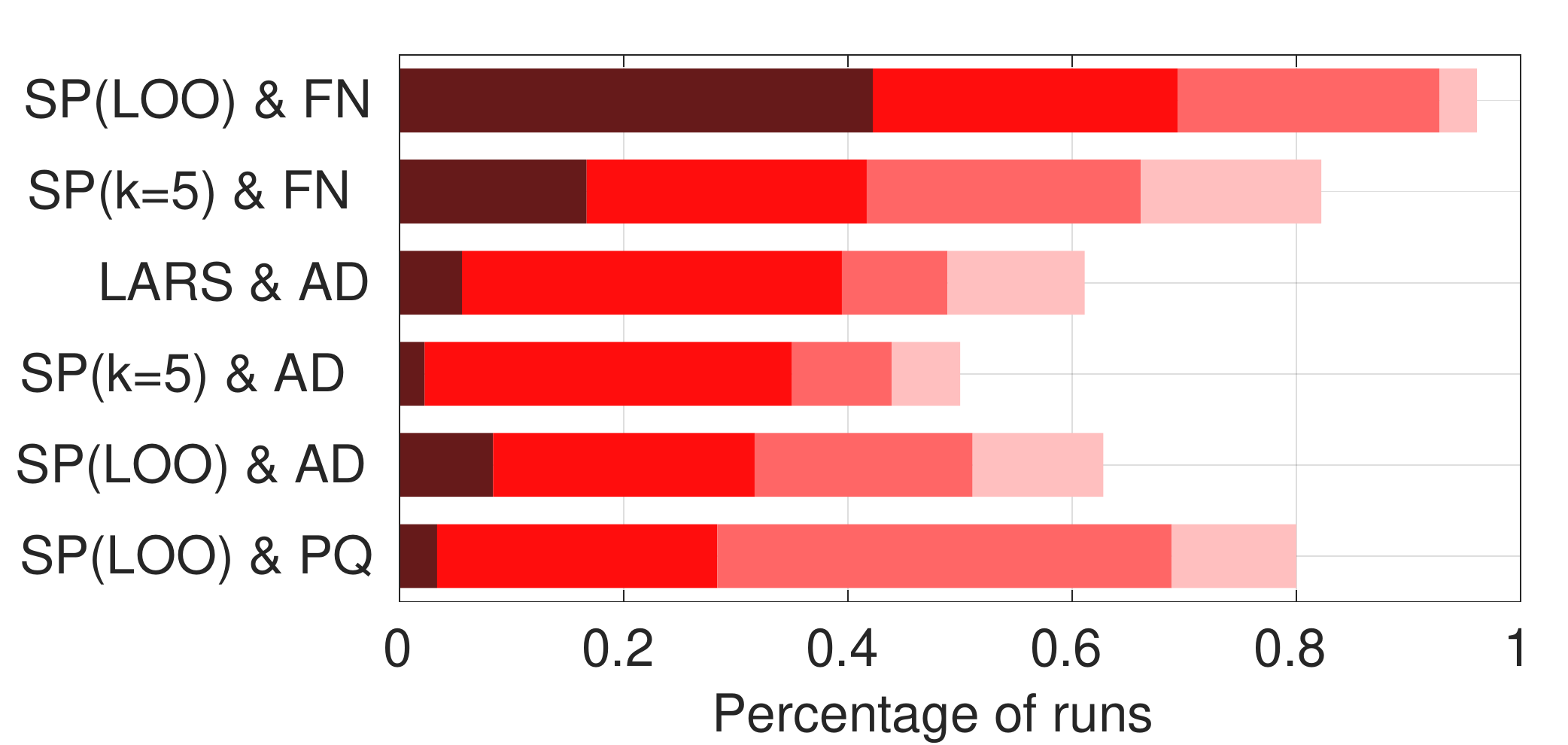}}
	\\
	\subcaptionbox{high-dimensional models, small ED {($5\times30$ runs)} \label{fig:aggregated_results_BA_c}}	
	{\includegraphics[width=.48\textwidth, trim=0.25cm 0cm 0.6cm 0.75cm, clip]{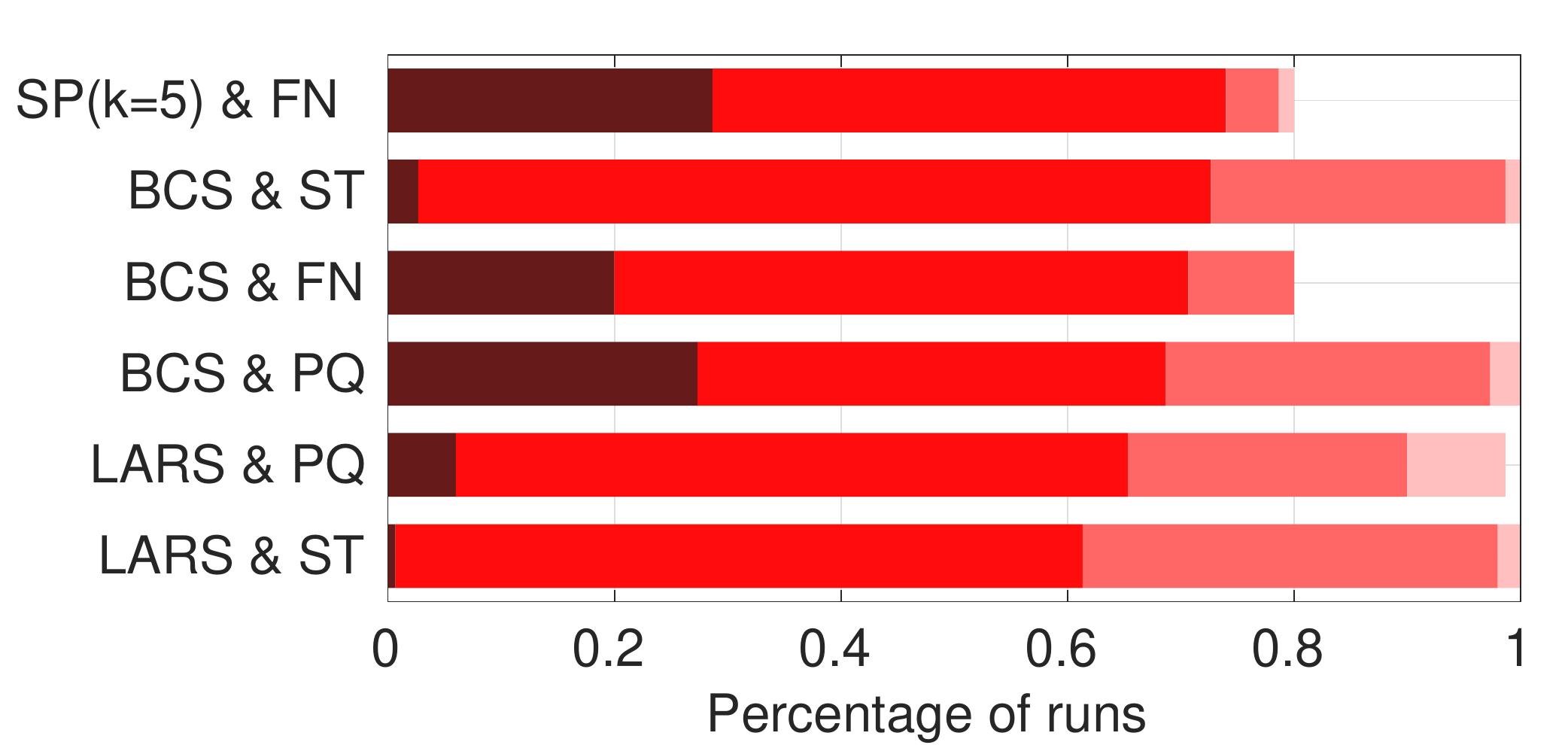}}
	\hfill
	\subcaptionbox{high-dimensional models, large ED {($5\times30$ runs)} \label{fig:aggregated_results_BA_d}}
	{\includegraphics[width=.48\textwidth, trim=0.25cm 0cm 0.6cm 0.75cm, clip]{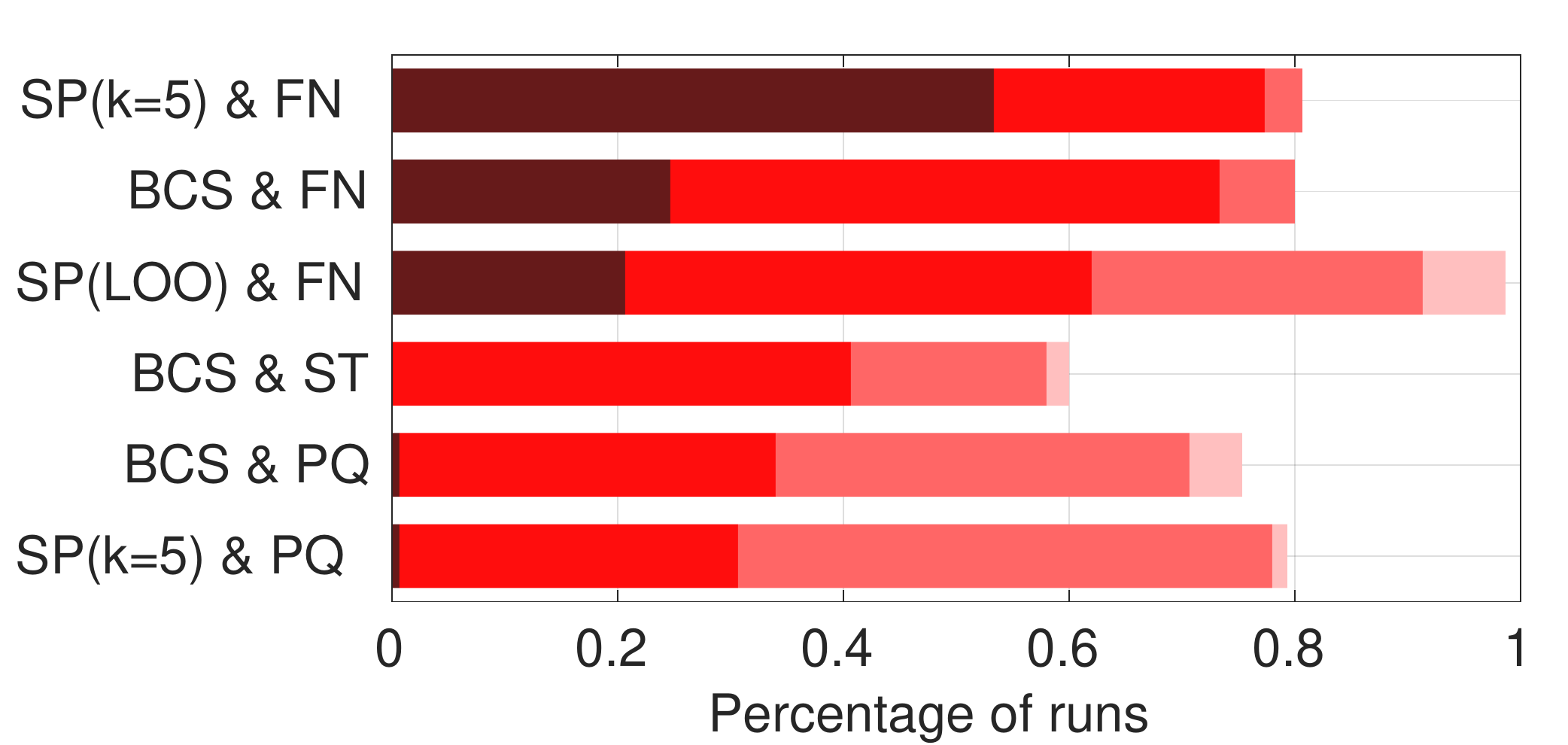}}
	\\[-4pt]
	\subcaptionbox{Aggregated results for all 11 models and both ED sizes ($11\times2\times30$ runs) \label{fig:aggregated_results_BA_all}}
	{\includegraphics[width=.63\textwidth]{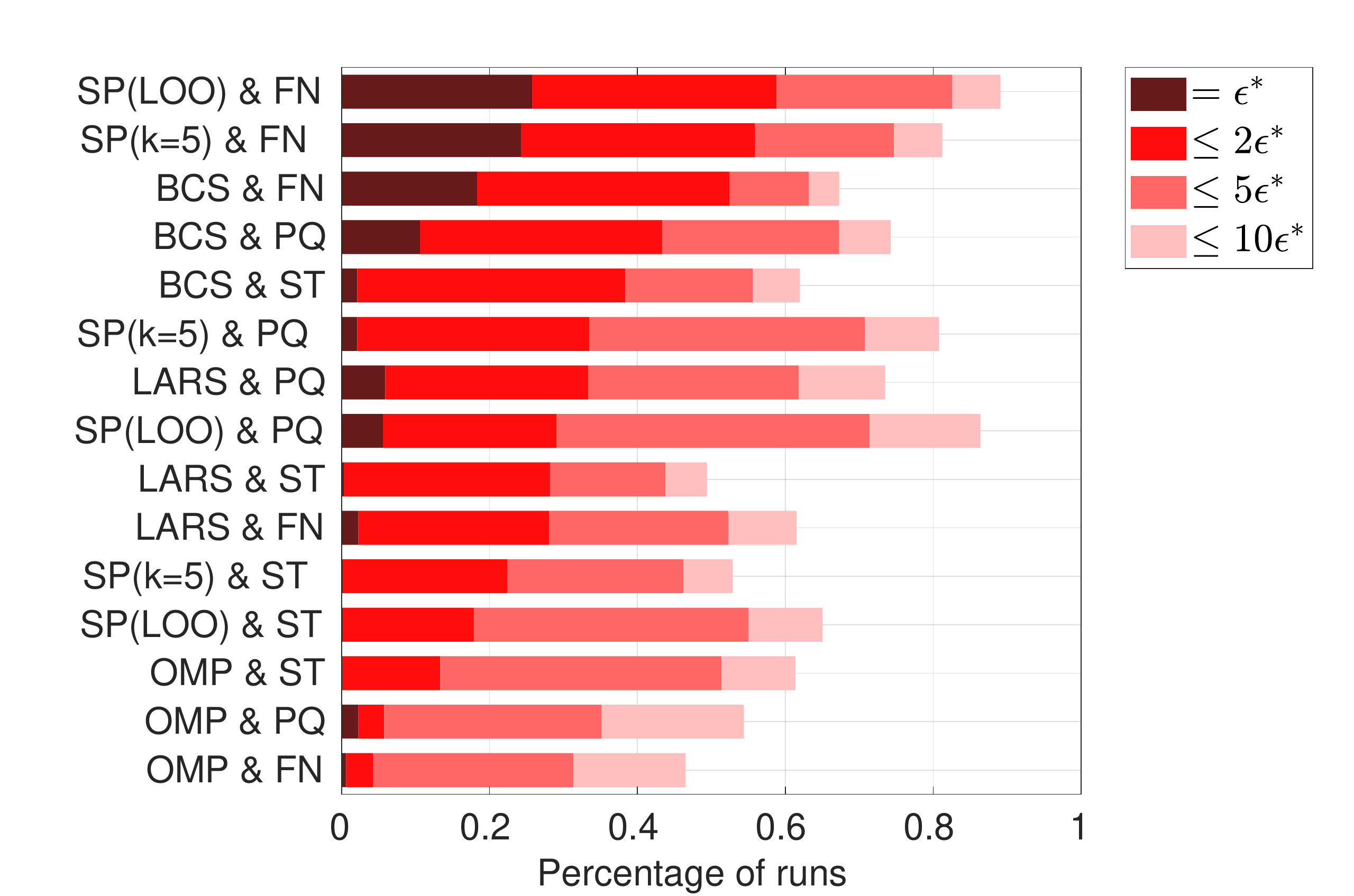}}
	\vspace{-6pt}
	\caption{ \footnotesize
			Aggregated results of our comparative study. For each of the 11 models, 2 ED sizes, and 30 replications (i.e., ED realizations), we have run each of the combinations of sparse solver and basis adaptivity scheme and computed the corresponding relative mean-squared errors (MSE) (see Fig.~\ref{fig:BA_ishigami} for results for the Ishigami function).
			We determine the relative performance of the method combinations \textit{separately on each ED realization}: denoting the best attained relative MSE on a specific ED by $\epsilon^*$, we record which of the method combinations reached this error or came within a factor of $\{2,5,10\}$ of $\epsilon^*$ on the same ED.\\
			We then aggregate the counts over all replications and models, separating by low/high dimensionality and small/large ED size (panels \ref{fig:aggregated_results_BA_a}--\ref{fig:aggregated_results_BA_d}). This results in $6\times30$ runs for the low-dimensional cases, and $5\times30$ runs for the high-dimensional cases (per method combination).
			For readability, in panels \ref{fig:aggregated_results_BA_a}--\ref{fig:aggregated_results_BA_d} only the six best combinations are shown (for the full results, see \ref{app:BA_bar_plots}).
			The results are displayed in the form of \textit{overlapping bars}, representing percentages. The dark red bar represents the percentage of runs where the method combination reached the best relative MSE on the specific ED realization. The other bars in shades of red denote the percentages of runs where the error of the particular method combination was within a factor of $\{2,5,10\}$ of the best error on the same ED realization.
			The method combinations are sorted by the length of the bright red bar (number of runs with an error $\leq 2\epsilon^*$).
			\\
			In panel \ref{fig:aggregated_results_BA_all}, we display the relative performance of the method combinations aggregated over all models, ED sizes and replications (i.e., $11\times2\times30$ runs).					
	}
	\label{fig:aggregated_results_BA}	
\end{figure}

Therefore, we propose an aggregated visualization of the results as displayed in Fig.~\ref{fig:aggregated_results_BA} and described in the following.
For every model and ED size, we determine 
on each unique ED realization 
which combination attained the smallest error $\epsilon^*$
(for example, for the ED realization visualized in Fig.~\ref{fig:BA_ishigami_lowdim} by the star-shaped markers, \SPloo{} \& PQ attains rank 1). 
We also determine \textit{on the same ED} for each combination whether it came within a factor of 2, 5, or 10 of 
$\epsilon^*$. 
For example, for the ED realization visualized in Fig.~\ref{fig:BA_ishigami_lowdim} by the star-shaped markers, the best error is attained by \SPloo{} \& PQ. All other combinations achieve an error that is more than a magnitude larger, and would therefore not get a count for this ED realization.

We then aggregate the counts over all repetitions and all models in four different classes: low dimensionality ($d \leq 10$) and small experimental design, low dimensionality and large experimental design, and their corresponding high dimensionality counterparts ($d \geq 20$).
In other words, for each of the four classes we count how often each combination 
achieved the best error or
came within a factor of 2, 5, or 10 of the best error, for each unique experimental design.
These counts, scaled as percentages of all runs%
\footnote{There are $6 \times 30$ realizations for the two low-dimensional cases (Figs.~\ref{fig:aggregated_results_BA_a} and \ref{fig:aggregated_results_BA_b}), and $5 \times 30$ realizations for the two high-dimensional cases (Figs.~\ref{fig:aggregated_results_BA_c} and \ref{fig:aggregated_results_BA_d}). Fig.~\ref{fig:aggregated_results_BA_all} aggregates over all 660 ED realizations.}%
, are then visualized in the form of overlapping horizontal bars in Figs.~\ref{fig:aggregated_results_BA_a}--\ref{fig:aggregated_results_BA_all} as follows:
The dark red bar shows the percentage of runs in which the particular combination of sparse solver and basis-adaptive scheme found the smallest relative MSE $\epsilon^*$.
The other three bars
in hues of red illustrate how often the respective combination was within a factor of 2, 5, or 10 of 
$\epsilon^*$ \textit{on the same ED}. 

These bars are therefore a measure of the dispersion of the accuracy of each method in terms of distance to the best attainable accuracy. This measure is more interesting than the absolute scatter of a method, since a large variability does not matter as long as the method always provides a close-to-optimal solution; and small variability is of no value if the method is always orders of magnitude away from the best possible solution.
We only show the six combinations of solver and basis adaptivity scheme whose relative MSE was most often within two times the best relative MSE (denoted by the bright red bar), and sort the combinations according to this criterion. 
(For the full list of combinations, see Figs.~\ref{fig:aggr_lowdim_small}--\ref{fig:aggr_highdim_large} in the appendix.)

Comparing the subplots \ref{fig:aggregated_results_BA_a}--\ref{fig:aggregated_results_BA_d}, we observe that the results for small and large ED sizes and low- and high-dimensional models are indeed quite different, which justifies analyzing the results separately.
As detailed in the previous paragraph, the plots show both which combinations of solver and basis adaptivity scheme attain the smallest relative MSE how often (length of the dark red bar), and how robust the combination is (lengths of the other three bars), i.e., how often the solution was within a small factor of the best solution. 
We observe that depending on the criterion considered to assess the combinations of solver and basis adaptivity scheme, different combinations turn out to perform best. 
In the following, we summarize the performance of solver-basis adaptivity combinations with three numbers $X$-$Y$-$Z$, where $X$ denotes the percentage of runs where the respective combination turned out best (i.e., the length of the dark red bar), $Y$ denotes the percentage of runs that were within a factor of two of the smallest error on that ED (length of the bright red bar), and $Z$ denotes the percentage of runs that were within 1 order of magnitude of the smallest error (light red bar). The numbers are rounded to integer values.
Italic numbers indicate that this is the best value achieved among all combinations.
The enumeration tags refer to the panels in Figure \ref{fig:aggregated_results_BA}.

\begin{itemize}
	\item[(\ref{fig:aggregated_results_BA_a})] \textit{Low-dimensional models, small ED}: BCS together with forward neighbor basis adaptivity (\textit{22}-\textit{56}-75) achieves the smallest error more often than any other combination. However, this combination is not the most robust: in $100 - 75 = 25\%$ of runs, its error is more than one magnitude worse than the best error.
	\SPloo{} together with p\&q adaptivity (11-32-\textit{89}) is the most robust instead. 
	\item[(\ref{fig:aggregated_results_BA_b})] \textit{Low-dimensional models, large ED}: by far the best combination in all three categories is \SPloo{} together with forward neighbor basis adaptivity (\textit{42}-\textit{69}-\textit{96}).
	\item[(\ref{fig:aggregated_results_BA_c})] \textit{High-dimensional models, small ED}: there are two  combinations that outperform the others: \SPorig{} with forward neighbor basis adaptivity (\textit{29}-\textit{74}-80) and BCS with a static basis (3-73-\textit{100}). A close third is BCS with p\&q adaptivity (27-69-\textit{100}).
	\item[(\ref{fig:aggregated_results_BA_d})] \textit{High-dimensional models, large ED}: the two best combinations are \SPorig{} with forward neighbor basis adaptivity (\textit{53}-\textit{77}-81) and \SPloo{} with forward neighbor basis adaptivity (21-62-\textit{99}).
\end{itemize}

We observe from this enumeration that only the solvers BCS, \SPorig{} and \SPloo{} are found among the best combinations. 
Considering the best six combinations (based on the second number $Y$) as in Fig.~\ref{fig:aggregated_results_BA}, only LARS is joining the list. OMP does not perform well in any of the cases (see also Appendix, Figs.~\ref{fig:aggr_lowdim_small}--\ref{fig:aggr_highdim_large}). This is likely due to its tendency to severely underestimate the generalization error, which might make the comparison between error estimates of different bases unreliable. Likewise, SPGL1 is never among the best six combinations for low-dimensional models. 

Regarding basis adaptivity schemes, the static basis is not among the best six combinations (based on the second number $Y$) for low-dimensional models. However for high-dimensional models, especially for the small ED case, it is among the six best combinations several times and performs well. 

Note that in the high-dimensional, small ED case (Figs.~\ref{fig:aggregated_results_BA_c} and \ref{fig:aggr_highdim_small}), the static basis shows the most robust behavior (in terms of the third number $Z$) while any solver with forward neighbor basis adaptivity is off by more than a magnitude in at least 20\% of runs. This is due to its bad performance for the 100D function (see also Fig.~\ref{fig:BA_100D_small}). 
The 100D function is also responsible for the peculiar results of the high-dimensional, large ED case (Figs.~\ref{fig:aggregated_results_BA_d} and \ref{fig:aggr_highdim_large}): only \SPloo{} together with forward neighbor basis adaptivity is able to find a very accurate solution, reaching a validation error of $10^{-10}$ which is several orders of magnitude smaller than what the other combinations reach. However, several other combinations are within one order of magnitude of the smallest error for all other high-dimensional models, reaching overall 80\% for $Z$, as can be seen from Fig.~\ref{fig:aggr_highdim_large}.

\subsubsection{Bar plots -- results aggregated over all models, ED sizes, and replications}
To get a complete picture of solver and basis adaptivity scheme performance, we display in Fig.~\ref{fig:aggregated_results_BA_all} the relative performance of all combinations of solvers and basis adaptivity schemes, averaged over all models and experimental designs regardless of their dimensionality and size. We exclude combinations involving SPGL1 and anisotropic degree basis adaptivity (AD), which were only tested for low-dimensional models. 
Therefore, Fig.~\ref{fig:aggregated_results_BA_all} contains $3 \times 5 = 15$ combinations.
We see that \SPloo{} together with forward neighbor basis adaptivity (\textit{26}-\textit{59}-\textit{89}) performs best in all three categories. \SPorig{} together with forward neighbor basis adaptivity (24-56-81) is on the second place in terms of the first two criteria ($X$ and $Y$), while
\SPloo{} together with p\&q basis adaptivity (6-29-86) is the second-best solver in terms of achieving an error within one order of magnitude of the best error most often ($Z$).
BCS together with any basis adaptivity scheme performs well, while all combinations involving OMP are on the bottom of the list. Combinations with a static basis are found more towards the end of the list, and those with forward neighbor basis adaptivity are found more towards the beginning of the list.

\subsubsection{Conclusion of the comparative study}
From these plots, we see clearly that there is no single combination of sparse solver and basis adaptivity scheme that always outperforms all others. While \SPloo{} together with forward neighbor basis adaptivity shows superior performance when averaged over all models and ED sizes, we identify better-performing combinations when we differentiate by model dimension and ED size. 
In some cases, we are faced with a trade-off between accuracy and robustness: e.g., 
for low-dimensional models and small EDs, the choice of BCS \& FN has a higher probability of yielding a near-optimal solution (within a factor of 2 to the best), while the choice of \SPloo{} \& PQ has a higher probability of not being more than a magnitude off from the optimal one.

It seems to be easier to identify trends regarding solvers than regarding basis adaptivity schemes. For example, BCS performs well when information is scarce (small ED sizes or high-dimensional models) while \SPloo{} performs well for large ED sizes, as already observed in \citep{LuethenSIAMJUQ2021}, and OMP generally underperforms. This might be because sparse regression solvers are based on quite different principles, from gradient descent over greedy stepwise regression to Bayesian reasoning, while the basis adaptivity schemes all work with variations of the same concept -- namely, gradually increasing the degree of the basis. 
Basis-adaptive schemes generally outperform the static basis, most likely because they offer more basis elements to choose from. However, it seems to not matter as much how the additional basis elements are generated, since none of the three basis-adaptive methods always finds the best basis.
Note that the selection of the bases is guided by the cross-validation error. If the cross-validation error does not correlate well with the real validation error, the scheme will select a suboptimal basis. OMP generally has an unreliable cross-validation error, which might explain its bad performance with basis-adaptive schemes.

\FloatBarrier

\subsection{Automatic selection of sparse solver and basis adaptivity scheme}
\label{sec:results_final_model_selection}

As we saw in the previous section, there is no single best-performing combination of sparse solver and basis adaptivity scheme. In this section, we investigate the question: is there any criterion which can help us choose the best combination of solver and basis adaptivity scheme for a given problem? One such criterion could be a deterministic rule based on dimensionality of the model and ED size, relying on the results from the previous section. 
Another option could be to use
cross-validation-based model selection criteria
as described in Section~\ref{sec:autom_selection} (see also Section~\ref{sec:model_selection}). 
We call this process of automatically choosing a well-performing combination from a number of candidate combinations using a model selection criterion \textit{automatic selection}. 

Due to the performance of the methods in the benchmark in the previous sections, we restrict our investigation to the solvers \SPorig{}, \SPloo{} and BCS, and to the basis adaptivity schemes
\begin{itemize}
	\item p\&q adaptivity (PQ), forward neighbor adaptivity (FN), and anisotropic degree adaptivity (AD) for low-dimensional models
	\item static basis (static), p\&q adaptivity (PQ), and forward neighbor adaptivity (FN) for high-dimensional models
\end{itemize}
resulting in 9 possible solutions for each model, ED size and repetition.

We use the following model selection criteria (see also Section~\ref{sec:autom_selection}):
\begin{enumerate}
	\item the oracle solution, i.e, the smallest relative MSE attained among \textit{all} 24 or 15 combinations of methods (for low-dimensional or high-dimensional models, respectively) -- i.e., among all methods tested in Section~\ref{sec:BA_benchmark}, not only among the 9 candidate solutions considered here -- on each ED realization, as an ideal lower bound. Of course, this information is not available in practice.
	\item the criterion used for basis and hyperparameter selection by the respective solver (see Section \ref{sec:model_selection})
	\item hybrid LOO, computed by OLS on the active basis functions only
	\item hybrid modified LOO, computed by OLS on the active basis functions only, and using the correction factor from Eq.\ \eqref{eq:modification_factor}
	\item hybrid 10-fold cross-validation error, computed by OLS on the active basis functions only
	\item A fixed rule dependent on dimensionality and ED size, according to the findings in Section \ref{sec:BA_benchmark}, choosing the solver that achieved the smallest error most often (i.e., having the longest dark red bar, and coincidentally also the longest bright red bar):
	\begin{itemize}
		\item low-dimensional models, small ED: BCS \& FN
		\item low-dimensional models, large ED: \SPloo{} \& FN 
		\item high-dimensional models, small ED: \SPorig{} \& FN
		\item high-dimensional models, large ED: \SPorig{} \& FN 
	\end{itemize}
	This implies that by design, the lengths of the bars for this selection criterion are identical with the results for the respective combination in Fig.~\ref{fig:aggregated_results_BA}.
	\item A randomly picked combination of solver and basis adaptivity scheme from the 9 available options as upper bound: any sensible model selection criterion must perform better than this.
\end{enumerate}

\begin{figure}[htbp]
	\centering
	\subcaptionbox{low-dimensional models, small ED}
	{\includegraphics[width=.49\textwidth, trim=0 0 2cm 0, clip]
		{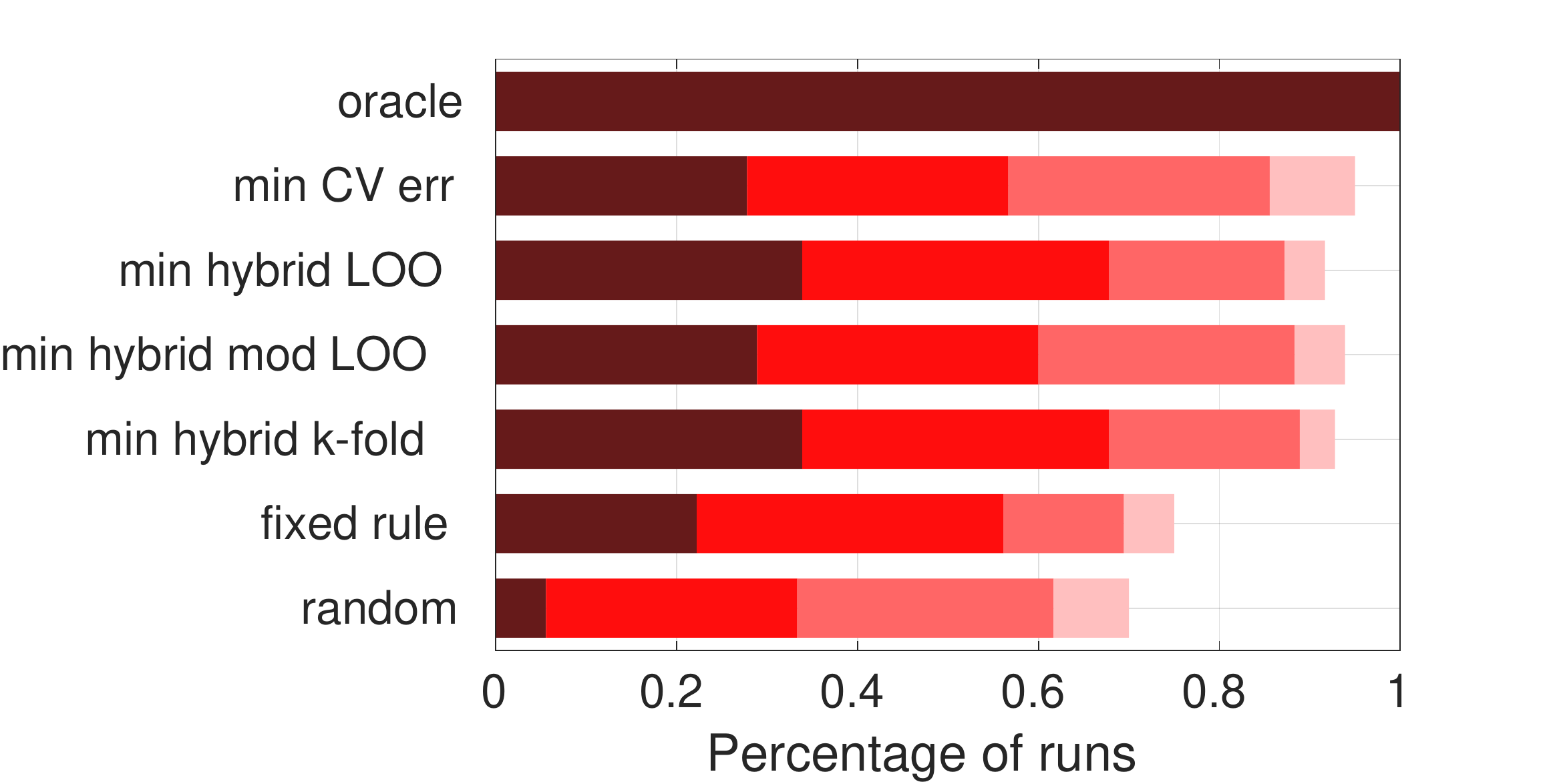}}
	\subcaptionbox{low-dimensional models, large ED}
	{\includegraphics[width=.49\textwidth, trim=0 0 2cm 0, clip]
		{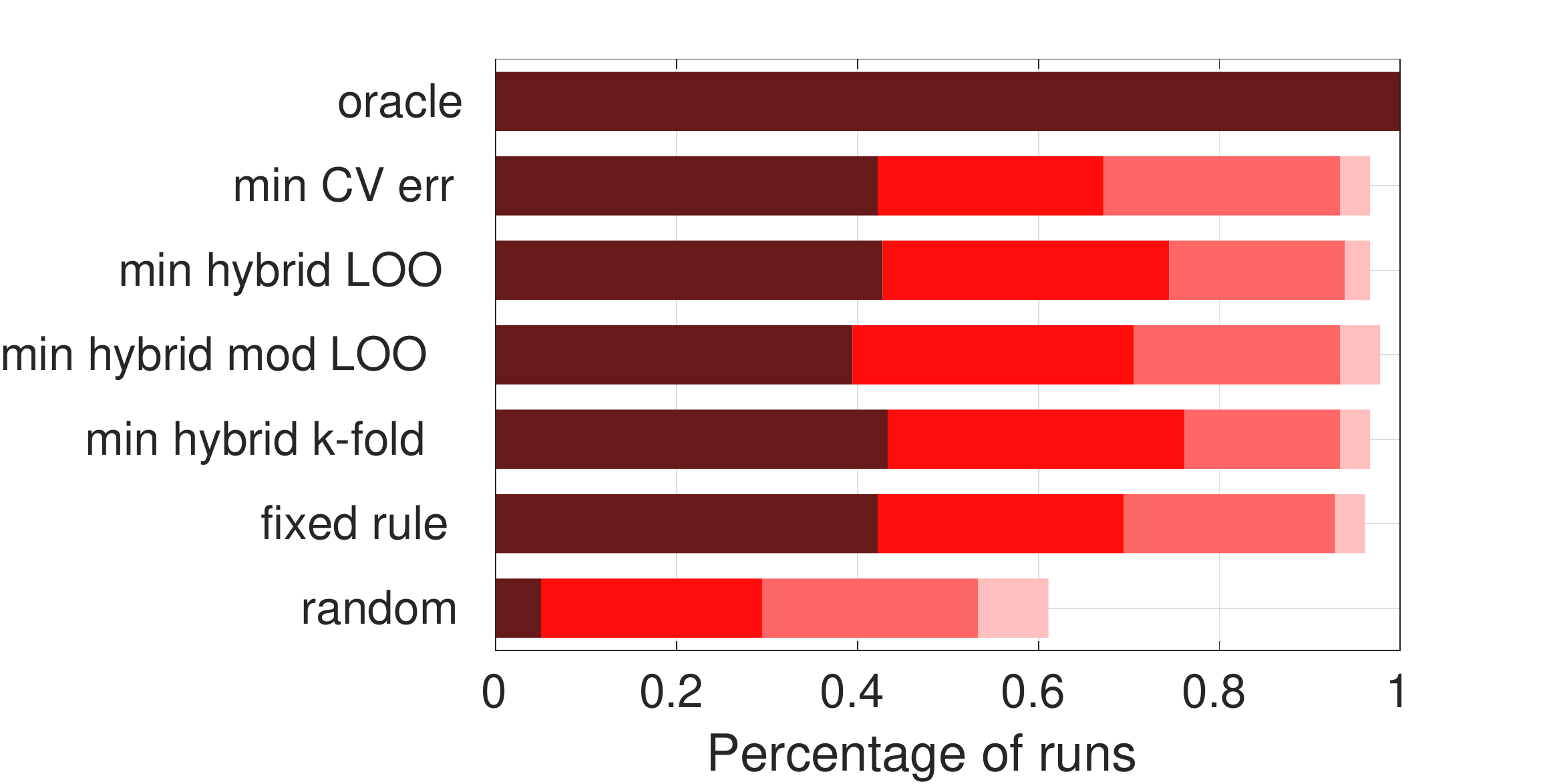}}
	\\[10pt]
	\subcaptionbox{high-dimensional models, small ED}
	{\includegraphics[width=.49\textwidth, trim=0 0 2cm 0, clip]
		{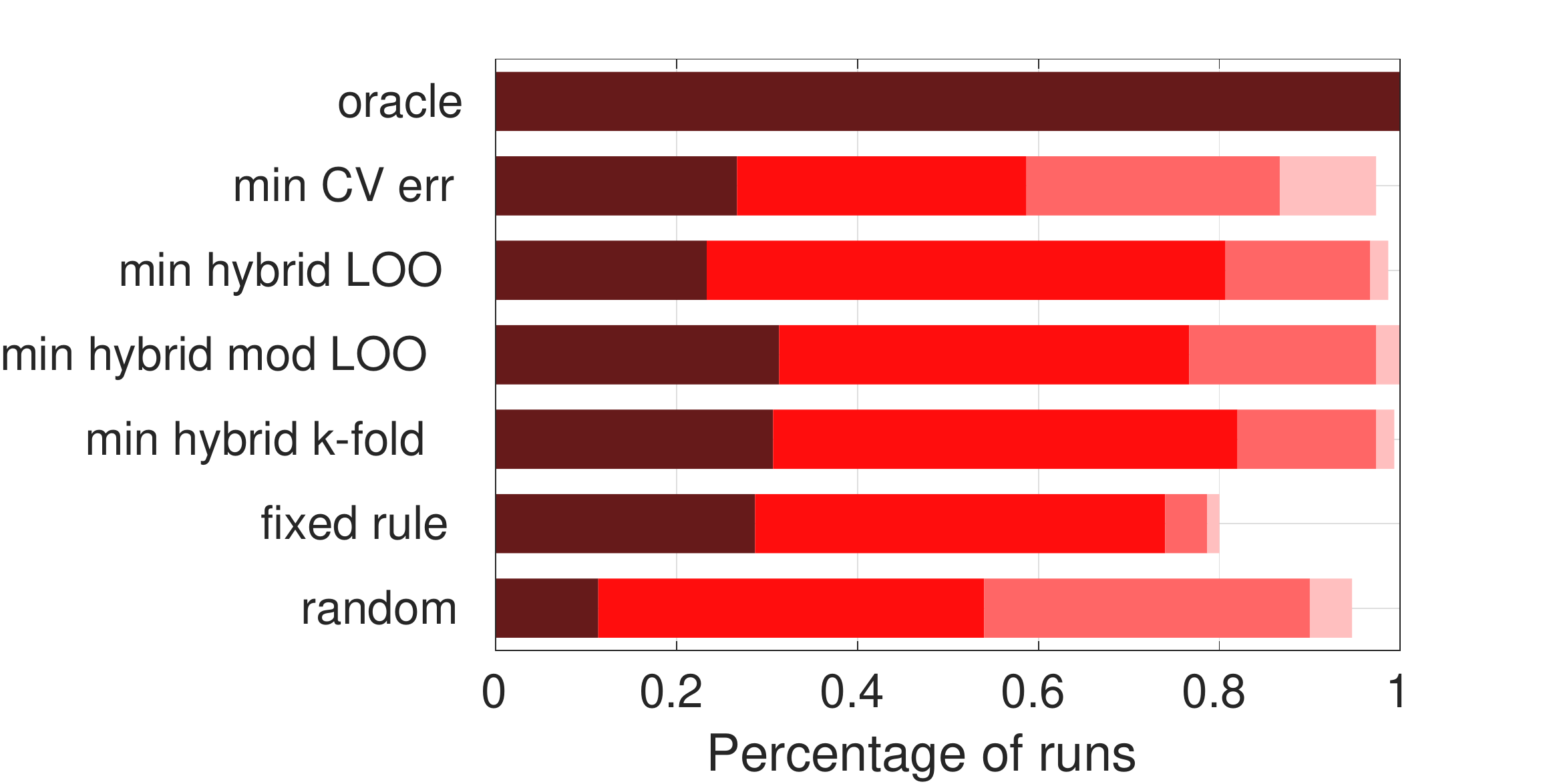}}
	\subcaptionbox{high-dimensional models, large ED\label{fig:automatic_selection-highdimlarge}}
	{\includegraphics[width=.49\textwidth, trim=0 0 2cm 0, clip]
		{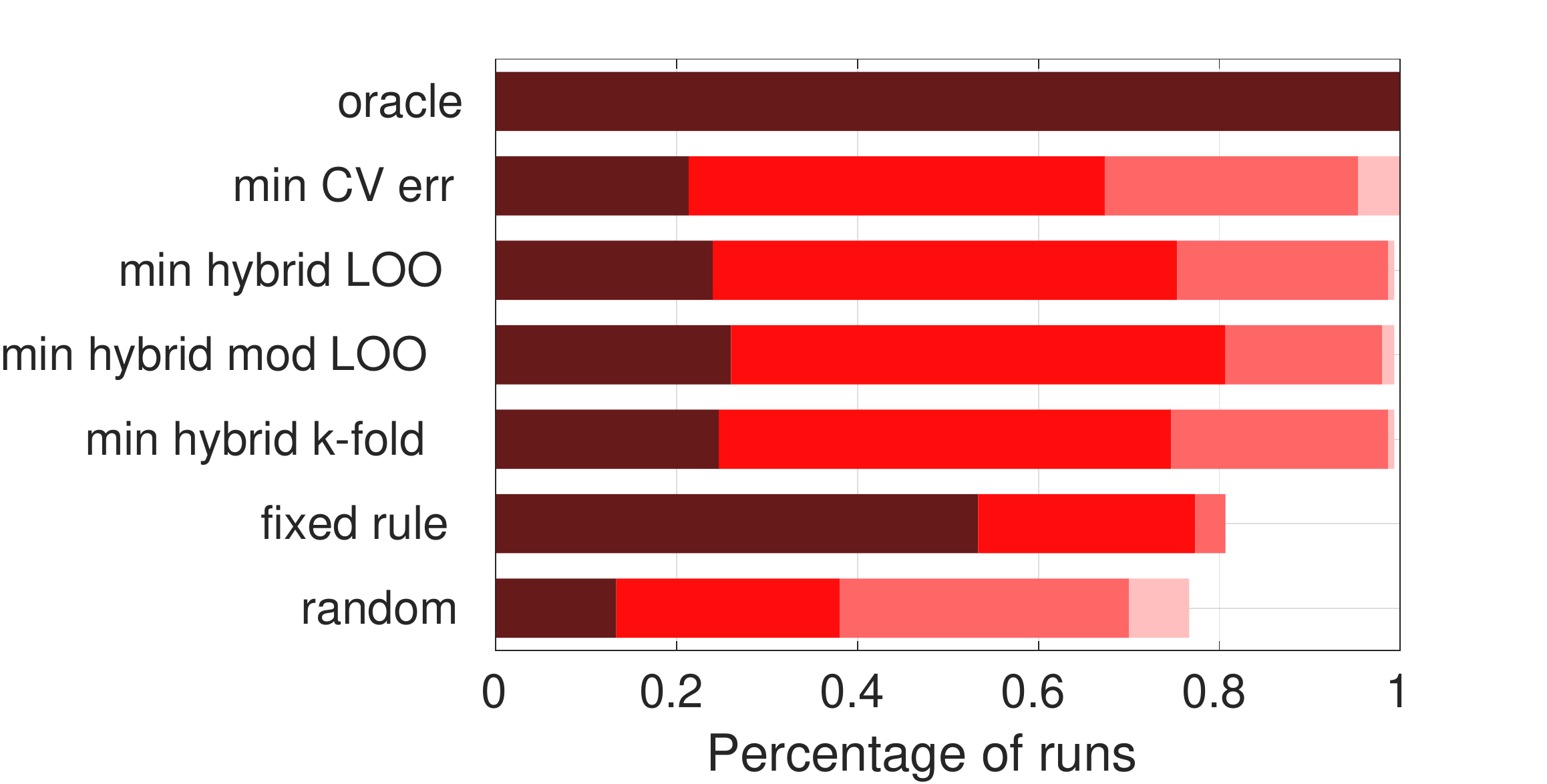}}
	\\[10pt]
	\subcaptionbox{Aggregated over all 11 models and both ED sizes \label{fig:automatic_selection-allinone}}
	{\includegraphics[width=.65\textwidth]{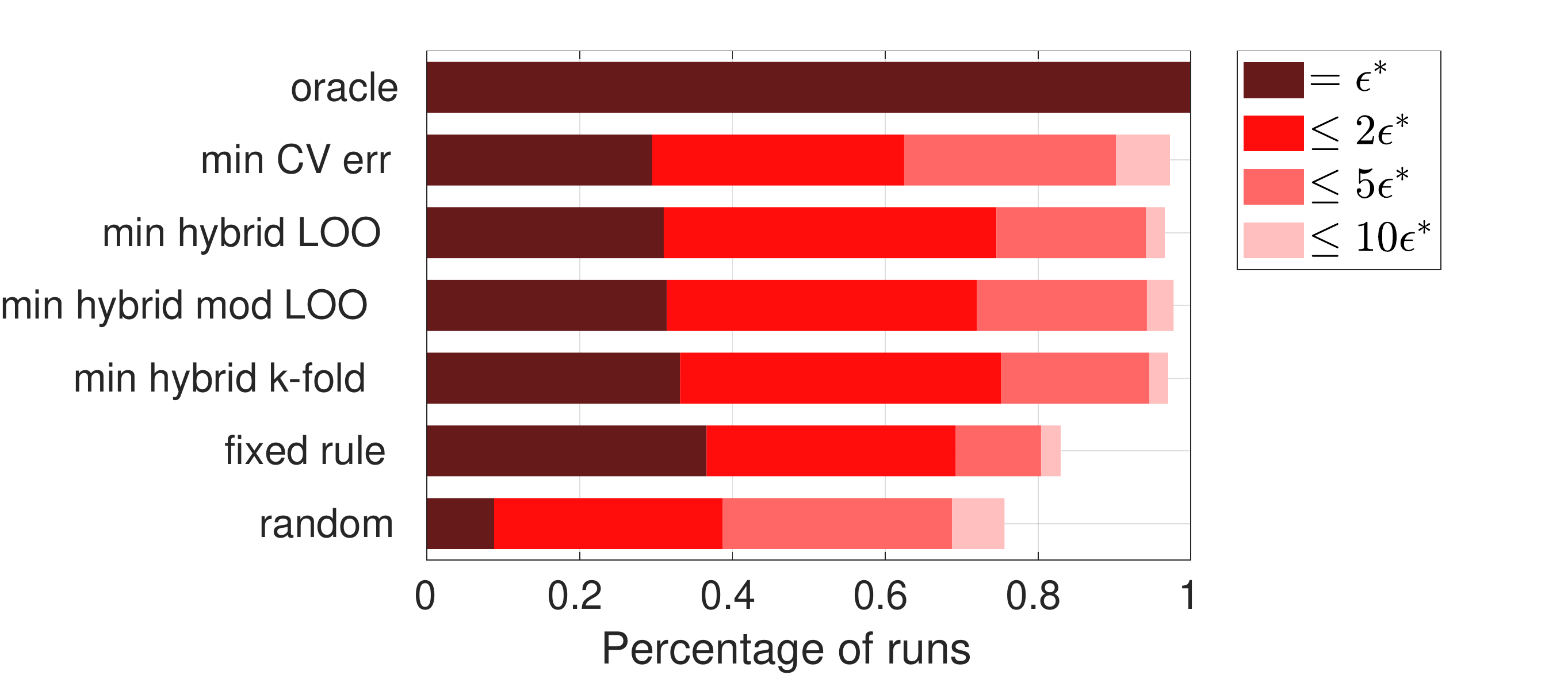}}
	\vspace{4pt}
	\caption{Testing different automatic selection strategies. 
	As before, the overlapping bar plot visualizes the percentage of runs where the respective strategy achieved the best error $\epsilon^*$ or was within a factor of $\{2,5,10\}$ of the best error $\epsilon^*$ among the seven selection strategies. 
	The bar labels are explained in the enumeration of model selection criteria in the text.}
	\label{fig:automatic_selection}
\end{figure}

The results are presented in Figure~\ref{fig:automatic_selection}.
{Note that we do not display which sparse solver and basis adaptivity scheme was chosen -- we only show how close the chosen solution comes to the best possible solution.}
\begin{itemize}
	\item By construction, the oracle selection performs best in all cases. 
	(We only include it into the plot to emphasize that all other criteria are measured against the best attained error on each ED.)
	\item The random selection is by far the worst selection criterion, which shows that automatic selection provides a statistically significant improvement over random selection.
	\item All three hybrid selection criteria are in general more robust than the fixed rule, attaining an error within 2 times of the best solution (bright red bar) similarly or more often than the fixed rule, and being within one order of magnitude of the best error (light red bar) even more than 90\% of runs (low-dimensional models) or 99\% of runs (high-dimensional models).
	Averaging over all runs (Fig.~\ref{fig:automatic_selection-allinone}), the hybrid criteria attain the best error almost as often as the fixed rule, which was chosen for this reason, i.e., because it attained the best error most often. 
	This shows that automatic selection is able to return a solution with almost the same accuracy as the best-performing method, while being more robust than a fixed rule or a random choice of methods.
	\item The criterion used for basis and hyperparameter selection, i.e., k-fold CV for \SPorig{} and BCS, and LOO for \SPloo{}, performs slightly worse than the three hybrid selection criteria in all cases shown in Fig.~\ref{fig:automatic_selection}. 
	The slightly worse performance of this selection criterion might be explained by selection bias: since the criterion was already used twice for selection, it is likely that it underestimates the generalization error. Another reason might be that the criteria used by different solvers (OLS-based LOO vs.\ k-fold CV) might not be comparable among each other (inconsistency). 
\end{itemize}

We conclude that automatic selection, i.e., using cross-validation to choose a PCE solution from a number of candidate solutions computed with different methods on the same experimental design, can achieve results that are as accurate as and more robust than results from a fixed rule for which combination of solver and basis adaptivity scheme to use, even if this rule is based on a thorough benchmark such as the one in Section~\ref{sec:BA_benchmark}.
While automatic selection leads to increased computational cost due to the need of training the surrogate with different solvers and basis adaptivity schemes on the same experimental design, 
it is a case-independent strategy applicable to models of any dimensionality and experimental designs of any size, and it finally results in a PCE that is both accurate and more reliable.

\section{Conclusion and discussion}
\label{sec:conclusion}

Our goal was to provide guidance for the choice of sparse PCE methods in engineering applications, by answering the two questions: (1) Is there a significant difference between different combinations of sparse solvers and basis adaptivity strategies, and does the proper choice matter in actual applications? (2) Is there a smart strategy to automatically select a good combination?

To answer these questions, we performed an extensive comparative study investigating several approaches for computing sparse PCE with the goal of surrogate modeling, using the relative mean squared error on a validation set as the main performance metric.
In particular, we studied the performance of combinations of basis adaptivity schemes and sparse solvers in order to identify combinations that yield the smallest generalization error.
Our investigations are based on 11 analytical and numerical benchmark models of varying input dimension and complexity, representative of a wide range of engineering problems.
We considered the sparse solvers least angle regression (LARS), orthogonal matching pursuit (OMP), subspace pursuit based on $k$-fold cross-validation (\SPorig{}), subspace pursuit based on LOO (\SPloo{}), Bayesian compressive sensing/FastLaplace (BCS), and spectral projected gradient-$\ell^1$ (SPGL1). The basis adaptivity schemes we compared were a fixed basis truncation scheme following the rule $N \approx \frac{10}{3}P$, degree- and q-norm adaptivity, forward neighbor basis adaptivity, and anisotropic degree basis adaptivity. 
We made a distinction between four cases, namely low- and high-dimensional models as well as small and large experimental design sizes. 

The comparative study revealed that it is important to carefully select the strategy, since the difference in generalization error between different combinations of methods can be large, even more than an order of magnitude. 
No single solver or basis-adaptive scheme significantly outperformed the others. However, by dividing the analysis into classes (low-/high-dimensional models and small/large ED sizes), some significant patterns can be identified.
The combinations that performed well always involved the solvers \SPloo{}, \SPorig{}, or BCS, but never SPGL1 or OMP. For the basis-adaptive schemes, the picture is less clear, except that combinations involving forward neighbor basis adaptivity achieved the best accuracy most often, and that the static basis (i.e., no basis adaptivity) was in nearly all cases outperformed by basis-adaptive schemes. 
This might imply that basis-adaptive schemes provide the opportunity to improve the solution by offering more regressors to choose from, but that none of them consistently generates the most suitable basis. The choice of the final basis is guided by the cross-validation error, which is computed based on the solution returned by the respective sparse solver. A good correlation of cross-validation and validation error is crucial for the scheme to make good choices -- which for example is not the case for the solver OMP.
Overall, there is no combination of solver and basis-adaptive scheme that was consistently best across all models and experimental design sizes, although averaging over all four cases, \SPloo{} together with forward neighbor basis adaptivity outperformed all other combinations.

Since no clear best combination of solver and basis adaptivity scheme emerged, we introduced an automatic selection step. Based on a suitable error estimate, automatic selection chooses one PCE solution out of several candidate solutions computed by various combinations of solvers and basis adaptivity schemes.
We found that automatic selection using any hybrid cross-validation error estimate performs better than the fixed rules obtained from the basis adaptivity benchmark: automatic selection attains the best possible relative MSE almost as often as the best solver-basis adaptivity combination, while being significantly more robust. 
An additional advantage of automatic selection is that it is 
independent of model dimension or size of the available experimental design, unlike the proposed fixed rules, which rely on the somewhat arbitrary, albeit simple, classification we applied (low/high dimension and small/large experimental design).

These findings demonstrate that when building a sparse PCE for an expensive black-box computational model, it is worth it to carefully select a sparse solver, and to apply a basis-adaptive scheme, because the difference in relative MSE between different combinations of methods on the same experimental design can be larger than one order of magnitude. While we could identify a number of methods that generally perform well, and others that should be avoided, as we described above, a superior strategy is to compute several PCE solutions and perform a final model selection using one of the presented hybrid cross-validation error estimators.

Further research could investigate the use of true cross-validation for automatic selection instead of the hybrid estimates which we used here. Also, it might be possible to identify other problem characteristics besides model dimension and the size of the available experimental design to guide the choice of methods in the sparse PCE framework. 
A promising class of methods combines basis adaptivity with the sequential enrichment of the experimental design, adapted to the current candidate or active basis \citep{FajraouiMarelli2017, Diaz2018, Hampton2018}, which might be able to further improve on the results obtained here.


\section*{Acknowledgements}
We thank John Jakeman (Sandia National Laboratories), Negin Alemazkoor (University of Virginia), Hadi Meidani (University of Illinois at Urbana-Champaign), and Jerrad Hampton (University of Colorado Boulder) for generously providing their code and explanations. We thank John Jakeman for his useful hints to relevant literature on basis adaptivity.


\bibliography{basis_adaptive_sparse_PCE_neutral.bbl}


\appendix

\section{Details on the settings for the basis adaptivity schemes in the benchmark}
\label{app:basis_adapt_details}

In Table~\ref{table:BAdetails}, we list the basis adaptivity settings for each of the 11 models in our benchmark. 
\begin{itemize}
	\item The rule for the static basis (ST) is to choose a total-degree basis with $p$ so that the number of basis functions $P$ is closest to $\frac{10}{3} N$, where $N$ is the number of ED points. Here, for low-dimensional models, the q-norm is chosen as $q = 1$, while for high-dimensional models, we use $q = 0.5$.
	
	\item For degree and q-norm basis adaptivity (PQ), we choose the ranges for degree and q-norm as large as possible while still keeping the number of basis functions computationally manageable (rule of thumb $\lesssim 10^4$).
	
	\item For forward neighbor degree adaptivity (FN), the degree of the initial basis is chosen so that the size of the basis is closest to $10N$ (as recommended in \citep{Jakeman2015}), while we set the q-norm to the maximum of the q-norm-range for PQ basis adaptivity.
	
	\item Finally, for anisotropic degree basis adaptivity, which is only used for low-dimensional models, we use $q = 1$ and $p = \lceil \frac{p_\text{max}}{2} \rceil$, where $p_\text{max}$ is the maximum of the degree range for PQ basis adaptivity.
\end{itemize}

\begin{table}[htbp]
	\footnotesize
	\centering
	\caption{Details on the initial bases and degree and q-norm ranges for the various basis adaptivity schemes}
	\label{table:BAdetails}
	\renewcommand{\arraystretch}{1.45}
	\begin{tabular}{p{2.8cm} c p{2.5cm} p{2.8cm} p{2.6cm} p{1.4cm}}
		\hline
		Model & dim $d$ & static basis & PQ range & FN initial & AD initial \\
		\hline
		Ishigami function & 3 
		& $p = 8$ (small ED)\newline / $p = 12$ (large), \newline $q = 1$ 
		& $p \in [1 \enum 25]$, \newline $q \in [0.5, 0.6 \enum 1]$
		& $p{} = 12$ (small ED) \newline / $p{} = 19$ (large) 
		& $p{} = 13$ \\
		
		Undamped oscillator & 6
		& $p = 4$ / $p = 4$ , \newline $q = 1$
		& $p \in [1 \enum 10]$, \newline $q \in [0.5, 0.6 \enum 1]$
		& $p{} = 5$  / $p{} = 6$ 
		& $p{} = 5$\\
		
		Borehole function & 8
		& $p = 4$ / $p = 4$ , \newline $q = 1$
		& $p \in [1 \enum 10]$, \newline $q \in [0.5, 0.6 \enum 1]$
		& $p{} = 5$  / $p{} = 6$ 
		& $p{} = 5$\\ 
		
		Damped oscillator & 8
		& $p = 4$ / $p = 5$ , \newline $q = 1$
		& $p \in [1 \enum 7]$, \newline $q \in [0.5, 0.6 \enum 1]$
		& $p{} = 5$  / $p{} = 6$ 
		& $p{} = 4$\\ 
		
		Wingweight function & 10
		& $p = 3$ / $p = 4$ , \newline $q = 1$
		& $p \in [1 \enum 7]$, \newline $q \in [0.5, 0.6 \enum 1]$
		& $p{} = 4$  / $p{} = 5$ 
		& $p{} = 4$\\ 
		
		\textit{Truss model} & 10
		& $p = 3$ / $p = 4$ , \newline $q = 1$
		& $p \in [1 \enum 6]$, \newline $q \in [0.5, 0.6 \enum 1]$
		& $p{} = 4$  / $p{} = 5$ 
		& $p{} = 3$\\ 
		
		\hline
		Morris function & 20
		& $p = 6$ / $p = 8$ , \newline $q = 0.5$
		& $p \in [1 \enum 8]$, \newline $q \in [0.4, 0.5, 0.6]$
		& $p{} = 6$  / $p{} = 8$, \newline $q = 0.6$
		& -- \\ 
		
		\textit{Structural frame}\newline \textit{model} & 21
		& $p = 5$ / $p = 8$ , \newline $q = 0.5$
		& $p \in [1 \enum 8]$, \newline $q \in [0.4, 0.5, 0.6]$
		& $p{} = 6$  / $p{} = 8$, \newline $q = 0.6$
		& -- \\ 
		
		\textit{2-dim diffusion}\newline \textit{model}  & 53
		& $p = 3$ / $p = 4$ , \newline $q = 0.5$
		& $p \in [1 \enum 6]$, \newline $q \in [0.4, 0.5, 0.6]$
		& $p{} = 4$  / $p{} = 5$, \newline $q = 0.6$
		& -- \\ 
		
		\textit{1-dim diffusion}\newline \textit{model} & 62
		& $p = 3$ / $p = 4$ , \newline $q = 0.5$
		& $p \in [1 \enum 5]$, \newline $q \in [0.4, 0.5, 0.6]$
		& $p{} = 3$  / $p{} = 4$, \newline $q = 0.6$
		& -- \\ 
		
		100D function  & 100
		& $p = 3$ / $p = 4$ , \newline $q = 0.5$
		& $p \in [1 \enum 5]$, \newline $q = 0.5$
		& $p{} = 4$  / $p{} = 5$, \newline $q = 0.5$
		& -- \\ 
		\hline	
	\end{tabular}
\end{table}

\section{Additional results}
\label{app:benchmark_basis_adaptivity}


\subsection{Basis adaptivity benchmark: raw data}
\label{app:raw_data}
Figs.~\ref{fig:benchmark_basis_adaptivity_lowdim} and \ref{fig:benchmark_basis_adaptivity_highdim} show the raw data of the basis adaptivity benchmark: for each model and ED size, we display the boxplots of resulting relative MSE for each combination of sparse solver and basis-adaptive scheme. 
The star-shaped markers visualize the attained error for one specific ED realization.
They illustrate that the resulting errors are not always well correlated: on the same ED, some methods achieve one of their smallest errors while others produce one of their largest. 
Therefore, we continue in Section~\ref{sec:BA_benchmark} by investigating the relative performance of the different methods.

\begin{figure}[htbp]
	\subcaptionbox{Ishigami function, $N = 50$}
	{\includegraphics[width=.48\textwidth, trim=0 .3cm 1.6cm .4cm, clip, height=.21\textheight, keepaspectratio]
		{boxplot_ishigami_relMSE_basis_adaptivity_N50.pdf}}
	\hfill%
	\subcaptionbox{Ishigami function, $N = 150$}
	{\includegraphics[width=.48\textwidth, trim=0 .3cm 1.6cm .4cm, clip, height=.21\textheight, keepaspectratio]
		{boxplot_ishigami_relMSE_basis_adaptivity_N150.pdf}}%
	\\[6pt]
	\subcaptionbox{Undamped oscillator, $N = 60$}
	 {\includegraphics[width=.48\textwidth, trim=0 .3cm 1.6cm .4cm, clip, height=.21\textheight, keepaspectratio]
		{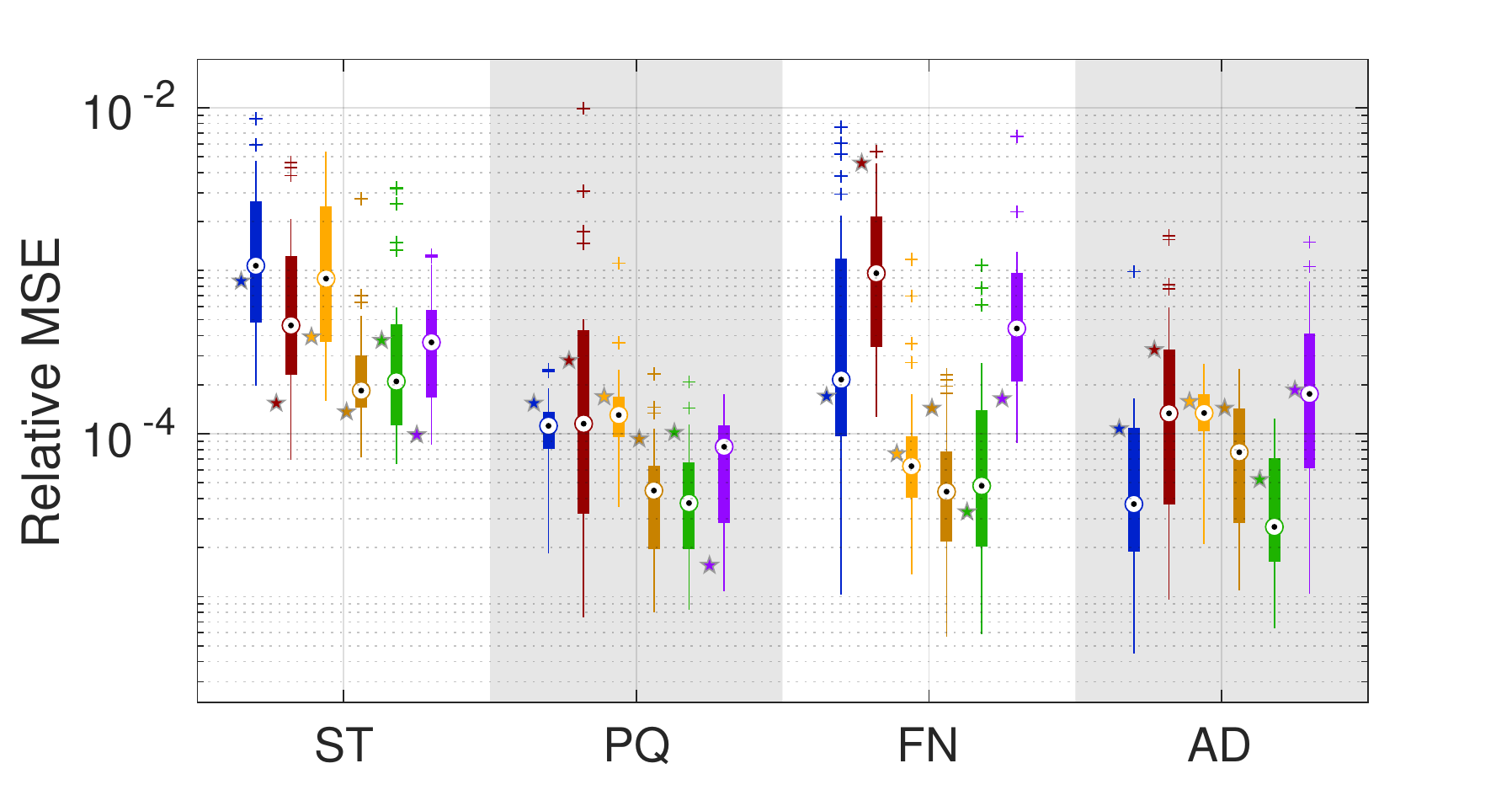}}
	\hfill%
	\subcaptionbox{Undamped oscillator, $N = 120$}
	 {\includegraphics[width=.48\textwidth, trim=0 .3cm 1.6cm .4cm, clip, height=.21\textheight, keepaspectratio]
	 	{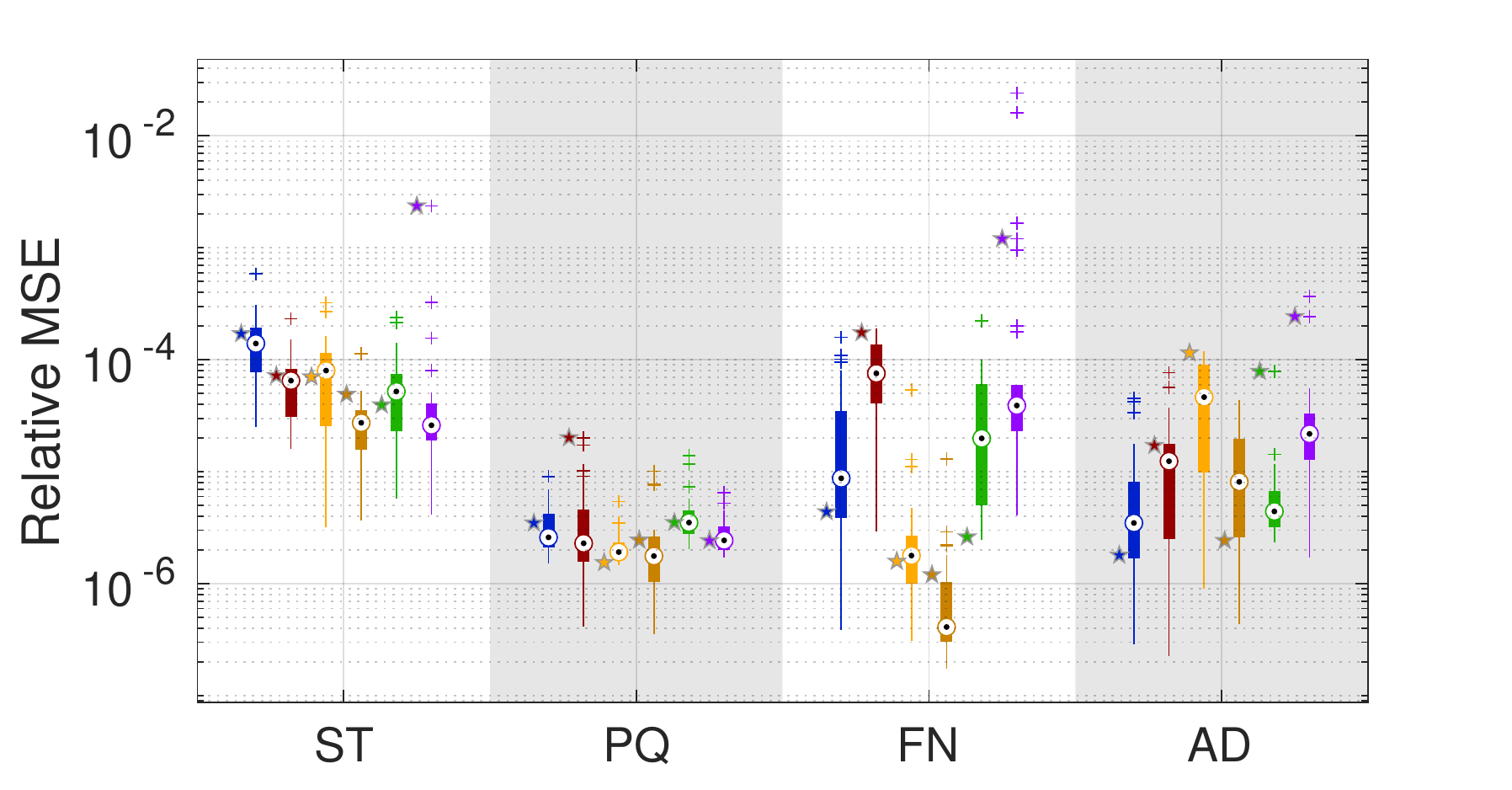}}%
	\\[6pt]
	\subcaptionbox{Borehole function, $N = 100$}
	 {\includegraphics[width=.48\textwidth, trim=0 .3cm 1.6cm .4cm, clip, height=.21\textheight, keepaspectratio]
		{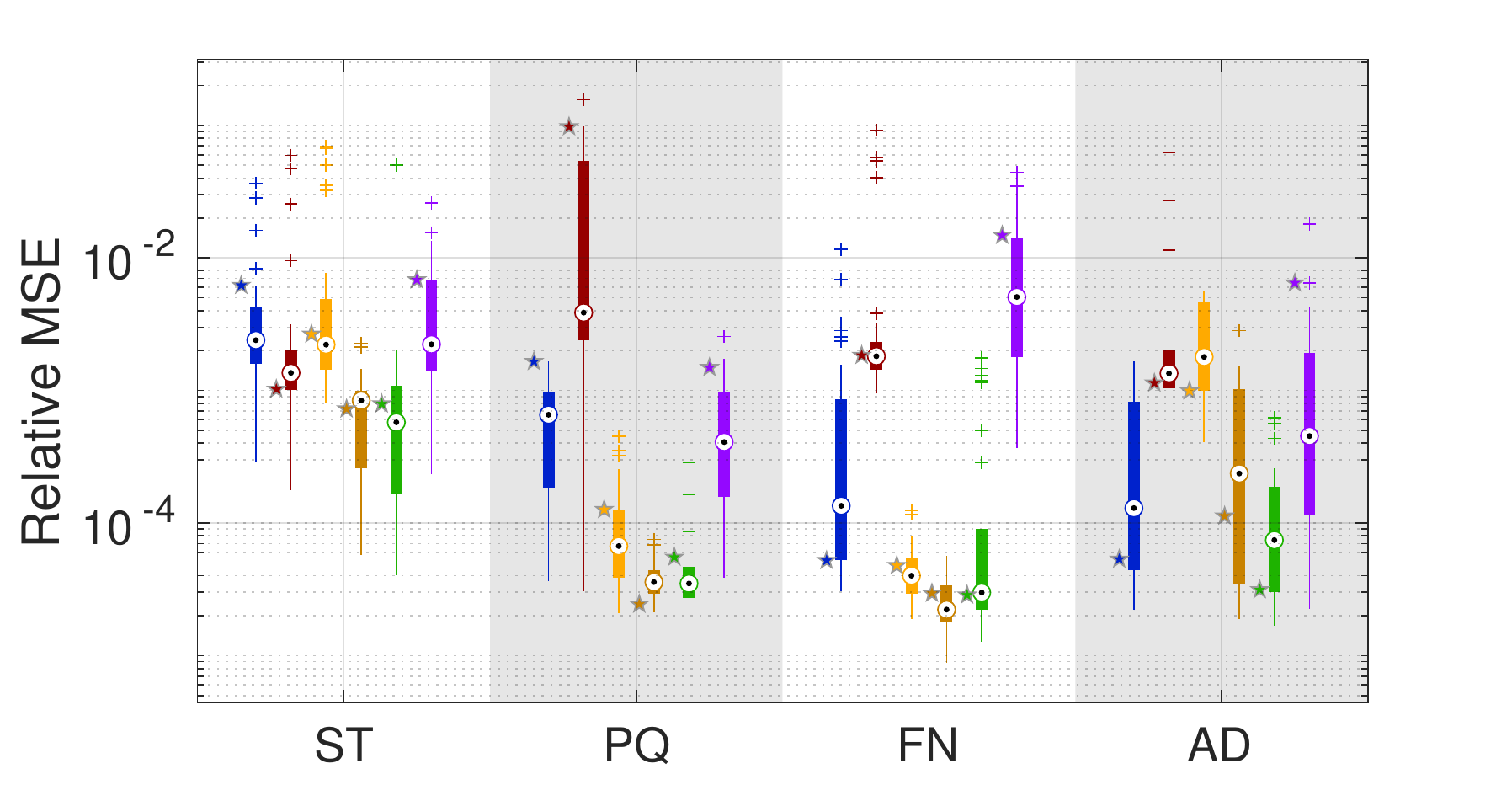}}
	\hfill%
	\subcaptionbox{Borehole function, $N = 250$}
	 {\includegraphics[width=.48\textwidth, trim=0 .3cm 1.6cm .4cm, clip, height=.21\textheight, keepaspectratio]
		{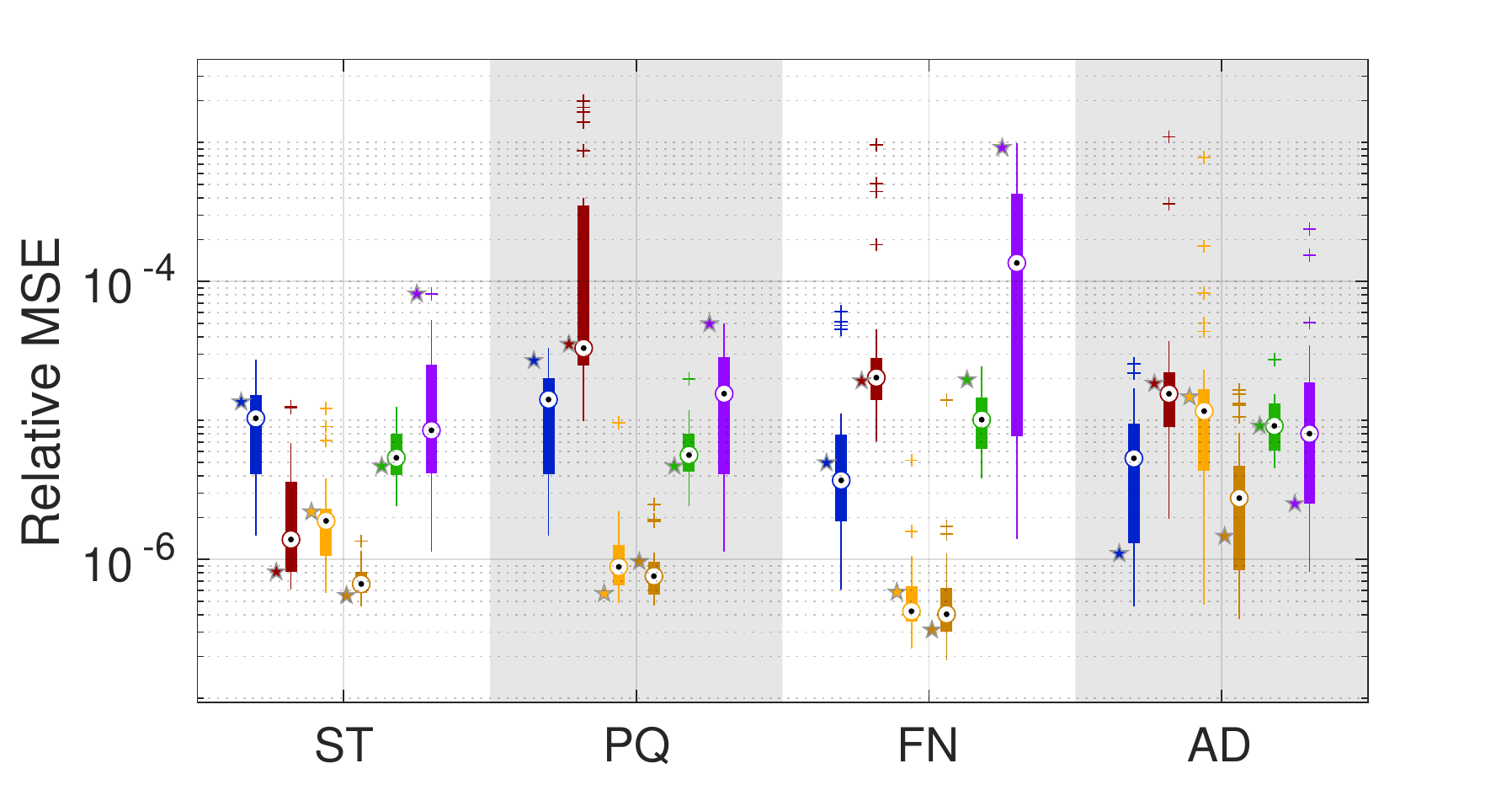}}%
	\\[6pt]
	\subcaptionbox{Damped oscillator, $N = 150$}
	 {\includegraphics[width=.48\textwidth, trim=0 .3cm 1.6cm .4cm, clip, height=.21\textheight, keepaspectratio]
	 	{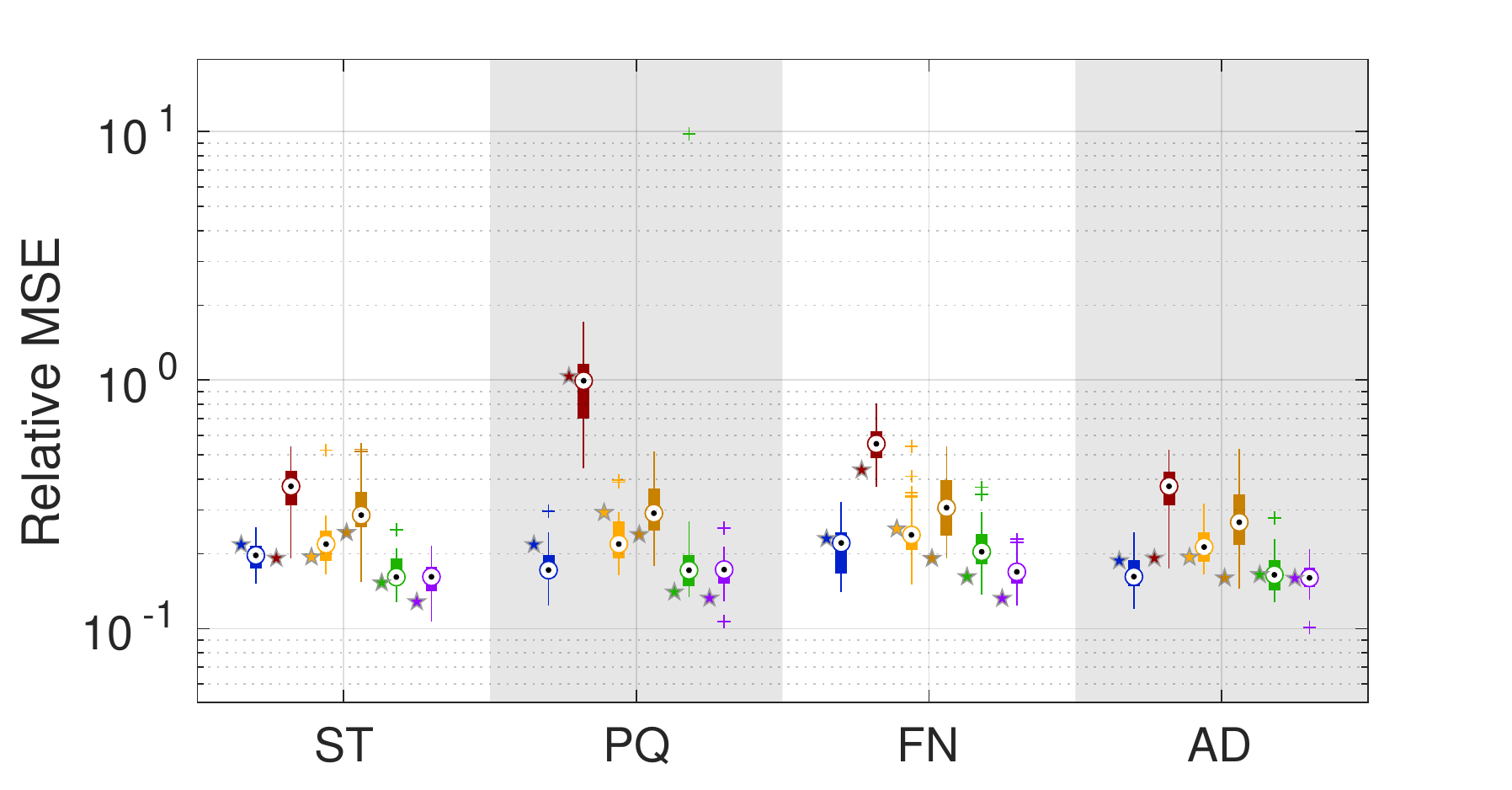}}
	\hfill%
	\subcaptionbox{Damped oscillator, $N = 350$} 
	{\includegraphics[width=.48\textwidth, trim=0 .3cm 1.6cm .4cm, clip, height=.21\textheight, keepaspectratio]
		{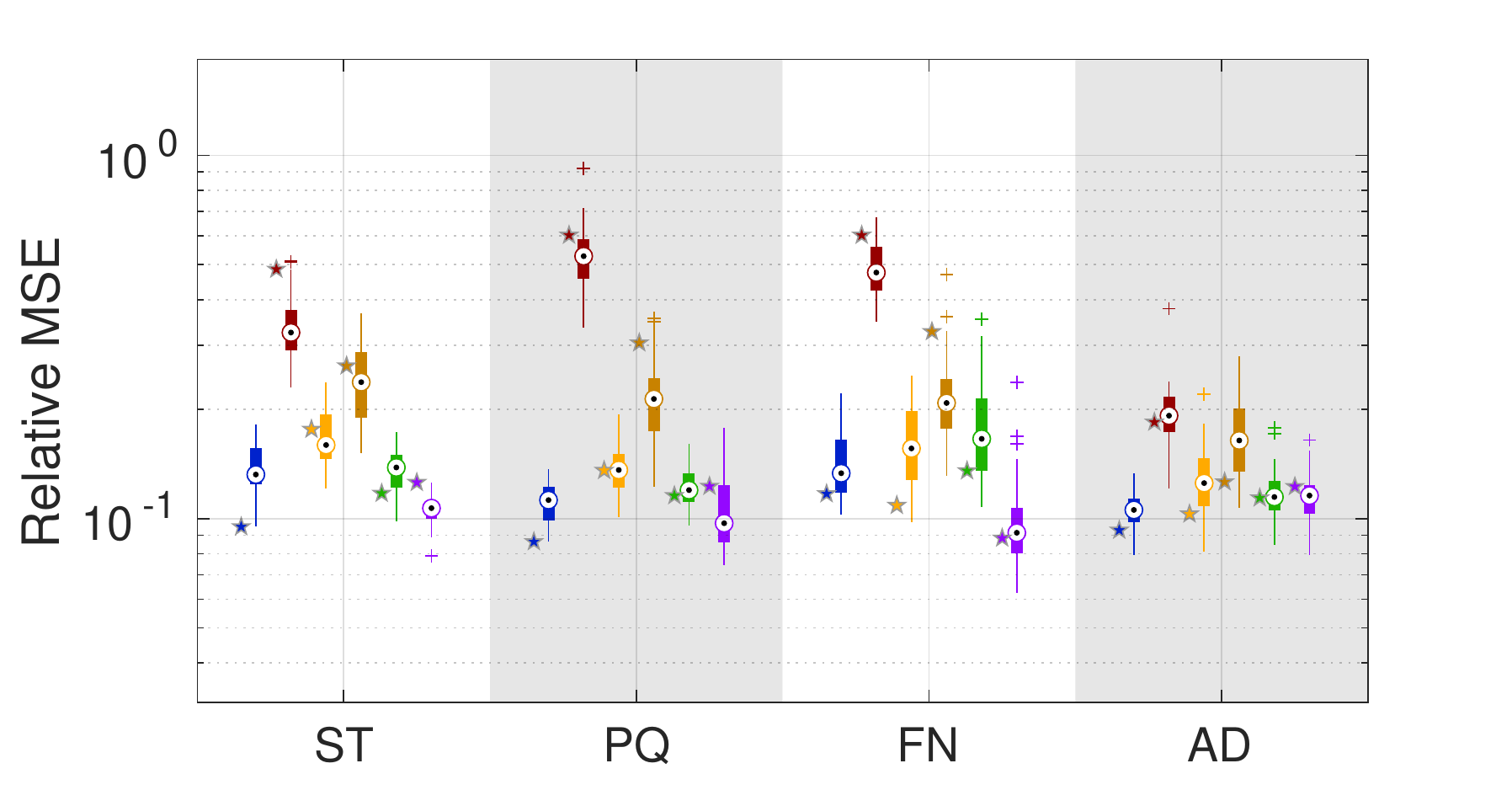}}
	\vspace{4pt}
	\caption{ Comparison of different combinations of solvers and basis adaptivity schemes for \textbf{low-dimensional models}.
	We show validation errors attained by all 24 combinations of methods on 30 realizations of experimental design. The boxplots visualize the attained errors across all 30 realizations. 
	The star-shaped markers denote the attained errors of all combinations for one selected ED realization. They highlight one set of data points which is also part of the larger set visualized by the boxplots.
	Left: small ED; right: large ED. Abbreviations of basis-adaptive schemes: ST: static basis; PQ: degree and q-norm; FN: forward neighbor; AD: anisotropic degree. Continued in next figure.
	}
\end{figure}
\clearpage
\begin{figure}[H]
	\ContinuedFloat	
	\subcaptionbox{Truss model, $N = 100$}
	 {\includegraphics[width=.48\textwidth, trim=0 .3cm 1.6cm .4cm, clip, height=.21\textheight, keepaspectratio]
	 	{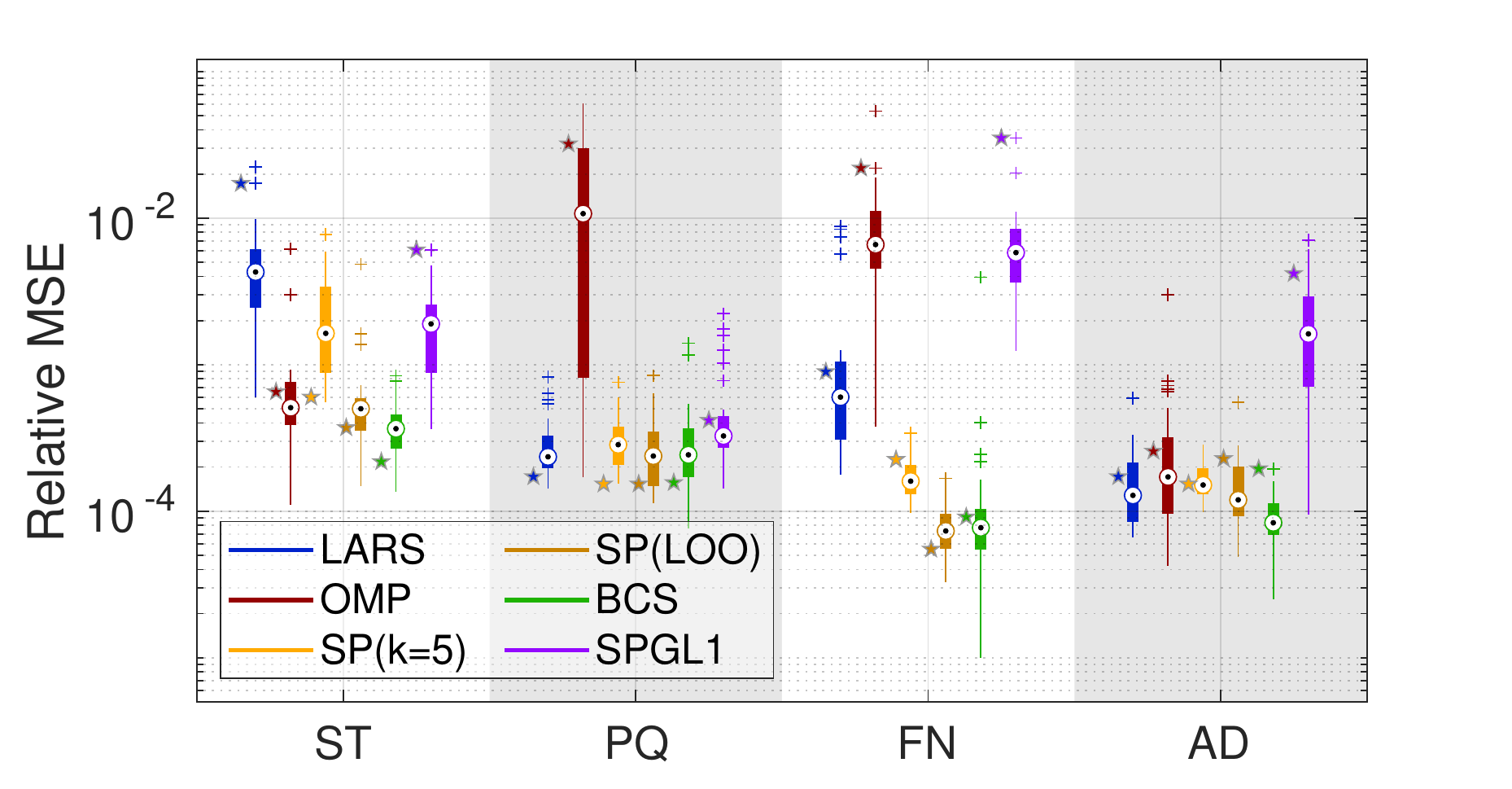}}
	\hfill%
	\subcaptionbox{Truss model, $N = 250$} 
	{\includegraphics[width=.48\textwidth, trim=0 .3cm 1.6cm .4cm, clip, height=.21\textheight, keepaspectratio]
		{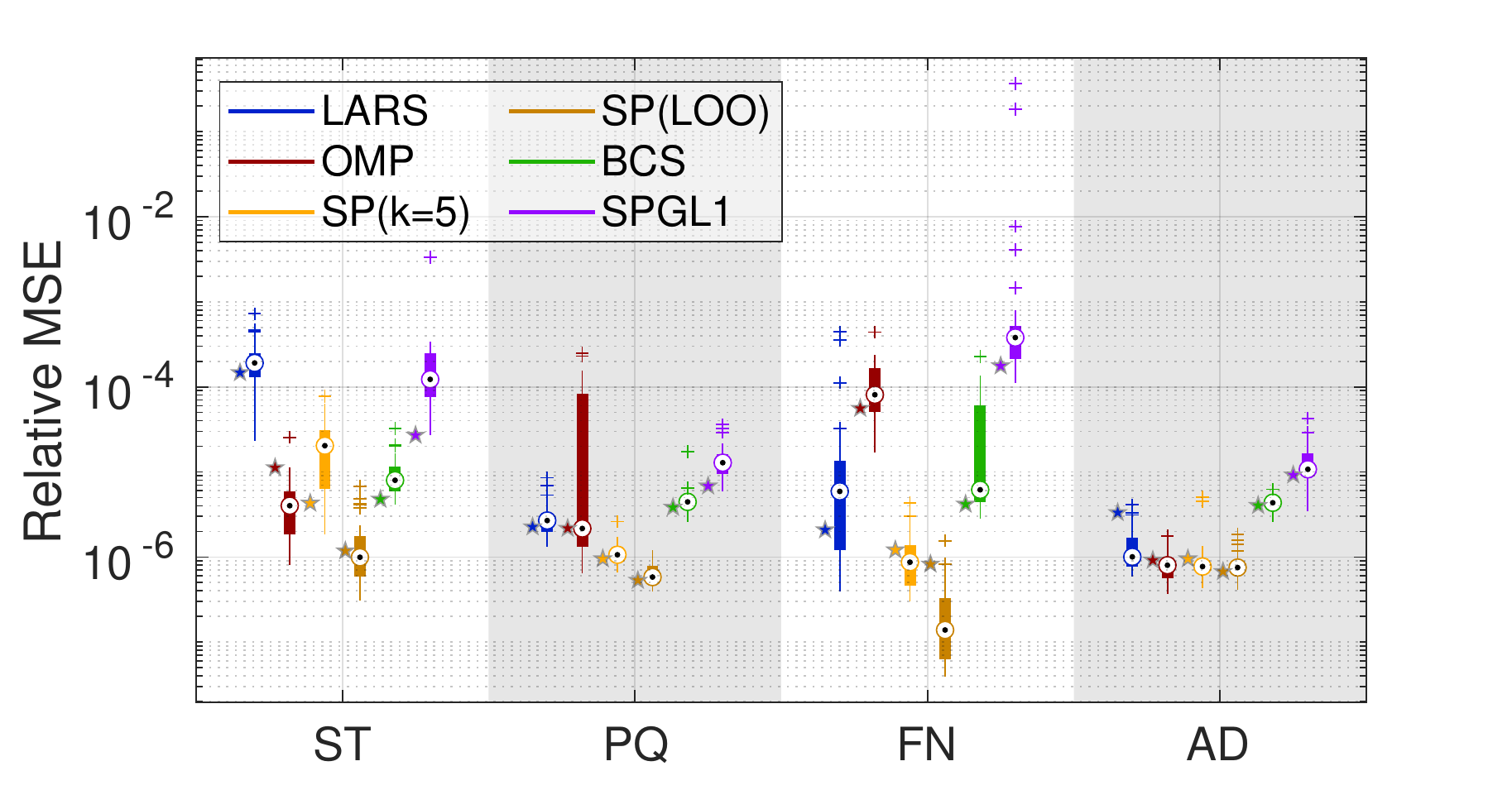}}%
	\\[6pt]
	\subcaptionbox{Wingweight function, $N = 100$}
	 {\includegraphics[width=.48\textwidth, trim=0 .3cm 1.6cm .4cm, clip, height=.21\textheight, keepaspectratio]
		{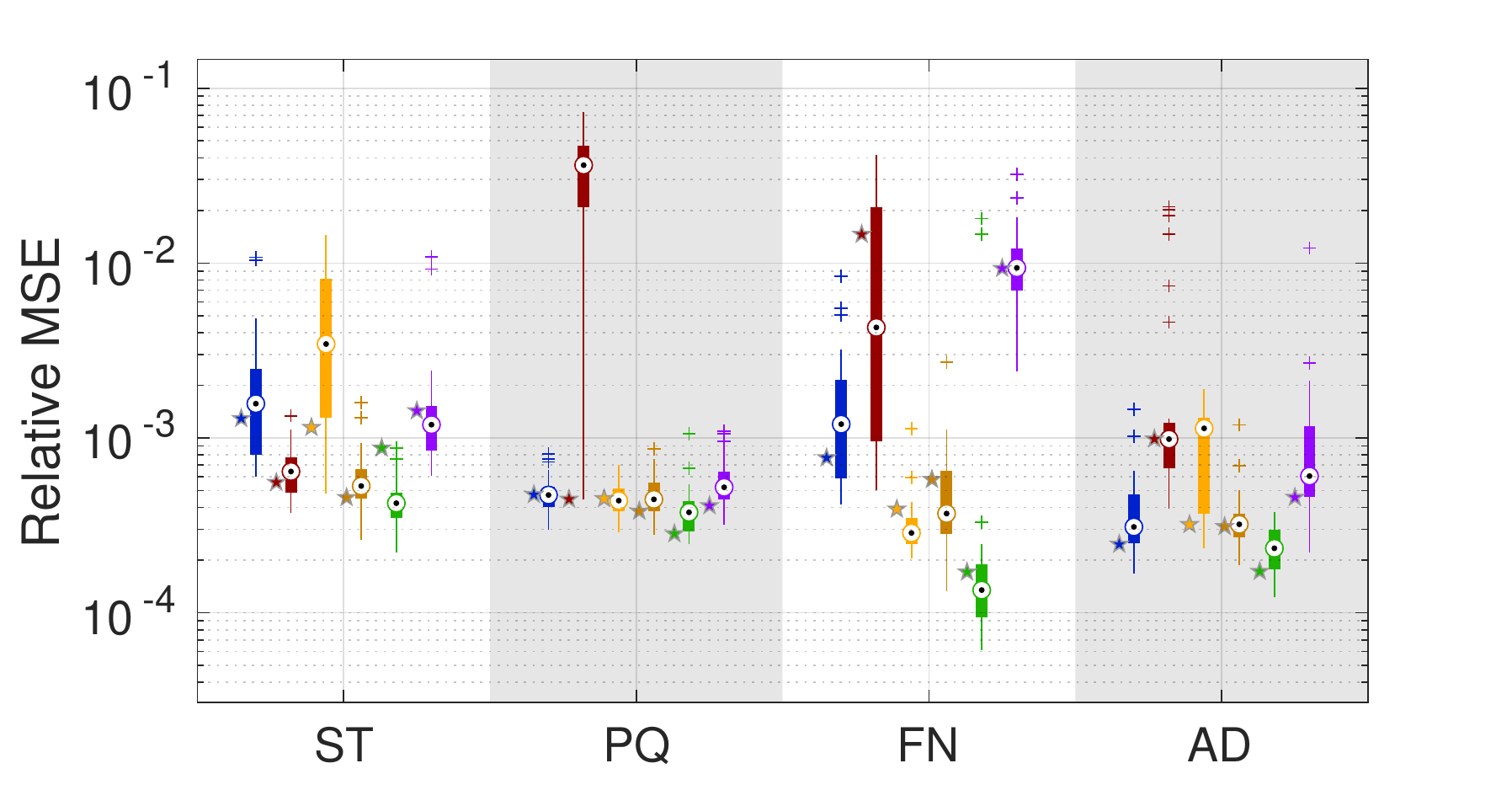}}
	\hfill%
	\subcaptionbox{Wingweight function, $N = 250$}
	 {\includegraphics[width=.48\textwidth, trim=0 .3cm 1.6cm .4cm, clip, height=.21\textheight, keepaspectratio]
	 	{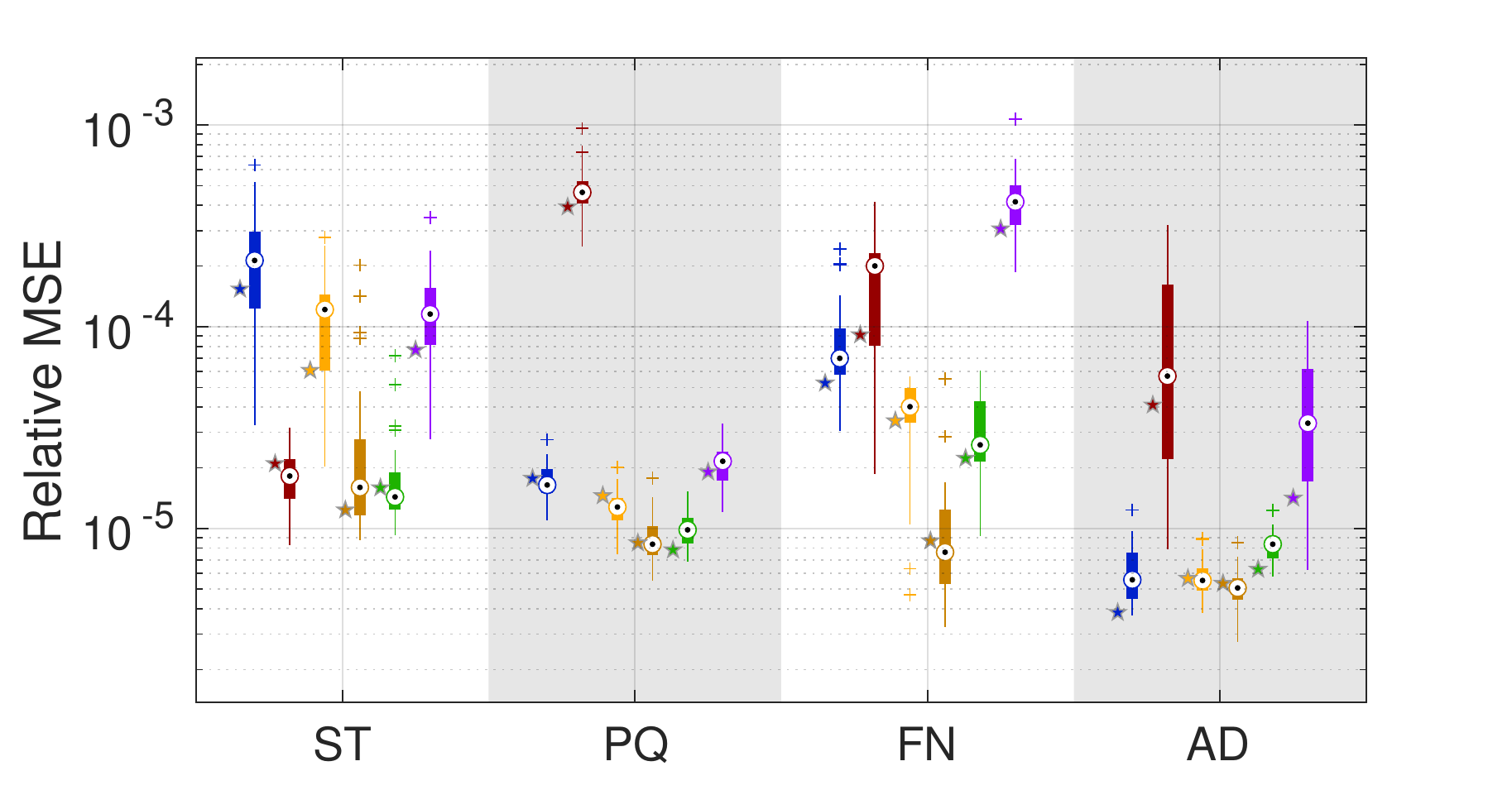}}%
	\vspace{4pt}
	\caption{Continued. Comparison of different combinations of solvers and basis adaptivity schemes for \textbf{low-dimensional models}.
	We show validation errors attained by all 24 combinations of methods on 30 realizations of experimental design. The boxplots visualize the attained errors across all 30 realizations. 
	The star-shaped markers denote the attained errors of all combinations for one selected ED realization. They highlight one set of data points which is also part of the larger set visualized by the boxplots.
	Left: small ED; right: large ED. Abbreviations of basis-adaptive schemes: ST: static basis; PQ: degree and q-norm; FN: forward neighbor; AD: anisotropic degree.
	}
	\label{fig:benchmark_basis_adaptivity_lowdim}	
\end{figure}

\begin{figure}[H]
	\subcaptionbox{Morris function, $N = 150$} {\includegraphics[width=.48\textwidth, trim=0 .3cm 1.6cm .4cm, clip, height=.21\textheight, keepaspectratio]{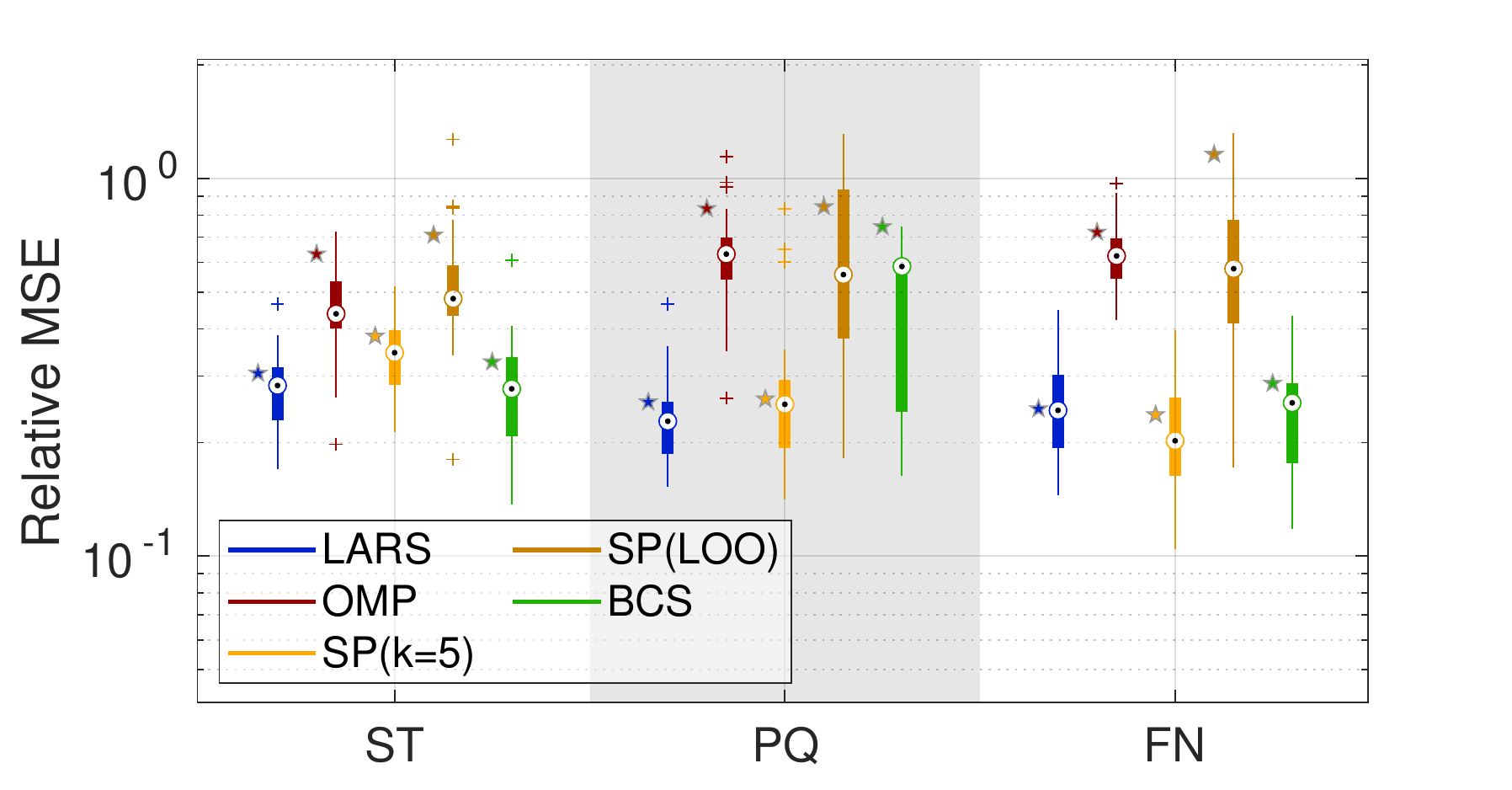}}
	\hfill%
	\subcaptionbox{Morris function, $N = 350$} {\includegraphics[width=.48\textwidth, trim=0 .3cm 1.6cm .4cm, clip, height=.21\textheight, keepaspectratio]{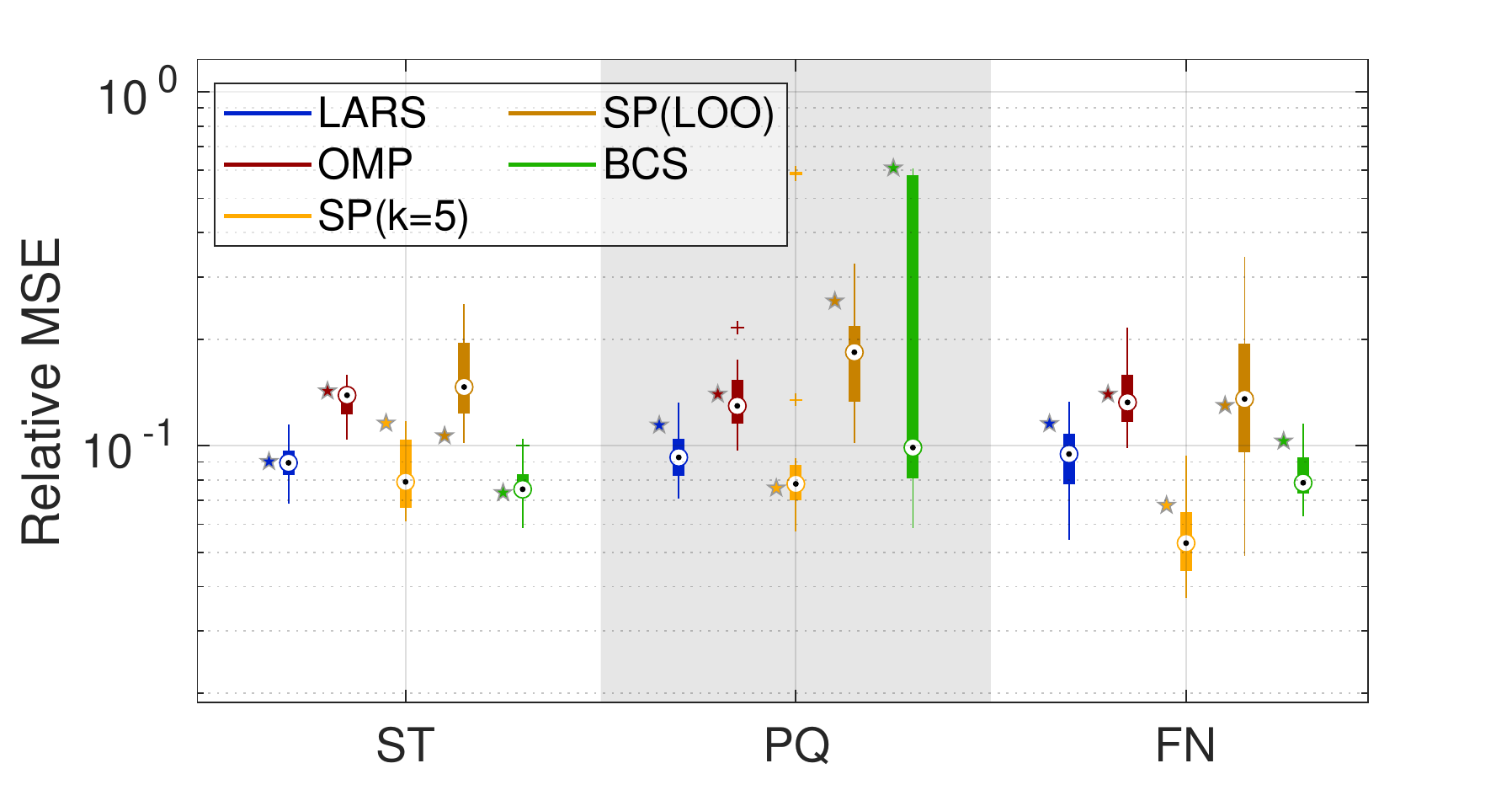}}
	\vspace{4pt}
	\caption{Comparison of different combinations of solvers and basis adaptivity schemes for \textbf{high-dimensional models}.
	We show validation errors attained by all 15 combinations of methods on 30 realizations of experimental design. The boxplots visualize the attained errors across all 30 realizations. 
	The star-shaped markers denote the attained errors of all combinations for one selected ED realization. They highlight one set of data points which is also part of the larger set visualized by the boxplots.
	Left: small ED; right: large ED. Abbreviations of basis-adaptive schemes: ST: static basis; PQ: degree and q-norm; FN: forward neighbor.
	Continued in next figure.
	}
\end{figure}
\begin{figure}[htbp]
	\ContinuedFloat
	\subcaptionbox{Structural frame, $N = 150$} {\includegraphics[width=.48\textwidth, trim=0 .3cm 1.6cm .4cm, clip, height=.21\textheight, keepaspectratio]{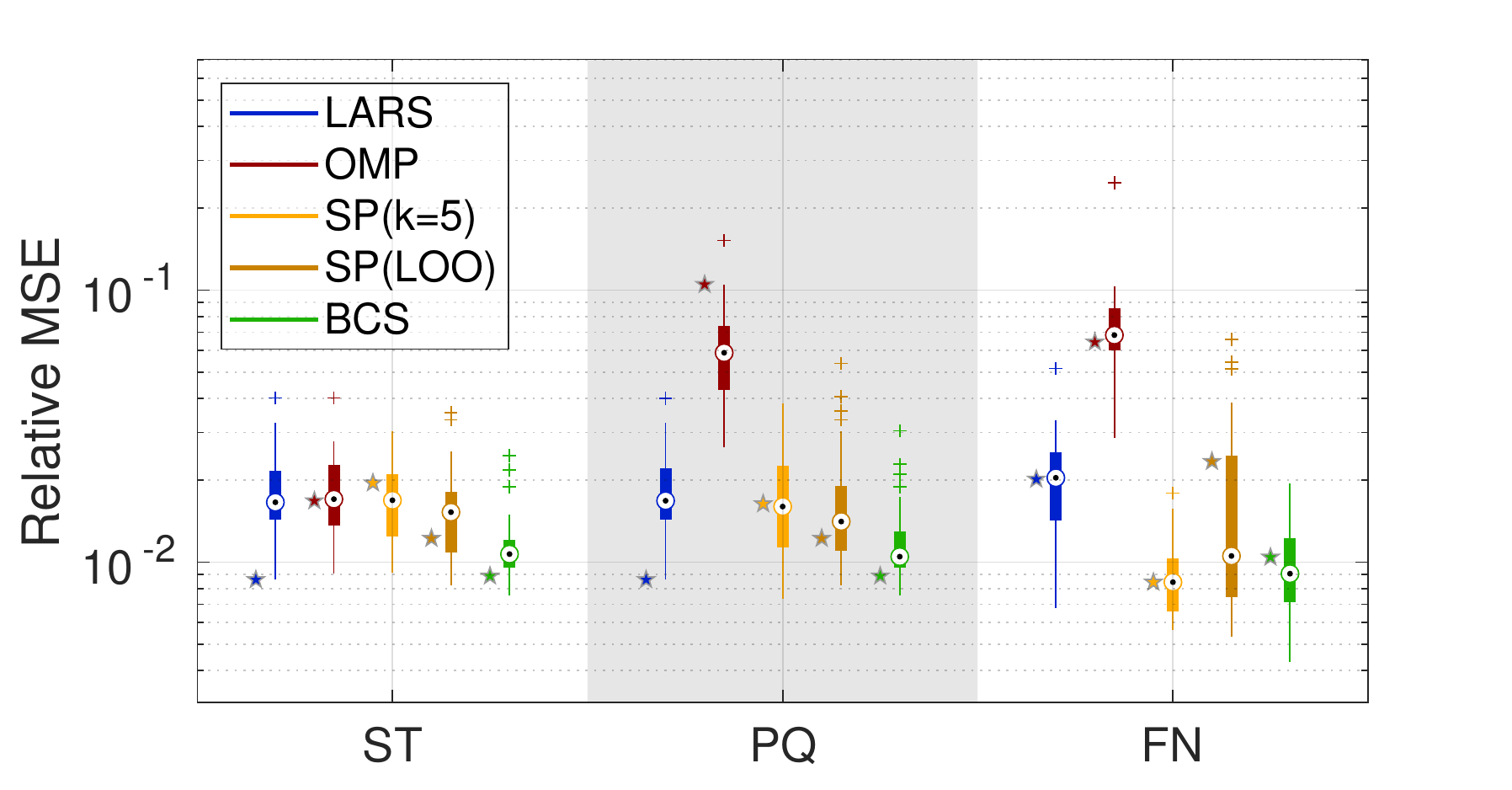}}
	\hfill%
	\subcaptionbox{Structural frame, $N = 350$} {\includegraphics[width=.48\textwidth, trim=0 .3cm 1.6cm .4cm, clip, height=.21\textheight, keepaspectratio]{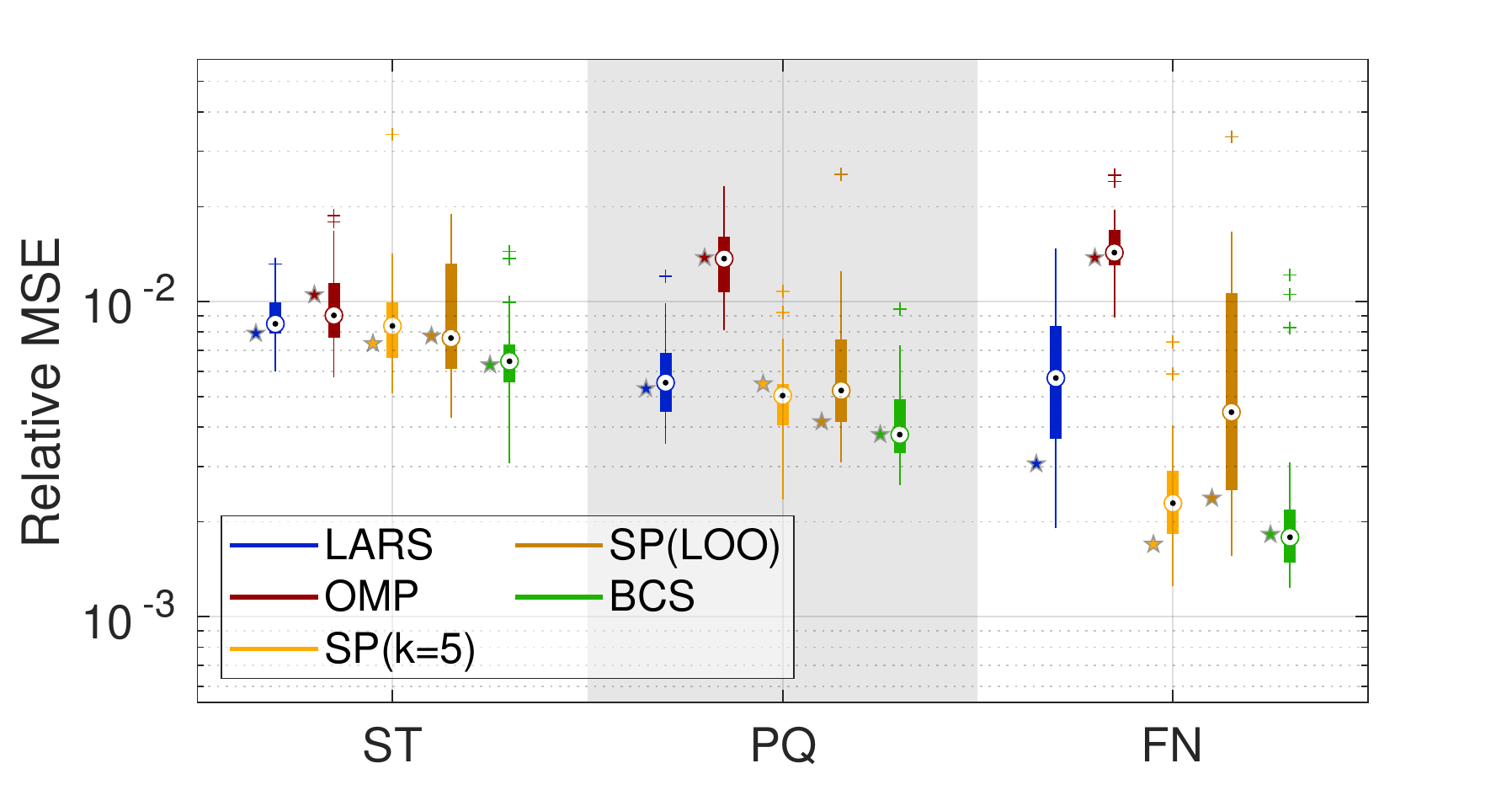}}%
	\\[6pt]
	\subcaptionbox{Diffusion 2D, $N = 100$} {\includegraphics[width=.48\textwidth, trim=0 .3cm 1.6cm .4cm, clip, height=.21\textheight, keepaspectratio]{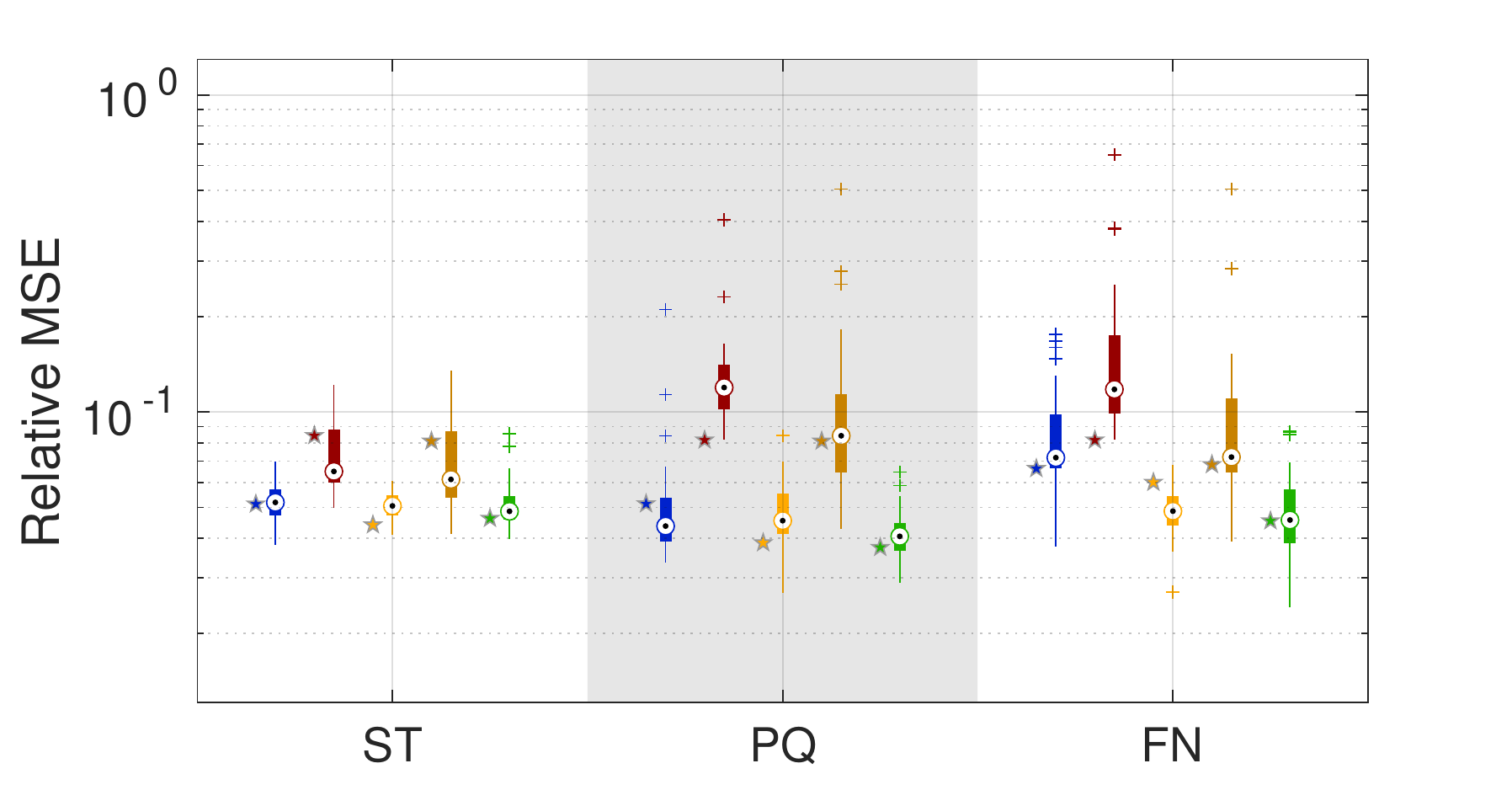}}
	\hfill%
	\subcaptionbox{Diffusion 2D, $N = 400$} {\includegraphics[width=.48\textwidth, trim=0 .3cm 1.6cm .4cm, clip, height=.21\textheight, keepaspectratio]{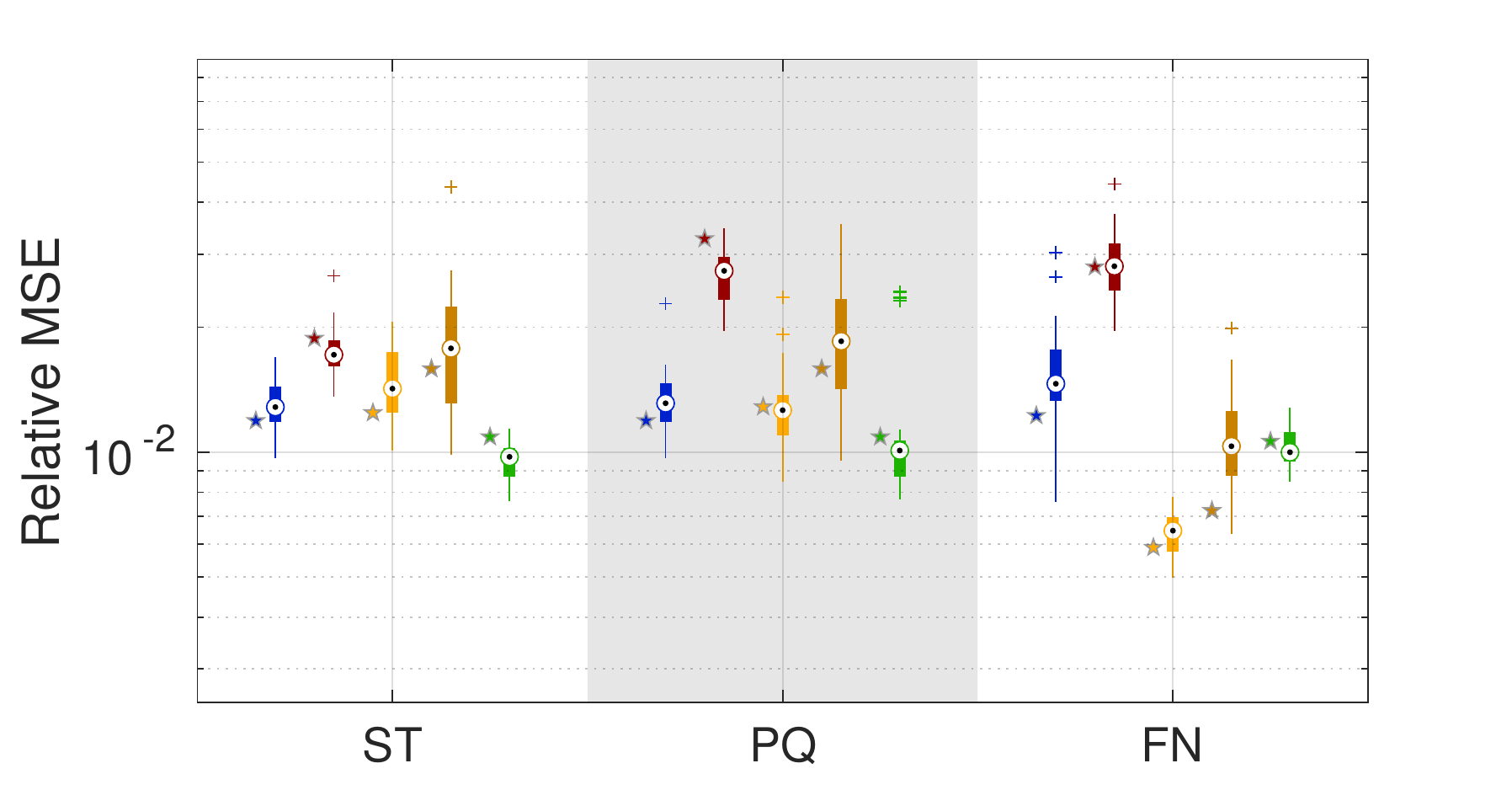}}%
	\\[6pt]
	\subcaptionbox{Diffusion 1D, $N = 100$} {\includegraphics[width=.48\textwidth, trim=0 .3cm 1.6cm .4cm, clip, height=.21\textheight, keepaspectratio]{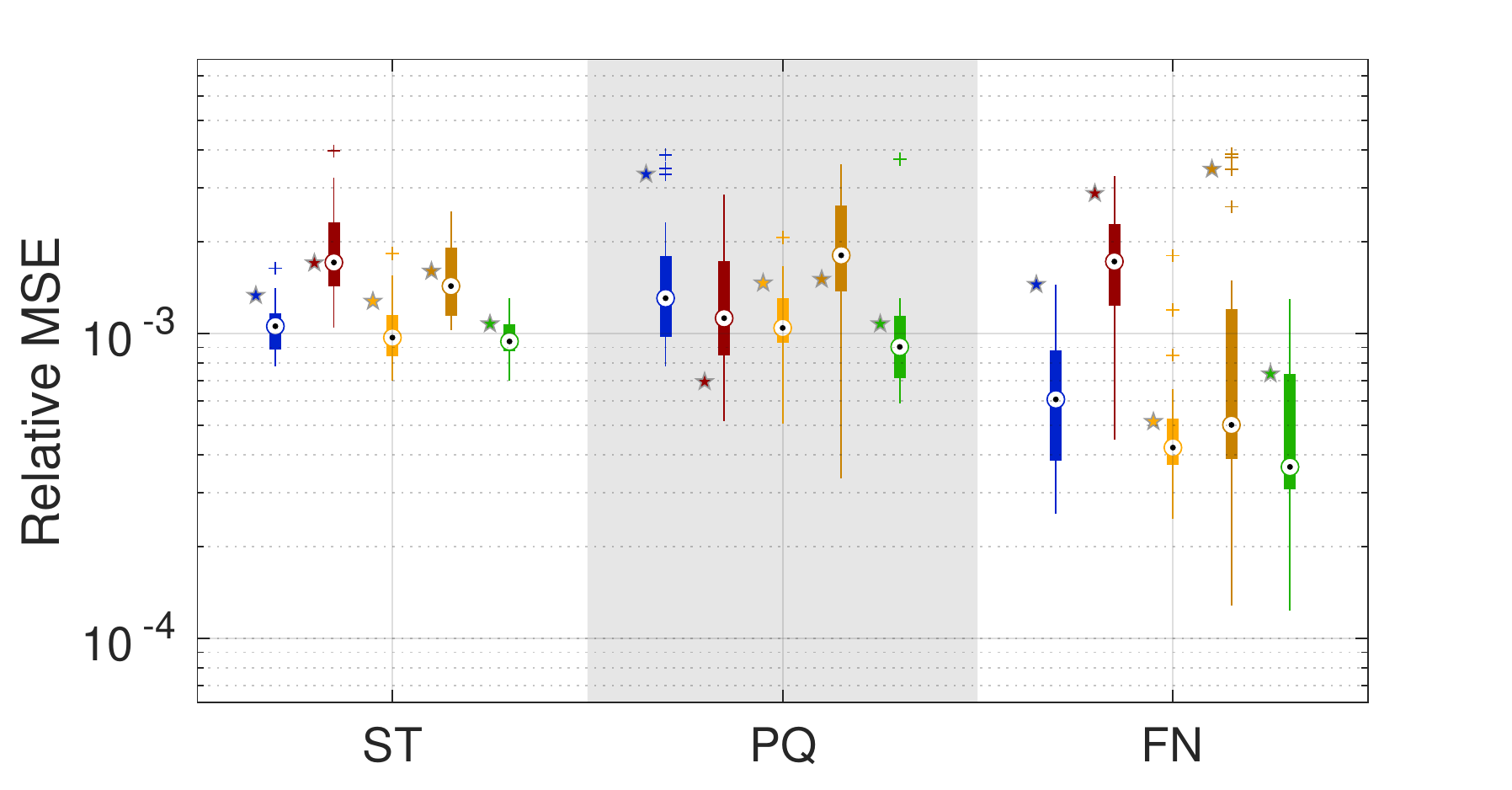}}
	\hfill%
	\subcaptionbox{Diffusion 1D, $N = 400$} {\includegraphics[width=.48\textwidth, trim=0 .3cm 1.6cm .4cm, clip, height=.21\textheight, keepaspectratio]{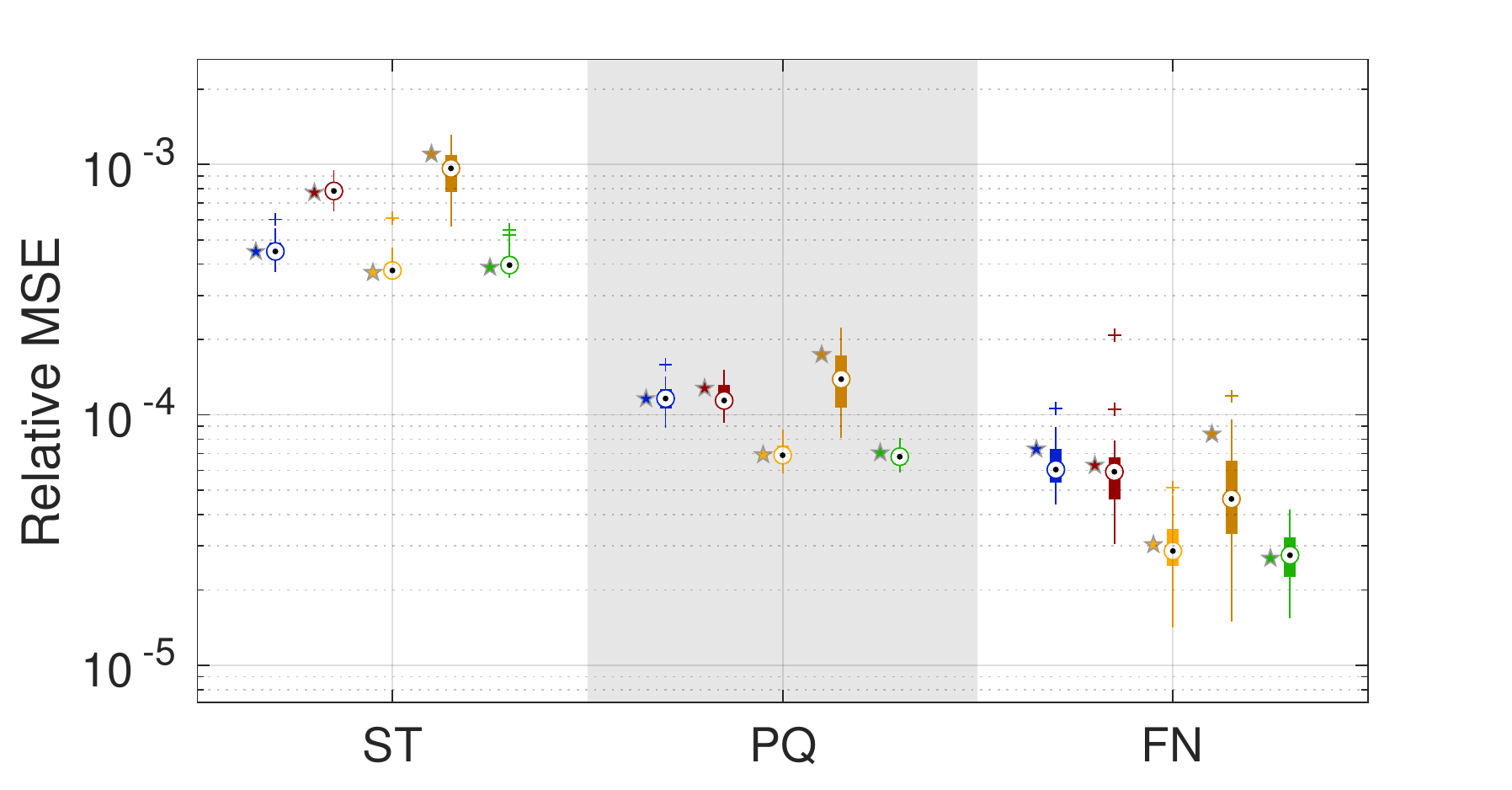}}%
	\\[6pt]
	\subcaptionbox{100D function, $N = 400$ \label{fig:BA_100D_small}} {\includegraphics[width=.48\textwidth, trim=0 .3cm 1.6cm .4cm, clip, height=.21\textheight, keepaspectratio]{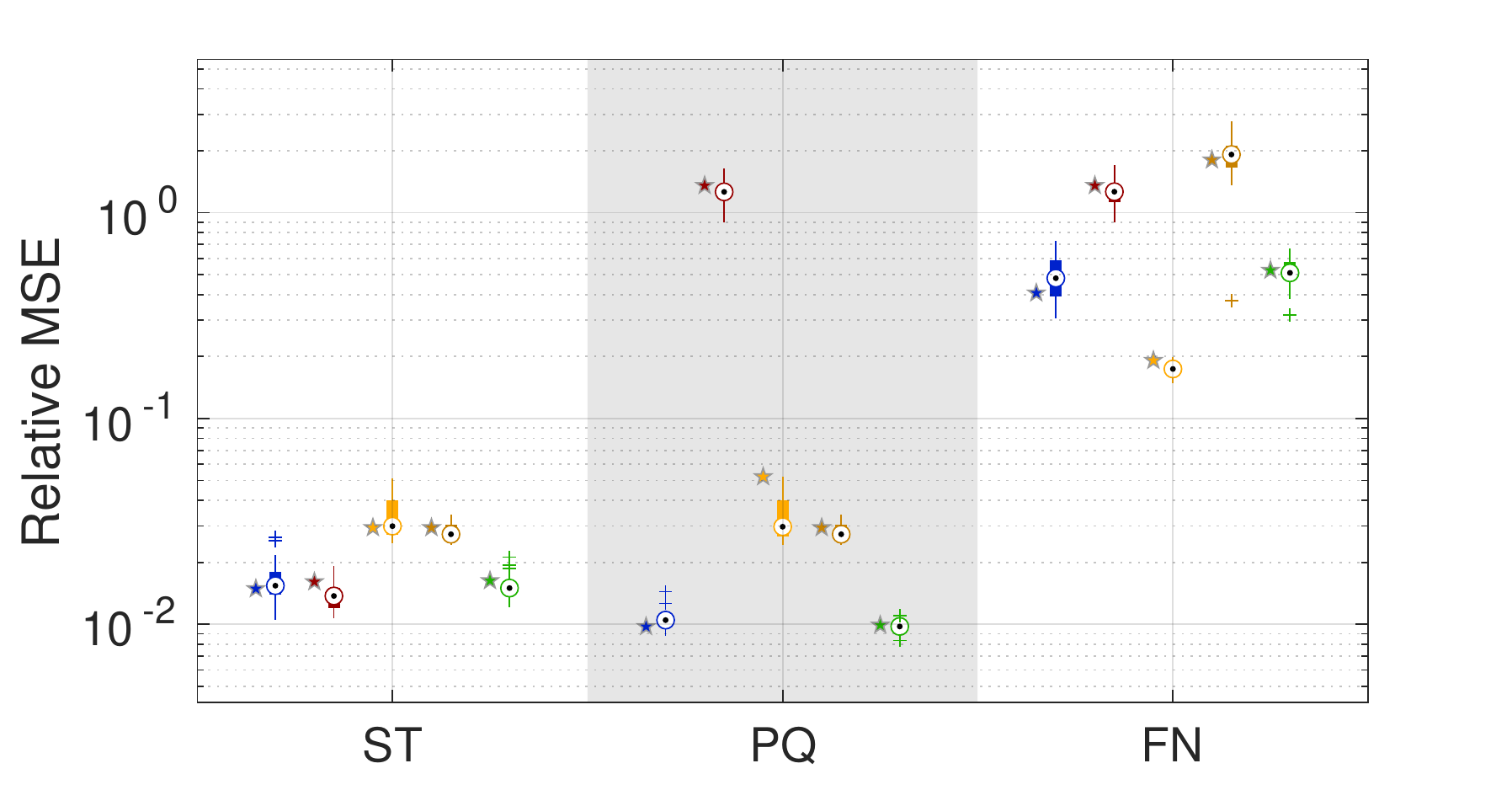}}
	\hfill%
	\subcaptionbox{100D function, $N = 1200$ \label{fig:BA_100D_large}} {\includegraphics[width=.48\textwidth, trim=0 .3cm 1.6cm .4cm, clip, height=.21\textheight, keepaspectratio]{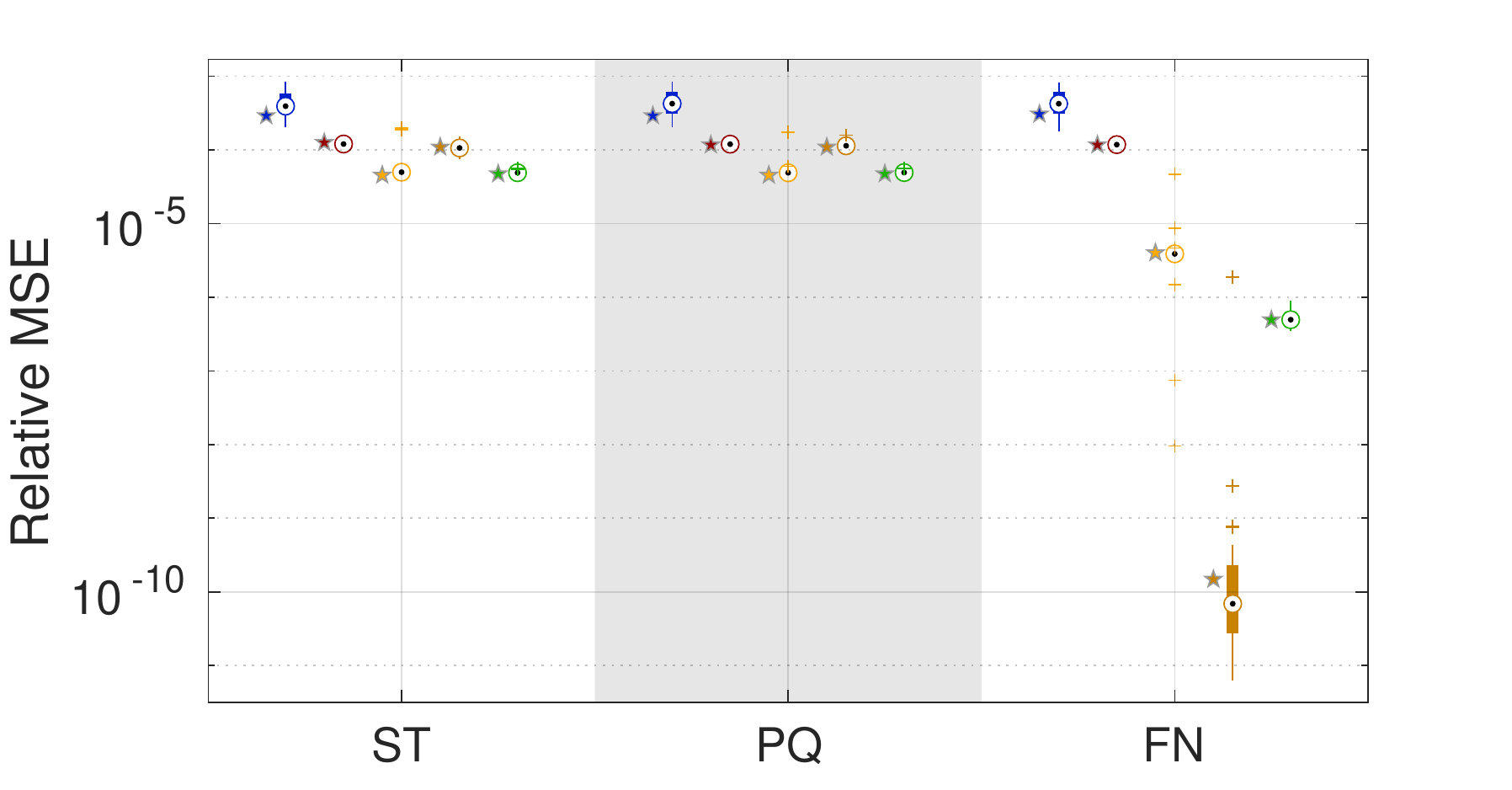}}%
	\vspace{4pt}
	\caption{Continued. 
	Comparison of different combinations of solvers and basis adaptivity schemes for \textbf{high-dimensional models}.
	We show validation errors attained by all 15 combinations of methods on 30 realizations of experimental design. The boxplots visualize the attained errors across all 30 realizations. 
	The star-shaped markers denote the attained errors of all combinations for one selected ED realization. They highlight one set of data points which is also part of the larger set visualized by the boxplots.
	Left: small ED; right: large ED. Abbreviations of basis-adaptive schemes: ST: static basis; PQ: degree and q-norm; FN: forward neighbor; AD: anisotropic degree.
	}
	\label{fig:benchmark_basis_adaptivity_highdim}	
\end{figure}

\subsection{Bar plots for all combinations of methods}
\label{app:BA_bar_plots}
\begin{figure}[htbp]
	\centering
	{\includegraphics[width=.7\textwidth]{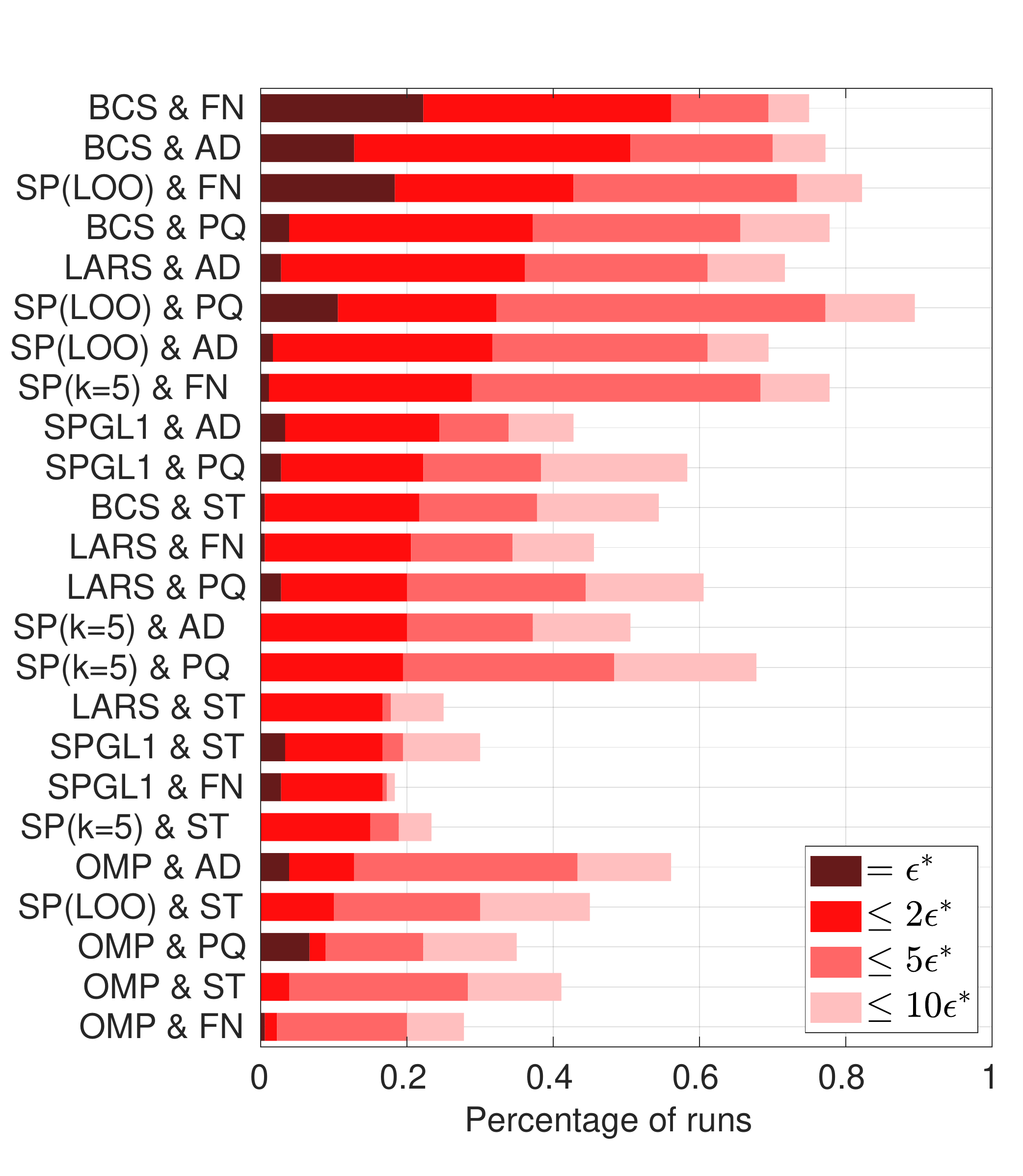}}
	\caption{\textbf{Small ED sizes}. Aggregated results for \textbf{low-dim} models.}
	\label{fig:aggr_lowdim_small}
\end{figure}

\begin{figure}[htbp]
	\centering
	{\includegraphics[width=.7\textwidth]{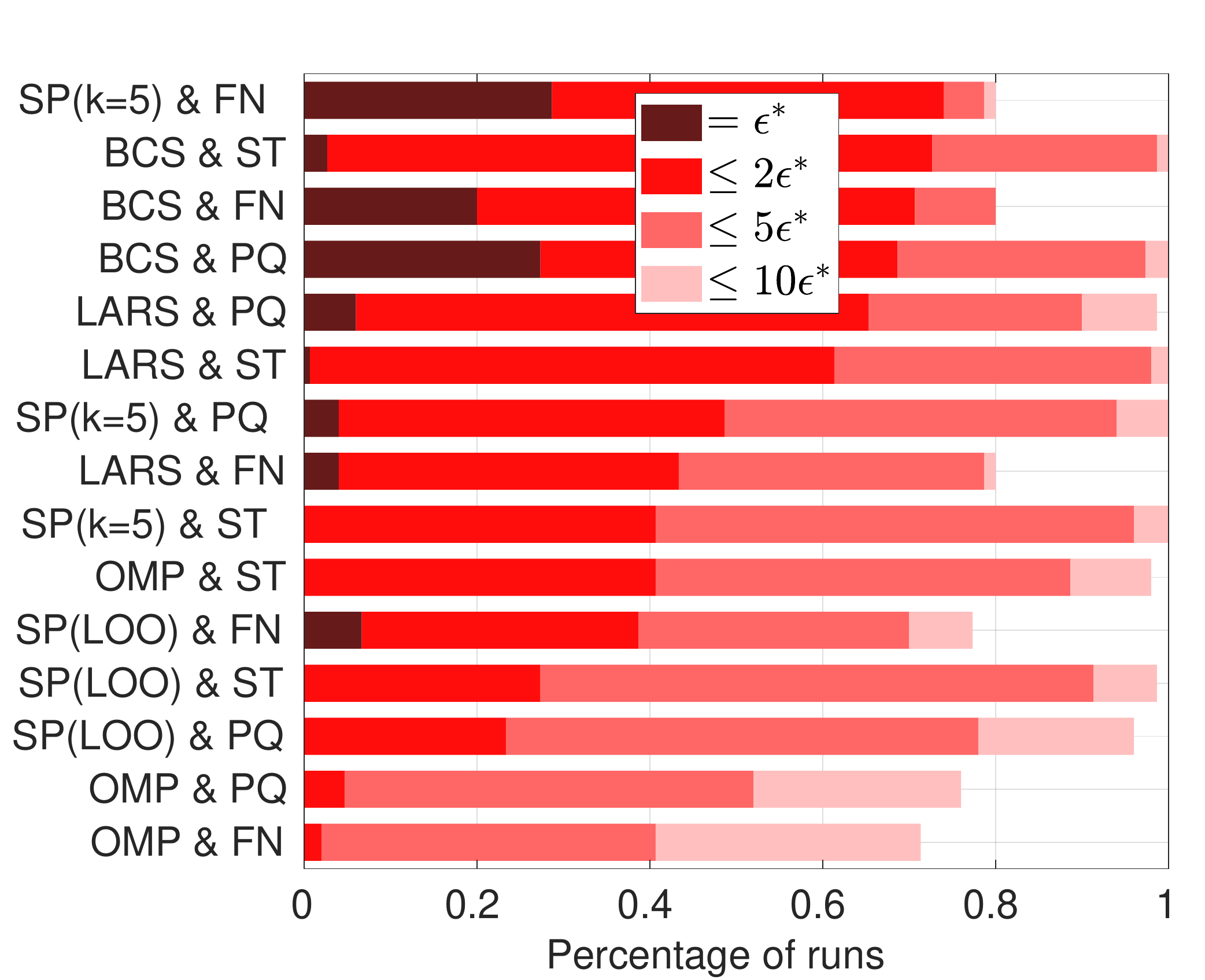}}
	\caption{\textbf{Small ED sizes}. Aggregated results for \textbf{high-dim} models. }
	\label{fig:aggr_highdim_small}
\end{figure}

\begin{figure}[htbp]
	\centering
	{\includegraphics[width=.7\textwidth]{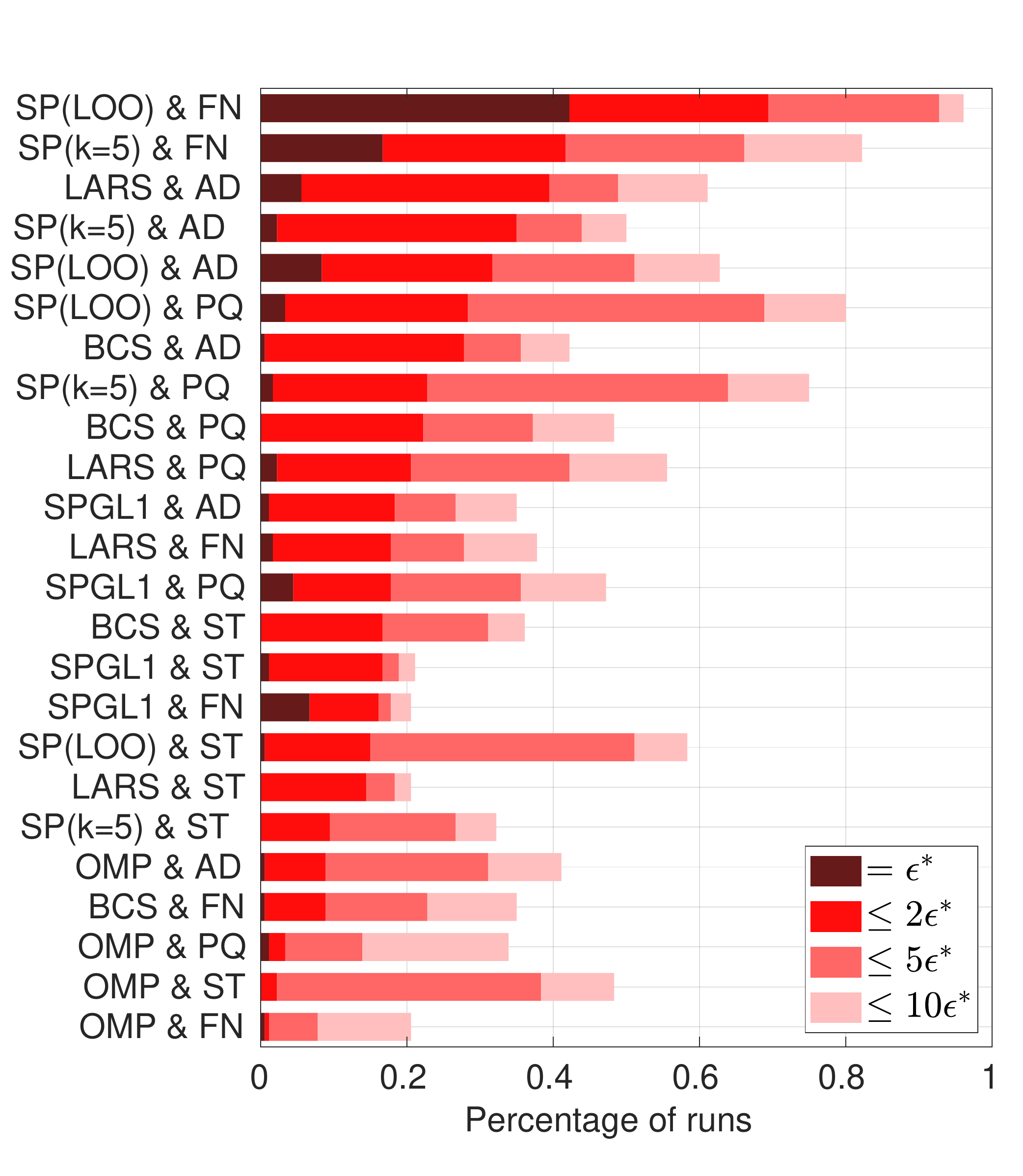}}
	\caption{\textbf{Large ED sizes}. Aggregated results for \textbf{low-dim} models.}
	\label{fig:aggr_lowdim_large}
\end{figure}

\begin{figure}[htbp]
	\centering
	{\includegraphics[width=.7\textwidth]{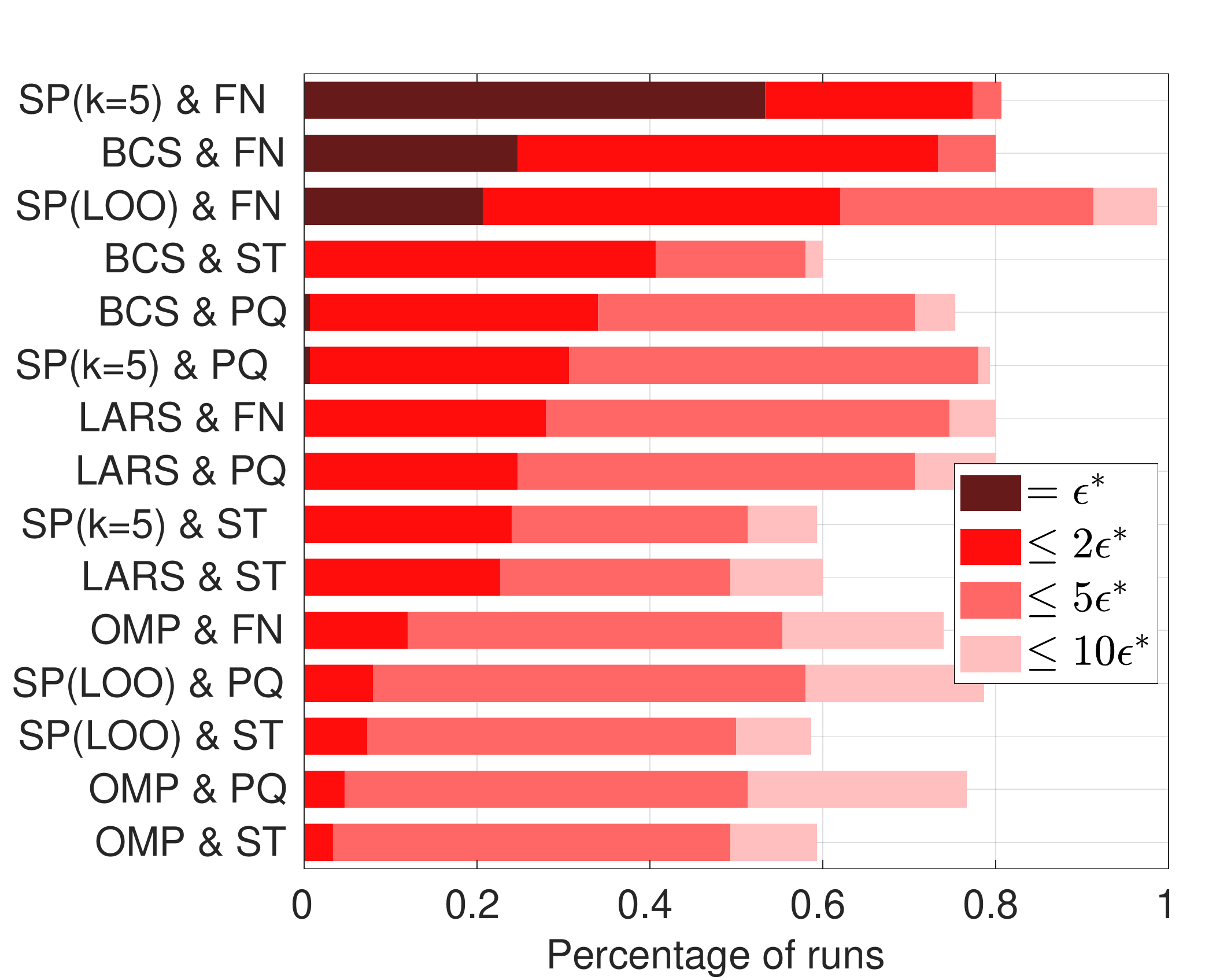}}
	\caption{\textbf{Large ED sizes}. Aggregated results for \textbf{high-dim} models. }
	\label{fig:aggr_highdim_large}
\end{figure}

In Figs.~\ref{fig:aggr_lowdim_small}-\ref{fig:aggr_highdim_large}, we display the aggregated results for all combinations of solvers and sampling schemes (while in Section \ref{sec:BA_benchmark}, we only displayed the six best combinations). We sort the method combinations according to how often they achieved an error within two times the best relMSE (bright red bar). 
For a detailed description of how to read this plot, we refer to Section \ref{sec:BA_benchmark}.

\end{document}